\documentclass[nofootinbib,aps,prl,reprint,
superscriptaddress]{revtex4-2}

\newif\ifprintauthors
\printauthorsfalse

\usepackage[dvipsnames]{xcolor}

\usepackage{orcidlink}
\usepackage{amsmath}
\usepackage{graphicx}
\usepackage{color}
\usepackage[normalem]{ulem}

\usepackage{siunitx}
\DeclareSIUnit\parsec{pc}
\DeclareSIUnit\Mpc{\mega\parsec}

\usepackage{txfonts}
\newlength{\capheight}
\newcommand\gwtc[1][?]{\mbox{GWTC\if#1?\else-#1\fi}}

\usepackage[markup=underlined,commentmarkup=uwave]{changes}
\definechangesauthor[name={Will Farr}, color={teal}]{wf}
\definechangesauthor[name={Alessandra Buonanno}, color={cyan}]{ab}
\definechangesauthor[name={Vasco Gennari}, color={brown}]{vg}
\definechangesauthor[name={Lorenzo Pompili}, color={orange}]{lp}
\definechangesauthor[name={Keefe Mitman}, color={purple}]{km}
\definechangesauthor[name={Geraint Pratten}, color={dodgerblue}]{gp}

\definecolor{dodgerblue}{HTML}{1E90FF}
\definecolor{rsred}{HTML}{BA0C2F}
\newcommand{\citeme}[1]{\textcolor{rsred}{CITE}}

\newcommand{\soft}[1]{\texttt{#1}}

\usepackage{acronym}
\newacro{BH}[BH]{black hole}
\newacro{BBH}[BBH]{binary black hole}
\newacro{GW}[GW]{gravitational-wave}
\newacro{PN}[PN]{post-Newtonian}
\newacro{GWTC-3.0}[GWTC-3.0]{third Gravitational-Wave Transient Catalog}
\newacro{GWTC-4.0}[GWTC-4.0]{fourth Gravitational-Wave Transient Catalog}
\newacro{HPD}{highest posterior density}
\newacro{SNR}[SNR]{signal-to-noise ratio}
\newacro{GR}[GR]{general relativity}
\newacro{NR}[NR]{numerical-relativity}
\newacro{IMR}[IMR]{inspiral--merger--ringdown}
\newacro{QNM}[QNM]{quasi-normal mode}
\newacro{QNMRF}[\soft{QNMRF}]{QNM rational filter}
\newacro{FAP}[FAP]{false-alarm probability}
\newacro{PCA}[\soft{PCA}]{Principal-Component Analysis}
\newacro{FTI}[\soft{FTI}]{Flexible Theory Independent}
\newacro{TIGER}[\soft{TIGER}]{Test Infrastructure for GEneral Relativity}
\newacro{LVK}[LVK]{LIGO-Virgo-KAGRA}

\newacro{2DS}[2DS]{two damped sinusoids}
\newacro{220}[220]{Kerr $\ell = m = 2$, $n=0$}
\newacro{220+221}[220+221]{Kerr $\ell = m = 2$, $n=0$ and $1$}
\newacro{220+221+222}[220+221+222]{Kerr $\ell = m = 2$, $n=0$, $1$, and $2$}
\newacro{220+p221}[220+p221]{Kerr $\ell = m = 2$, $n=0$ and $1$ with the $n=1$ frequency and damping time perturbed}

\newcommand{\eventname}{GW250114\xspace}
\newcommand{\eventnamefull}{GW250114\_082203\xspace}

\newcommand{\ringdown}{{\soft{ringdown}}\xspace}
\newcommand{\pyring}{{\soft{pyRing}}\xspace}
\newcommand{\cpnest}{\soft{cpnest}\xspace}
\newcommand{\Asimov}{\soft{Asimov}\xspace}
\newcommand{\BayesWave}{\soft{BayesWave}\xspace}
\newcommand{\bilby}{\soft{Bilby}\xspace}
\newcommand{\bilbyTGR}{\soft{Bilby\_TGR}\xspace}
\newcommand{\dynesty}{\soft{dynesty}\xspace}
\newcommand{\arviz}{\soft{arviz}\xspace}
\newcommand{\astropy}{\soft{astropy}\xspace}
\newcommand{\jupyter}{\soft{jupyter}\xspace}
\newcommand{\ipython}{\soft{ipython}\xspace}
\newcommand{\matplotlib}{\soft{matplotlib}\xspace}

\newcommand{\pesummary}{\soft{pesummary}\xspace}
\newcommand{\sxs}{\soft{sxs}\xspace}

\newcommand{\hfivepy}{\soft{h5py}\xspace}
\newcommand{\lalsuite}{\soft{lalsuite}\xspace}
\newcommand{\gwpy}{\soft{gwpy}\xspace}
\newcommand{\pandas}{\soft{pandas}\xspace}
\newcommand{\scipy}{\soft{scipy}\xspace} 
\newcommand{\seaborn}{\soft{seaborn}\xspace}
\newcommand{\tqdm}{\soft{tqdm}\xspace} 
\newcommand{\numpy}{\soft{numpy}\xspace}
\newcommand{\python}{\soft{python}\xspace}
\newcommand{\pyseobnr}{\soft{pySEOBNR}\xspace}
\newcommand{\qnmcode}{\soft{qnm}\xspace}

\newcommand{\pseobnr}{\soft{pSEOBNR}\xspace}
\newcommand{\seobnrphm}{\soft{SEOBNRv5PHM}\xspace}
\newcommand{\phenomxphm}{\soft{IMRPhenomXPHM\_SpinTaylor}\xspace}
\newcommand{\nrsur}{\soft{NRSur7dq4}\xspace}
\newcommand{\seobnrhmrom}{\soft{SEOBNRv5HM\_ROM}\xspace}
\newcommand{\kerrpostmerger}{\soft{KerrPostmerger}\xspace}
\newcommand{\sxsThreeSixOneSeven}{\soft{SXS:BBH:3617}\xspace}

\usepackage{xspace}

\newcommand\GWTwoFiveZeroOneOneFourdCSPolarQninety{40.1}
\newcommand\GWTwoFiveZeroOneOneFourdCSAxialQninety{32.2}

\newcommand{\PrimaryMass}{33.6^{+1.2}_{-0.8}}
\newcommand{\SecondaryMass}{32.2^{+0.8}_{-1.3}}
\newcommand{\PrimarySpinMag}{\leq0.24}
\newcommand{\SecondarySpinMag}{\leq0.26}

\newcommand{\mulitpoleSNR}{3.6^{+1.4}_{-1.5}}
\newcommand{\Eccentricity}{\leq0.03}

\newcommand{\EventpSEOBFreqFourFour}{-0.06^{+0.25}_{-0.35}}
\newcommand{\EventpSEOBTauFourFour}{0.20^{+0.53}_{-0.69}}
\newcommand{\EventpSEOBFreqTwoTwo}{0.02^{+0.02}_{-0.02}}
\newcommand{\EventpSEOBTauTwoTwo}{-0.01^{+0.10}_{-0.09}}
\newcommand{\HierpSEOBFreqTwoTwo}{0.00^{+0.06}_{-0.06}}
\newcommand{\HierpSEOBTauTwoTwo}{0.16^{+0.18}_{-0.16}}
\newcommand{\EventpSEOBFreqFourFourFull}{503^{+130}_{-185}}
\newcommand{\EventpSEOBTauFourFourFull}{4.7^{+2.1}_{-2.7}}
\newcommand{\EventpSEOBFreqTwoTwoFull}{251.7^{+5.1}_{-5.0}}
\newcommand{\EventpSEOBTauTwoTwoFull}{4.09^{+0.42}_{-0.38}}
\newcommand{\EventpSEOBQuantile}{54.2\%}
\newcommand{\HierpSEOBQuantile}{85.1\%}
\newcommand{\EventThetaJNFolded}{0.78^{+0.19}_{-0.23}}

\newcommand{\GWTwoFiveZeroOneOneFourIMRCTdMf}{{{0.02^{+0.07}_{-0.06}}}}
\newcommand{\GWTwoFiveZeroOneOneFourIMRCTdchif}{{{-0.01^{+0.11}_{-0.11}}}}
\newcommand{\GWTwoFiveZeroOneOneFourIMRCTGRQuantile}{{{49.5\%}}}
\newcommand{\GWTCFourIMRCTdMf}{{{0.01^{+0.07}_{-0.06}}}}
\newcommand{\GWTCFourIMRCTdchif}{{{-0.04^{+0.07}_{-0.07}}}}
\newcommand{\GWTCFourIMRCTGRQuantile}{{{51.1\%}}}

\newcommand{\dfOvertoneThreeM}{\ensuremath{-0.13^{+0.61}_{-0.16}}}
\newcommand{\dfOvertoneSixM}{\ensuremath{0.09^{+0.29}_{-0.30}}}
\newcommand{\dfOvertoneNineM}{\ensuremath{-0.07^{+0.72}_{-0.53}}}

\newcommand{\GWEventFTIDphiZero}{\ensuremath{0.00^{+0.03}_{-0.03}}}

\newcommand{\GWEventFTIDphiThree}{\ensuremath{-0.01^{+0.03}_{-0.02}}}

\newcommand{\GWTCFourFTIDphiZero}{\ensuremath{-0.00^{+0.09}_{-0.09}}}

\newcommand{\GWTCFourFTIDphiThree}{\ensuremath{0.00^{+0.07}_{-0.07}}}

\newcommand{\GWPCAoneStatsTIGER}{\ensuremath{0.02^{+0.05}_{-0.05}}}
\newcommand{\GWPCAoneStatsFTI}{\ensuremath{-0.01^{+0.02}_{-0.02}}}

\newcommand{\GWPCAtwoStatsFTI}{\ensuremath{0.04^{+0.13}_{-0.13}}}

\newcommand{\TEOBlogBF}{{0.54^{+0.18}_{-0.18}}}
\newcommand{\TEOBFreqTwoTwo}{{0.09^{+0.34}_{-0.22}}}
\newcommand{\TEOBTauTwoTwo}{{-0.14^{+0.25}_{-0.23}}}

\newcommand{\nFTIEventsOFourA}{18}
\newcommand{\nTIGEREventsOFourA}{24}
\newcommand{\nIMRCTEventsOFourA}{12}
\newcommand{\nIMRCTEventsGWTCFour}{30}

\newcommand{\NRConfidence}{38}

\newcommand{\QNMModelOneTime}{11}
\newcommand{\QNMModelTwoTime}{6}
\newcommand{\QNMModelThreeTime}{3}

\newcommand{\QNMModelOneStabilityTolerance}{14}
\newcommand{\QNMModelTwoStabilityTolerance}{24}
\newcommand{\QNMModelThreeStabilityTolerance}{40}

\newcommand{\QNMModelOneStabilityRegimeLeft}{11}
\newcommand{\QNMModelOneStabilityRegimeRight}{40}

\newcommand{\QNMModelTwoStabilityRegimeLeft}{6}
\newcommand{\QNMModelTwoStabilityRegimeRight}{34}

\newcommand{\QNMModelThreeStabilityRegimeLeft}{3}
\newcommand{\QNMModelThreeStabilityRegimeRight}{22}
\newcommand{\QNMModelThreeStabilityRegimeRightMTen}{12}

\newcommand{\ResidualsSNR}{6.86}
\newcommand{\ResidualsPValue}{0.34}
\newcommand{\ResidualsFF}{0.996}

\newcommand{\pyRingSixM}{{0.56^{+0.26}_{-0.26}}}
\newcommand{\pyRingEightM}{{2.38^{+0.31}_{-0.31}}}
\newcommand{\pyRingNineM}{{-0.53^{+0.31}_{-0.31}}}
\newcommand{\ringdownSixM}{{2.51^{+0.36}_{-0.26}}}
\newcommand{\ringdownEightM}{{3.47^{+2.02}_{-0.50}}}
\newcommand{\ringdownNineM}{{0.20^{+0.09}_{-0.08}}}

\newcommand{\TwoDSSignificanceSixM}{3.5}

\newcommand{\IMRCTAreaSigmaGauss}{4.8}
\newcommand{\IMRCTdAreaSigmaGauss}{3.7}

\newcommand{\SNRInspiralNRSur}{65}
\newcommand{\SNRPostInspiralNRSur}{40}
\newcommand{\SNRFullNRSur}{76}

\begin{document}

\title{Black Hole Spectroscopy and Tests of General Relativity with \eventname}

\date{\today}
\ifprintauthors
\input{LSC-Virgo-KAGRA-Authors-Feb-2025-prd.tex}
\else
\collaboration{The LIGO Scientific Collaboration, the Virgo Collaboration, and the KAGRA Collaboration}
\email{lvc.publications@ligo.org}
\noaffiliation{}
\fi

\begin{abstract} 
The binary black hole signal GW250114, the loudest gravitational
  wave detected to date, offers a unique opportunity to test  Einstein's general
  relativity (GR) in the high-velocity, strong-gravity regime and probe whether the remnant 
conforms to the  Kerr metric. Upon perturbation, black holes emit a spectrum of damped sinusoids  
with specific, complex frequencies.
Our analysis of the post-merger signal shows that at
  least two quasi-normal modes are required to explain the
  data, with the most damped remaining statistically significant for
  about one cycle. We probe the
    remnant's Kerr nature by constraining the spectroscopic pattern of
    the dominant quadrupolar ($\ell = m = 2$) mode and its first
    overtone to match the Kerr prediction to tens of percent at
    multiple post-peak times. The measured
  mode amplitudes and phases agree with a numerical-relativity
  simulation having parameters close to \eventname. By fitting a
  parameterized waveform that incorporates the full
  inspiral--merger--ringdown sequence, we constrain the fundamental
  $(\ell=m=4)$ mode to tens of percent and bound the quadrupolar frequency
  to within a few percent of the GR prediction. We perform a suite of
  tests---spanning inspiral, merger, and ringdown---finding constraints
  that are comparable to, and in some cases 2--3 times more stringent
  than those obtained by combining dozens of events in the fourth
  Gravitational-Wave Transient Catalog. These results constitute the
  most stringent single-event verification of GR and the Kerr
  nature of black holes to date, and outline the power of black-hole
  spectroscopy for future gravitational-wave observations.
\end{abstract}

 \keywords{KEYWORDS}

\maketitle

\textit{Introduction---} On January 14, 2025, the LIGO detectors~\cite{TheLIGOScientific:2014jea} recorded the loudest \ac{GW} signal to date, \eventnamefull (hereafter \eventname)~\cite{GW250114}. The Virgo~\cite{TheVirgo:2014hva} and KAGRA~\cite{KAGRA:2020tym} interferometers were offline at the time. The high network \ac{SNR} of $\SNRFullNRSur$ makes \eventname an especially powerful probe 
of whether Einstein's theory of \ac{GR}~\cite{Einstein:1915}, and, in particular, its rotating \ac{BH} solution~\cite{Kerr:1963ud}, accurately describe the observed gravitational radiation.

The non-linearity of Einstein's field equations, coupled with the interdependence of metric and matter, the inherent 
gauge freedom, and the complexity of the initial-value problem~\cite{Arnowitt:1962hi,Foures-Bruhat:1952grw} have made solving these equations 
notoriously challenging. Following the spherically symmetric solution~\cite{Schwarzschild:1916uq} in 1916, the search for an exact rotating axisymmetric solution in vacuum 
  spanned nearly 50 years, until the Kerr metric
  breakthrough~\cite{Kerr:1963ud}. This was followed by efforts to establish the
  uniqueness of static and stationary solutions, including their full
  characterization by conserved quantities, such as mass, spin, and
  charge~\cite{Newman:1965my,Israel:1967za,Carter:1971zc,Hawking:1971vc,Robinson:1975bv,Mazur:1982db}. The Kerr 
  metric's simplicity has enabled the derivation of
  unexpected properties, including integrability for geodesic motion~\cite{Carter:1968rr},
  the Penrose process~\cite{Penrose:1969pc,Penrose:1971uk}, and the four laws of \ac{BH} mechanics~\cite{Bardeen:1973gs}. 
The solution's application to rotating \acp{BH} has had a profound impact on
  astrophysics, particularly once quasars were discovered~\cite{Schmidt:1963wkp}. 
The Kerr solution underpins waveform models used to detect
  \acp{GW} and infer properties of dark objects. Thus,
  finding that these objects do not conform to Kerr \acp{BH} would
  have far-reaching implications for both astrophysics and fundamental
  physics.

From shortly after its inception, the theory of \ac{GR} has withstood a broad array of experimental probes. 
Nevertheless, there are open questions associated to \acp{BH}, such as their stability~\cite{Dafermos:2016uzj,TeixeiradaCosta:2019skg, 
Klainerman:2021qzy,Dafermos:2021cbw}, the existence of singularities 
inside their event horizon~\cite{Hawking:1970zqf}, and Hawking's information-loss paradox~\cite{
Almheiri:2020cfm,Raju:2020smc}. Furthermore,  
\ac{GR} is known to be incomplete in the quantum domain and requires a dark 
sector (dark matter and dark energy) to explain cosmological 
observations, motivating continued searches for possible deviations and
viable extensions of the theory~\cite{Will:2014kxa}. Some gravity theories alternative to \ac{GR} admit the 
Kerr metric as a solution~\cite{Berti:2015itd,Yunes:2024lzm}, thus tests of \ac{GR} and tests of Kerr spacetime are complementary.

In the last ten years, \ac{GW} observations from \ac{BBH} coalescences~\cite{LIGOScientific:2025hdt,LIGOScientific:2025slb} (as well as, from binary neutron stars and mixed binaries) have provided a
unique laboratory for testing \ac{GR} in the strong-gravity, dynamical and high-velocity regime, where
potential departures from \ac{GR} are expected to be most pronounced~\cite{1995PhRvL..74.1067B,Will:2014kxa,Berti:2015itd,Yunes:2024lzm}. Since the first detection of a
\ac{BBH} coalescence~\cite{GW150914_paper,LIGOScientific:2016lio}, the growing catalog of \ac{GW} events~\cite{GWTC1,GWTC2,GWTC2p1,GWTC3,LIGOScientific:2025slb} has enabled increasingly
stringent bounds of the inspiral, merger, and ringdown phases~\cite{LIGOScientific:2018dkp,LIGOScientific:2019fpa,LIGOScientific:2020tif,LIGOScientific:2021sio}. 
These results complement other \ac{GR} investigations---Solar System tests, binary-pulsar experiments, observations of massive \acp{BH} at galactic centers, and cosmological measurements~\cite{Will:2014kxa,Freire:2012mg,Kramer:2021jcw,GRAVITY:2018ofz,Do:2019txf,EventHorizonTelescope:2019dse,Clifton:2011jh}---which span low-velocity, quasi-static, weak-field regimes and, in some cases, strong-field environments with self-gravitating bodies.

In vacuum, \acp{BH} in binaries adiabatically and steadily approach each other during the inspiral until they merge, driven by \ac{GW} emission, a purely tensorial radiation in \ac{GR}, dominated by the quadrupolar multipole~\cite{Einstein:1918btx}. 
According to \ac{GR}, after the two \acp{BH} merge, a highly distorted remnant \ac{BH} is formed, which equilibrates by emitting gravitational radiation~\cite{Pretorius:2005gq,Campanelli:2006fy,Baker:2005vv}. In the 1970s, Vishveshwara and Press~\cite{Vishveshwara:1970zz,Press1971}, 
using results from Regge, Wheeler, and Zerilli~\cite{Regge:1957td,Zerilli:1970se}, 
made a significant discovery. In response to an incoming pulse of radiation, \acp{BH} ring the
spacetime, emitting a superposition of damped sinusoids with discrete frequencies and decay times, depending solely on the 
intrinsic properties of the \ac{BH}, notably its mass and spin. This follows from the no-hair theorem~\cite{Kerr:1963ud,Newman:1965my,Israel:1967za,Carter:1971zc,Hawking:1971vc,Robinson:1975bv,Mazur:1982db}, which states that in four-dimensional vacuum \ac{GR}, a stationary \ac{BH} that is non-singular outside the horizon  
is fully characterized by its mass and spin. 
Since those pioneering works~\cite{Vishveshwara:1970zz,Press1971}, using sophisticated analytical and numerical methods in \ac{BH} perturbation
theory~\cite{Teukolsky:1973ha, Chandrasekhar:1975zza, Detweiler:1977gy,Detweiler:1980gk,
Leaver:1985}, the full first-order \ac{BH} spectrum has been computed for rotating \acp{BH}, 
revealing also the presence of \ac{GW} tails~\cite{Price:1971fb, Price:1972pw} at
late times. It was also noted~\cite{Detweiler:1980gk} that with the advent of \ac{GW} astronomy, detecting \ac{BH}'s \ac{QNM} frequencies 
could confirm their existence with a certainty comparable to the way the 21 cm line unequivocally identifies interstellar hydrogen.
This started the \ac{BH}-spectroscopy program~\cite{Detweiler:1980gk,Nollert:1999ji,Kokkotas:1999bd,Dreyer:2003bv,Berti:2005ys,Gossan:2011ha,Meidam:2014jpa,Berti:2025hly}. A measurement of the frequency and damping rate of a single mode suffices to constrain the final-state mass and spin. Measurements of multiple mode frequencies and damping times enable a test of the no-hair theorem~\cite{Penrose:1969pc, Penrose:2002col, 2002nmgm.meet...28K,Chrusciel:2012jk} through the consistency of the modes' properties with the Kerr prediction~\cite{Kerr:1963ud}. In principle, the \acp{QNM} are affected by electromagnetic 
charges, but we expect the latter to be negligible for astrophysical \acp{BH}~\cite{Gibbons:1975kk, Blandford:1977ds, 1982PhRvD..25.2509H, Carullo:2021oxn}. Several studies in the last few years have investigated the presence of \acp{QNM} in \ac{GW} data claiming different levels of significance~\cite{LIGOScientific:2016lio,Carullo:2019flw,Isi:2019aib,Capano:2021etf,LIGOScientific:2021sio,Cotesta:2022pci,Siegel:2023lxl,Gennari:2023gmx}.

In this \textit{Letter}, we perform several studies of \eventname, aimed at constraining deviations from the \ac{GR} predictions 
throughout the inspiral, merger and ringdown; the Kerr nature of the components in the binary, and of the remnant via \ac{BH} spectroscopy.
Extending the recent work in~\citet{GW250114}, we investigate the post-merger stage with different ringdown models and methods. We  
corroborate our findings for the amplitudes and phases of the quadrupolar \acp{QNM} with results of a \ac{NR} simulation 
having parameters close to \eventname. We  bound the spectroscopic pattern of the dominant quadrupolar mode and its first overtone to match the Kerr prediction at multiple post-peak times, while constraining, for the first time, the hexadecapolar fundamental-mode frequency.  
We set the most stringent bounds on the \ac{PN} parameters determining the \ac{GW} phasing during the inspiral, perform signal consistency tests and 
use them to assess the increase of the \acp{BH} area from the inspiral to the ringdown at high credibility. Overall, \ac{GR} and the Kerr metric 
once again remain empirically unshaken.

\textit{\eventname---} Using the \ac{IMR} quasi-circular, spin-precessing \nrsur model~\cite{Varma:2019csw}, \citet{GW250114} found that the wave morphology is consistent with a \ac{BBH} with component masses $\PrimaryMass\,M_\odot$ and $\SecondaryMass\,M_\odot$ and dimensionless spin magnitudes $\PrimarySpinMag$ and $\SecondarySpinMag$ ($90 \%$ credible intervals). Its eccentricity is constrained to $e\,\Eccentricity$ at a reference frequency of $13.33~\mathrm{Hz}$, using eccentric aligned-spin models~\cite{Gamboa:2024hli, Nagar:2024dzj}.

\textit{\ac{BH} spectroscopy of the remnant alone---} In \ac{GR}, after a dynamical phase surrounding a \ac{BBH} merger, the post-merger signal is dominated by a superposition of exponentially damped sinusoids corresponding to QNMs of the
final Kerr remnant with redshifted mass $M_{\mathrm{f}}\left( 1 + z\right)$, where $z$ is the cosmological redshift, 
and dimensionless spin $\chi_{\mathrm{f}}$~\cite{Pretorius:2005gq,Campanelli:2006fy,Baker:2005vv}. 

\begin{figure}
	\includegraphics[width=\linewidth]{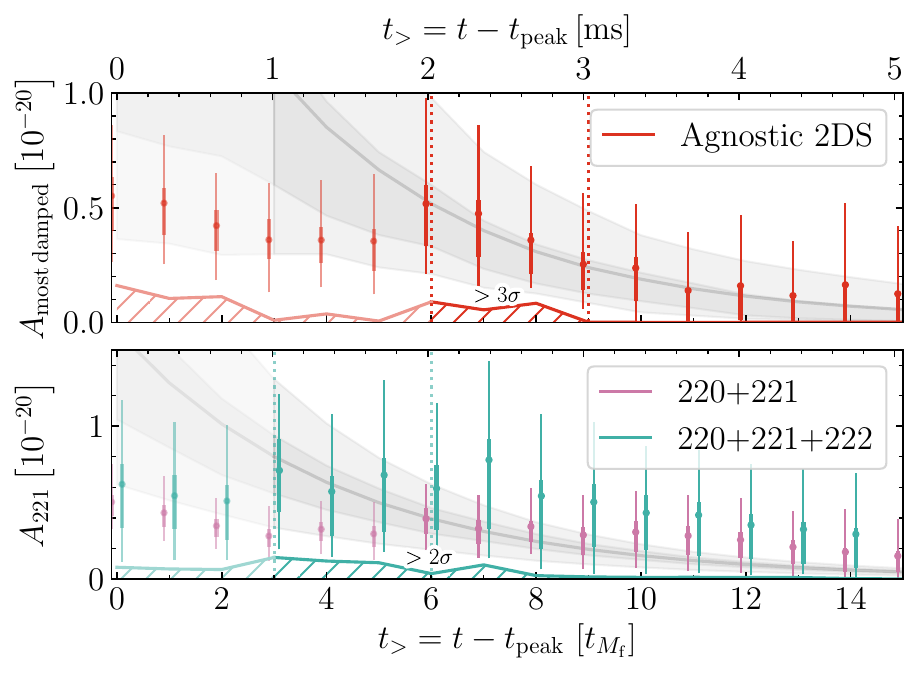}
	\caption{\label{fig:221_over_time} Consistency of post-merger data with two \acp{QNM}. \emph{Top:} The inferred amplitude of the
			most rapidly decaying damped sinusoid for the \acs{2DS} model at different fit times as measured by \ringdown. The dots
			indicate posterior medians, while the thick (thin) bars indicate the 50\% (90\%) credible
			interval; for each time, they have been offset slightly for clarity.
			The gray bands predict the median, 50\%, and 90\% credible ranges of amplitudes
			over time from the $t_{>} = 6 t_{M_{\mathrm{f}}}$ fit, marked by the first dotted line. The second dotted line at $t_{>} = 9 t_{M_{\mathrm{f}}}$ indicates the latest time that the 90\% credible range of the amplitude is distinct from zero. The hatched region shows the greater than $3\sigma$-equivalent credible region for the amplitude being strictly positive.
			\emph{Bottom:} The inferred amplitude of the 221 mode for the 220+221
			(pink) and 220+221+222 (green) models as measured by \pyring. The first dotted line indicates
			$t_{>} = 3 \, t_{M_{\mathrm{f}}}$, beyond which 220+221+222 fits yield
			221 amplitudes consistent with that at $t_{>} = 6 t_{M_{\mathrm{f}}}$. The second dotted line indicates $t_{>} = 6 \, t_{M_{\mathrm{f}}}$. The hatched curve shows the $>2\sigma$-equivalent credible region.}
\end{figure}

While the prompt response~\cite{Leaver:1986gd,Nollert:1999ji,Andersson:1996cm},
dynamical effects~\cite{Zhu:2024dyl,Chavda:2024awq,DeAmicis:2025xuh},
higher-order perturbative terms~\cite{London:2014cma, Mitman:2022qdl,
Cheung:2022rbm, Ma:2024qcv}, and nonlinearities~\cite{Pretorius:2005gq} are
expected to contribute to the early post-merger signal, they are subdominant to
the QNMs at sufficiently late times~\cite{Buonanno:2006ui,Berti:2007fi}. 
Gravitational-wave tails~\cite{Price:1971fb, Price:1972pw,Ching:1995tj, Ma:2024hzq, DeAmicis:2024not, DeAmicis:2024eoy} dominate at much later times. Here, because of theoretical modeling uncertainties, we neglect these other contributions, focusing on the exponentially decaying sinusoidal \ac{QNM} component 
and assume that the plus and cross polarizations of the post-merger signal at
the detectors take the form~\cite{Berti:2005ys,LIGOScientific:2016lio}
\begin{align}
    \label{eq:h-from-modes}
     h_{+} - i h_{\times} =\sum_{\substack{\ell \geq 2 \\ 0\leq m \leq \ell \\ n \geq 0}} e^{-t/\tau_{\ell m n}} \left ( A_{\ell m n}^{\rm R} e^{-2\pi i f_{\ell m n} t} 
 + A_{\ell m n}^{\rm L} e^{2 \pi i f_{\ell m n} t} \right ),
\end{align}
where the complex numbers $A_{\ell m n}^{\rm R}$ and $A_{\ell m n}^{\rm L}$ encode the
amplitudes and phases of the right- and left-circularly polarized components of
the mode, and depend on the excitations imprinted on the spacetime by the progenitors' dynamics~~\cite{Kamaretsos:2012bs,London:2018gaq,JimenezForteza:2020cve,Cheung:2023vki,MaganaZertuche:2024ajz,Nobili:2025ydt}.
The frequency $f_{\ell m n}$ and the damping time $\tau_{\ell m n}$ correspond to the Kerr \ac{QNM} frequencies and damping
times, and are indexed by the angular-mode numbers $\ell$ and $m$, and the radial-overtone
number $n$~\cite{Teukolsky:1973ha, Chandrasekhar:1975zza,Detweiler:1977gy}. The amplitude of the
elliptically polarized mode at time $t=0$, which we take to be the starting time
of fits of Eq.~\eqref{eq:h-from-modes} to \eventname, is $A_{\ell m n} =
|A_{\ell m n}^{\rm R} | + | A_{\ell m n}^{\rm L}|$~\cite{Isi:2021iql}.
Given \eventname's properties, in Eq.~\eqref{eq:h-from-modes} we have neglected retrograde modes,
which are known to be less excited and less important than prograde modes for this type of system~\cite{JimenezForteza:2020cve, Dhani:2020nik, Li:2021wgz}. 

We employ \ringdown \cite{Isi:2019aib,Isi:2021iql} and \pyring \cite{pyRing,Carullo:2019flw} to fit Kerr models of the form 
Eq.~\eqref{eq:h-from-modes}, as well as models with agnostic complex frequencies, to the post-peak 
signal from \eventname, starting at a range of times, $t_{>}$, after $t_{\mathrm{peak}}$: the time at
which \nrsur's maximum likelihood strain magnitude over the two-sphere achieves its 
maximum~\cite{GW250114,Carullo:2018sfu,Carullo:2019flw,Isi:2019aib,Isi:2021iql,Varma:2019csw}.
We adopt the reference peak time, mass, and sky location from the ringdown fits
in~\citet{GW250114}, with $t_{M_{\mathrm{f}}} = (1+z) G M_{\mathrm{f}} / c^3 =
0.337 \, \mathrm{ms}$. Henceforth, we refer to the
Kerr modes in Eq.~\eqref{eq:h-from-modes} as $\ell m n$, and a sum of multiple
modes is denoted as $\ell m n + \ell' m' n'$. The \ringdown and \pyring codes have different native priors and run settings, and therefore produce results for this analysis that differ~\cite{GW250114}.  The results from both codes are sufficiently qualitatively similar, however, that in most cases we show results from only one or the other code throughout.

\begin{figure}
	\includegraphics[width=\linewidth]{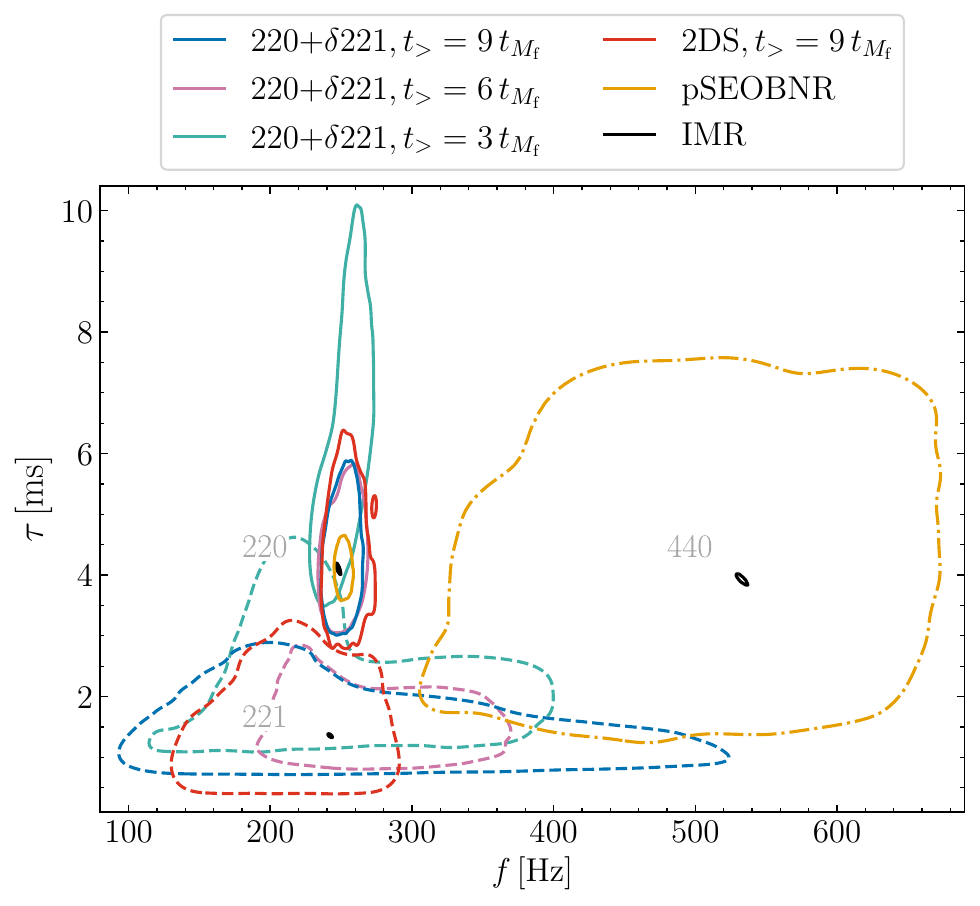}
	\caption{\label{fig:freq_tau_over_time}
		\ac{BH} spectroscopy with different \acp{QNM}. The $90 \%$ credible regions for the 220 (solid) and 221 (dashed) \ac{QNM} frequencies and damping times as measured when fitting the 220+$\delta$221 model at $t_>\in\{3,6,9\}t_{M_{\mathrm{f}}}$ (green, pink, blue) and the \acs{2DS} model at $t_>=9t_{M_{\mathrm{f}}}$ (red). In orange, the constraints from the \pseobnr analysis for the 220 (solid) and 440 (dot--dashed) \acp{QNM}. Black curves indicate the 220, 221, and 440 frequencies and damping times inferred from the \ac{IMR} remnant mass and spin posteriors.}
\end{figure}

As shown in~\citet{GW250114}, the post-peak data of \eventname is
consistent with the 220 and 221 QNMs. Here, we further motivate this identification, 
extend this study using other ringdown models and methods, and test the remnant's Kerr nature at different post-peak times. In the Supplemental Material, we also validate the use of Eq.~(\ref{eq:h-from-modes}) at the times at which we apply our QNM models by comparing to \ac{NR} waveforms. 
We start by adopting an agnostic sum of \ac{2DS} whose complex amplitudes, frequencies, and damping
times are arbitrary~\cite{LIGOScientific:2016lio, LIGOScientific:2020tif,
LIGOScientific:2021sio}. In the top panel of Fig.~\ref{fig:221_over_time}, we
show that the amplitude of the more rapidly decaying damped sinusoid at various
fit start times $t_>$ is bounded away from zero at $>3\sigma$ until $t_> > 9 \,
t_{M_{\mathrm{f}}}$, and it is non-zero at $\TwoDSSignificanceSixM \sigma$ at $t_> = 6 \,t_{M_{\mathrm{f}}}$.
By examining the frequencies and damping times of the two damped sinusoids shown by the red contours in Fig.~\ref{fig:freq_tau_over_time} (the other contours will be discussed later), we find that they are broadly consistent with the 220 and 221 \acp{QNM} predicted from 
the remnant mass and spin inferred from the \ac{IMR} analysis of \eventname in \ac{GR}.
This motivates fitting \eventname with the 220+221 model. In the bottom panel of Fig.~\ref{fig:221_over_time}, which
shows the amplitude of the 221 \ac{QNM} at various fit start times, we
see that for the 220+221 model the 221 \ac{QNM}'s amplitude is not only bounded away from zero, but is consistent
with the exponential decay expected within the error bounds of the 221 \ac{QNM} fit at $t_> = 6t_{M_{\rm f}}$ for $t_> \geq 6t_{M_{\rm f}}$. At earlier times, the amplitude deviates from its expected value, suggesting a breakdown of the 220+221 model.
Motivated by analysis of \ac{NR} simulations~\cite{Giesler:2024hcr}, we find that if the 222 \ac{QNM} is added to
the fit, then consistency at the 90\% level with the amplitude at
$t_> = 6t_{M_{\rm f}}$ is obtained until $t_> \geq 3t_{M_{\rm f}}$, even though the amplitude 
of the 222 \ac{QNM} is never confidently measured
away from zero in these fits. These findings suggest that the data is indeed consistent with the 220 and 221 QNMs 
over a range of times.

In the Supplemental Material, we corroborate these results using a \ac{NR} simulation in \ac{GR}. Specifically, 
we demonstrate that the results in Fig.~\ref{fig:221_over_time} are broadly consistent with that of a
simulated signal of a \ac{NR} simulation with parameters close to
  \eventname; we show that the relative amplitudes and phases of \eventname's 220 and 221 \acp{QNM} are
consistent with those of the \ac{NR} simulation at the $\geq\NRConfidence\%$ credible level for $t_{>}\in[3,9]t_{M_{\mathrm{f}}}$. Separately, using \eventname we also find consistency among the final mass and spin computed
with the \ac{IMR} analysis and the various \ac{QNM} models we fit.

In Fig.~\ref{fig:freq_tau_over_time}, we probe the Kerr nature of 
the remnant. We examine the fit to the post-peak data of \eventname with a
220+$\delta$221 model whose 221 \ac{QNM}'s frequency and damping time
($f_{221}$ and $\tau_{221}$) are allowed to vary from their Kerr values
($f_{221,\rm Kerr}$ and $\tau_{221,\rm Kerr}$) by
$\delta\hat{f}_{221}=\ln(f_{221}/f_{221,\rm Kerr})$ and
$\delta\hat{\tau}_{221}=\ln(\tau_{221}/\tau_{221,\rm Kerr})$.
As suggested by the \ac{2DS} fit, the data are 
particularly consistent with the 220 and 221 \acp{QNM} predicted by the remnant mass and spin 
inferred by the full \ac{IMR} analysis. More specifically, we constrain
$\delta\hat{f}_{221}=\dfOvertoneThreeM$ at $t_>=3t_{M_{\rm f}}$,
$\delta\hat{f}_{221}=\dfOvertoneSixM$~\cite{GW250114} at $t_>=6t_{M_{\rm f}}$, and
$\delta\hat{f}_{221}=\dfOvertoneNineM$ at $t_>=9t_{M_{\rm f}}$, all at the 90\%
credible level. However, the recovered amplitude of the 221 \ac{QNM} is
not consistent with values at later times in the $3 t_{M_\mathrm{f}}$ fit, indicating that the model may be fitting other content at these times. 
Previously, analysis using the \mbox{\ac{GWTC-3.0}}~\cite{LIGOScientific:2021sio} set $\delta\hat{f}_{221}= 0.01^{+0.27}_{-0.28}$ by 
analyzing the data from the peak onward using \pyring, and hierarchically combining results from 21 
events~\cite{Isi:2019asy}. In the following section, we will return to this figure to discuss the \pseobnr analysis.

Additionally, we perform an analysis that does not measure the mode amplitudes, but 
filters out successive Kerr \acp{QNM} from the data in the frequency domain: the \ac{QNMRF}~\cite{Ma:2022wpv,Ma:2023cwe,Ma:2023vvr}.
We adopt a hybrid Bayesian-like approach, with a detection
statistic $\mathcal{D}$ that is analogous to a logarithmic Bayes factor, yet
differs from the Bayes factors used in other time-domain ringdown analyses (see
details in Supplemental Material). Figure~\ref{fig:qnmrf} shows the difference
between the \ac{QNMRF} detection statistic and the detection statistic
corresponding to a 1\% \ac{FAP} for the 220+221 and 220+221+222 models.  Like
the other analyses, the \ac{QNMRF} finds strong support for the 220+221 model
over the single-mode 220 model for times $t_{>} \leq 10 \, t_{M_{\mathrm{f}}}$.
Additionally, the three-mode 220+221+222 model is more weakly preferred over the
220+221 model from $t_{<} = 1 \, t_{M_\mathrm{f}}$ to $t_{<} = 5 \,
t_{M_{\mathrm{f}}}$. This provides independent evidence that \eventname
is consistent with the 220 and 221 \acp{QNM}.

\begin{figure}
	\includegraphics[width=\linewidth]{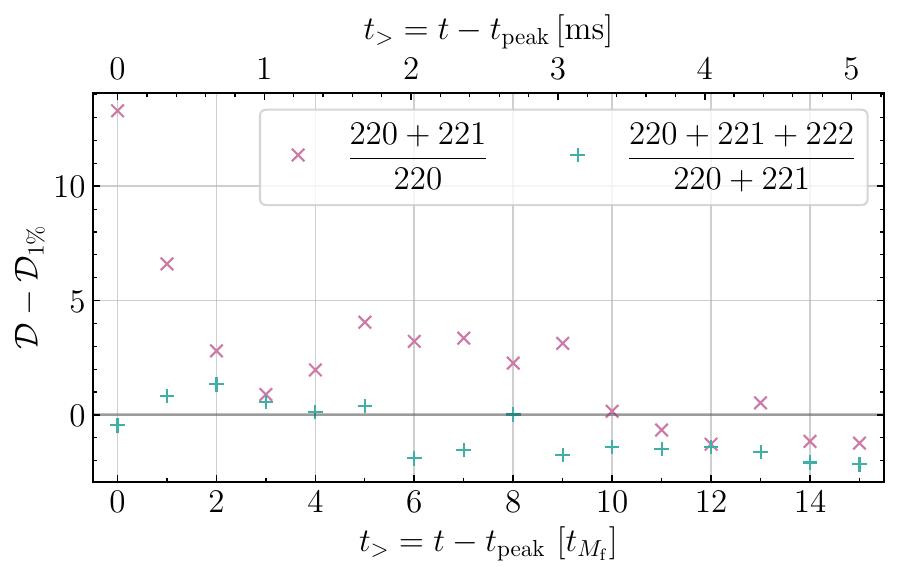}
	\caption{\label{fig:qnmrf} Filtering out two \acp{QNM} in the post-merger data. Detection statistics for the \ac{QNMRF} analysis
	with varying mode content relative to the 1\% \ac{FAP} threshold over time
	\cite{Lu:2025mwp}.  At times $t_{>} \leq 10 \, t_{M_{\mathrm{f}}}$, the
	\ac{QNMRF} finds strong support for the 220+221 model over the single-mode
	220 model.  The 220+221+222 model is weakly preferred over the 220+221 model from
	$t_{<} = 1 \, t_{M_\mathrm{f}}$ to $t_{<} = 5 \, t_{M_{\mathrm{f}}}$.}
\end{figure}

\textit{\ac{BH} spectroscopy with full signal--} 
So far, we have treated the complex amplitudes of the \acp{QNM} in Eq.~\eqref{eq:h-from-modes} as free
parameters, and directly constrained them from the data using only the post-peak signal. When these amplitudes
are instead predicted from the binary's properties using \ac{NR} simulations in
\ac{GR}, consistency tests are feasible even with a single
mode~\cite{LIGOScientific:2021sio, Brito:2018rfr, Gennari:2023gmx}. This results
in a more stringent test of \ac{GR}, at the cost of stronger assumptions about
the emission process: most existing amplitude models assume the perturbed
\ac{BH} originates from a binary merger, and are restricted to quasi-circular
orbits~\cite{Kamaretsos:2012bs,London:2018gaq,JimenezForteza:2020cve,Cheung:2023vki,MaganaZertuche:2024ajz,Nobili:2025ydt}.

Including additional pre-merger information, one can test for deviations in \ac{QNM} frequencies by analyzing a full \ac{IMR} waveform calibrated to \ac{NR} simulations.
The \pseobnr analysis~\cite{Pompili:2025cdc, Maggio:2022hre, Ghosh:2021mrv, Brito:2018rfr} introduces fractional deviations $(\delta \hat{f}_{\ell m 0}$, $\delta \hat{\tau}_{\ell m 0})$ to the frequency and decay time of the fundamental \acp{QNM} in the ringdown description of \seobnrphm~\cite{Ramos-Buades:2023ehm,Pompili:2023tna} as:
\begin{equation}
    \label{eq:nongr_freqs}
f_{\ell m 0} = f_{\ell m 0}^{\text{GR}}\, (1 + \delta \hat{f}_{\ell m 0})\,,\quad \tau _{\ell m 0} = \tau _{\ell m 0}^{\text{GR}}\, (1 + \delta \hat{\tau}_{\ell m 0})\,.
\end{equation}

The \ac{GR} predictions for these quantities are obtained using the final mass and spin of the remnant \ac{BH}, estimated using \ac{NR} fits based on the measured component masses and spins~\cite{Jimenez-Forteza:2016oae,Hofmann:2016yih}. Rather than isolating the post-merger stage, excluding the inspiral and merger phases, the analysis uses the full \ac{IMR} signal, assuming \ac{GR} holds up to the merger. The merger--ringdown model is based on a factorized ansatz: the contributions of the fundamental \acp{QNM}, modified via the parametrization in Eq.~\eqref{eq:nongr_freqs}, are multiplied by phenomenological, time-dependent amplitudes calibrated to \ac{NR}~\cite{Baker:2008mj,Damour:2014yha, Pompili:2023tna}. These amplitudes aim to capture the ringdown prompt response~\cite{Leaver:1986gd,Nollert:1999ji,Andersson:1996cm} and dynamical phase~\cite{Zhu:2024dyl,Chavda:2024awq,DeAmicis:2025xuh}. As for now, this approach enables constraints on fundamental \acp{QNM}, but not on overtones, whose effects are implicitly absorbed by the time-dependent amplitudes rather than being parameterized explicitly.

We first perform an analysis allowing for deviations in the dominant 220 \ac{QNM} only, which has been the focus of previous constraints~\cite{LIGOScientific:2020tif, Ghosh:2021mrv, LIGOScientific:2021sio, Pompili:2025cdc}.
Owing to \eventname's high \ac{SNR} ($\sim \SNRInspiralNRSur$ up to merger and $\sim \SNRPostInspiralNRSur$ post-merger), we also extend the analysis to probe higher fundamental \acp{QNM}. The nearly equal masses and low spins of the binary's components imply that multipoles with odd $m$ are suppressed due to rotational symmetry, while even-$m$ multipoles are expected to be more prominent~\cite{Blanchet:2013haa, Berti:2007fi, Cotesta:2018fcv}. 
The inclination ($\Theta=\EventThetaJNFolded$ rad~\cite{GW250114} at a reference frequency of $20~\mathrm{Hz}$, when folded to $[0, \pi/2]$) and azimuthal phase inferred from \eventname favor the excitation of the $(\ell, |m|)=(4,4)$ multipoles~\cite{GW250114}, which contribute an \ac{SNR} of $\mulitpoleSNR$ to the full \ac{IMR} signal~\cite{Mills:2020thr}.
Therefore, we perform an analysis including deviations in both the 220 and $440$ \acp{QNM}. 
We find minimal correlation between the deviation parameters of the two modes, and in the following we report constraints on $(\delta \hat{f}_{220}, \delta \hat{\tau}_{220})$ from the joint fit. 

\begin{figure}
    \includegraphics[width=\linewidth]{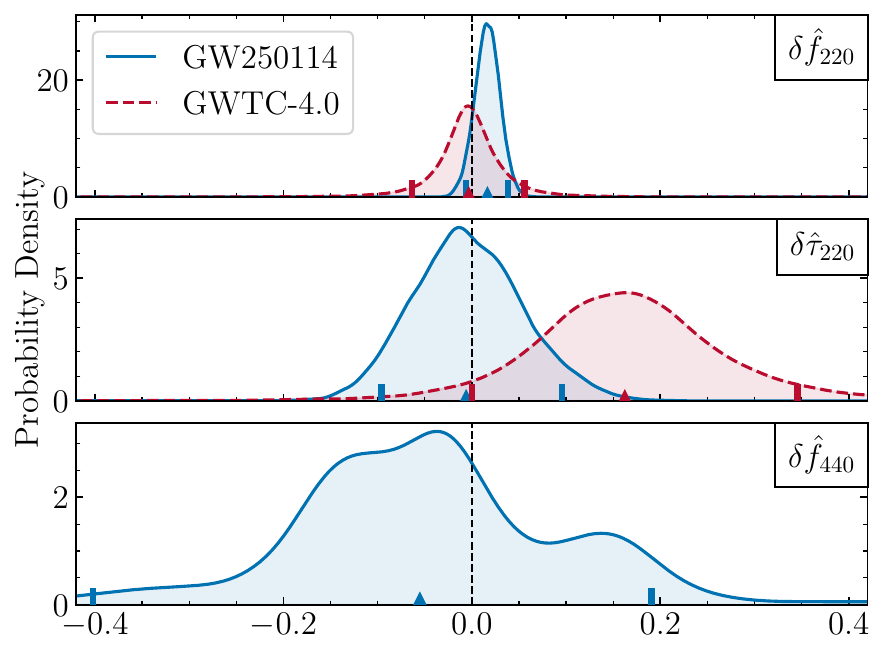}
    \caption{Constraining the 220 and 440 \acp{QNM} using the full signal. 
        Marginalized posterior distributions for fractional deviations in the frequency and damping time of the 220 \ac{QNM} $(\delta \hat{f}_{220},\delta \hat{\tau}_{220})$, 
        and in the frequency of the $440$ \ac{QNM} $(\delta \hat{f}_{440})$, from the \pseobnr analysis of \eventname. Hierarchically combined results from \mbox{\ac{GWTC-4.0}}~\cite{TGR-GWTC4} are also shown. Triangles mark the median values and vertical bars the symmetric $90$\% credible interval.}
\label{fig:ridgeline_domega_dtau}
\end{figure}

The results are summarized in Fig.~\ref{fig:ridgeline_domega_dtau}. The frequencies and damping times of both modes are consistent with the predictions of \ac{GR}, based on the Kerr remnant parameters inferred from the inspiral. The dominant 220 \ac{QNM} is especially well constrained, with $\delta \hat f_{220} = \EventpSEOBFreqTwoTwo$ and $\delta \hat \tau_{220} = \EventpSEOBTauTwoTwo$. Owing to the exceptional \ac{SNR} of this signal, these constraints are roughly twice as stringent as those obtained by hierarchically combining~\cite{Isi:2019asy, Zhong:2024pwb} results from 17 events in the \ac{GWTC-4.0} \cite{TGR-GWTC4}, which have \ac{SNR} above $8$ in the inspiral and post-inspiral stages. That analysis yielded $\delta \hat{f}_{220} = \HierpSEOBFreqTwoTwo$ and $\delta \hat{\tau}_{220} = \HierpSEOBTauTwoTwo$. 
A measure of consistency with \ac{GR} is provided by the \ac{GR} quantile $\mathcal{Q}_{\textrm{GR}}$~\cite{LIGOScientific:2020tif}, which corresponds to the cumulative posterior probability enclosed by the isoprobability surface passing through the \ac{GR} prediction. 
A lower (higher) value of $\mathcal{Q}_{\textrm{GR}}$ indicates better (worse) consistency with \ac{GR}.
For \eventname, the \ac{GR} quantile is $\EventpSEOBQuantile$, lower than for the combined constraints from \mbox{\ac{GWTC-4.0}}~\cite{TGR-GWTC4}, $\HierpSEOBQuantile$. The \mbox{\ac{GWTC-4.0}} results show a mild tension with \ac{GR}~\cite{TGR-GWTC4}, potentially due to non-Gaussian or non-stationary noise~\citep{LIGOScientific:2020tif, Ghosh:2021mrv}, parameter correlations amplified by unrealistic astrophysical priors~\citep{LIGOScientific:2021sio, Payne:2023kwj}, intrinsic variance due to the limited number of events in the catalog~\citep{Pacilio:2023uef}, or unmodeled selection effects that could systematically influence which signals are included in the analysis.

We also constrain, for the first time, the frequency of the subdominant $440$ \ac{QNM}, obtaining
$\delta \hat{f}_{440} = \EventpSEOBFreqFourFour$. The damping time remains
weakly constrained, with $\delta \hat{\tau}_{440} = \EventpSEOBTauFourFour$. Since the \pseobnr method employs 
the entire \ac{IMR} signal, it enforces continuity across the waveform and does not allow the mode amplitudes to vanish.
As a result, the analysis cannot by itself establish whether the $440$ QNM is present in the data.
However, similar constraints do not appear in simulated signals that exclude this mode or
in lower-\ac{SNR} events (see details in the Supplemental Material), lending
support to the interpretation that the constraint is driven by the presence of
the $440$ \ac{QNM} in the ringdown signal of \eventname.

Beyond testing whether each mode is individually consistent with \ac{GR}, one can also test whether both are consistent with originating from the same Kerr remnant, as in classical no-hair-theorem tests~\cite{Detweiler:1980gk,Nollert:1999ji,Kokkotas:1999bd,Dreyer:2003bv, Berti:2005ys, Gossan:2011ha}. In the \pseobnr framework, this is done by reconstructing the complex frequencies of the two modes from Eq.~\eqref{eq:nongr_freqs}, giving $f_{220} = \EventpSEOBFreqTwoTwoFull~\mathrm{Hz}$, $\tau_{220} = \EventpSEOBTauTwoTwoFull~\mathrm{ms}$, $f_{440}=\EventpSEOBFreqFourFourFull~\mathrm{Hz}$, and $\tau_{440}=\EventpSEOBTauFourFourFull~\mathrm{ms}$. 
Their 90\% credible regions are shown as the orange solid and dot--dashed curves in Fig.~\ref{fig:freq_tau_over_time}.
These are then inverted to obtain two separate estimates of the remnant's mass and spin using fitting formulas~\cite{Berti:2005ys}, shown in Fig.~\ref{fig:M_chi_over_time}. We find that the estimates are mutually consistent, and agree with both the \ac{IMR}-inferred values and with results from ringdown remnant-alone analyses.

The presence of the $(4,4)$ multipole is independently supported by isolating the post-peak data in time domain, with the \kerrpostmerger model in \pyring~\cite{Gennari:2023gmx}.
This model has also amplitudes calibrated to \ac{NR} simulations at merger~\cite{Damour:2014yha, DelPozzo:2016kmd, Nagar:2020pcj}. Including the $(4,4)$ mode is mildly favored by a $\log_{10}$ Bayes factor of $\TEOBlogBF$ compared to a model with only the dominant $(2,2)$ mode.
Allowing for deviations from \ac{GR} with the parametrization in Eq.~\eqref{eq:nongr_freqs}, we constrain the 220 \ac{QNM} as $\delta \hat{f}_{220} = \TEOBFreqTwoTwo$ and $\delta \hat{\tau}_{220} = \TEOBTauTwoTwo$, while the 440 mode remains unconstrained. While less stringent than \pseobnr, this analysis quantifies the constraints achievable when isolating the remnant's relaxation.

The stringent \pseobnr results can improve current constraints on gravity theories beyond \ac{GR}~\cite{Yunes:2016jcc,Maselli:2019mjd,Silva:2022srr,Sanger:2024axs, Chung:2025wbg}, and constrain properties of exotic compact objects~\cite{Cardoso:2016oxy,Maggio:2020jml}. As a concrete example, we consider dynamical Chern--Simons (dCS) gravity~\cite{Alexander:2009tp}, a parity-violating extension of GR in which the \ac{QNM} spectrum receives corrections controlled by a coupling length $\sqrt{\alpha_{\rm dCS}}$~\cite{Chung:2025gyg}. Mapping the bound on the 220 \ac{QNM} frequency to the predicted dCS correction yields an approximate constraint on the dCS coupling length of $\sqrt{\alpha_{\rm dCS}} < \GWTwoFiveZeroOneOneFourdCSAxialQninety~(\GWTwoFiveZeroOneOneFourdCSPolarQninety)~\mathrm{km}$, assuming purely axial (polar) perturbations~\cite{Chung:2025gyg} (using the conventions therein). These single-event bounds are competitive with recent ringdown-only analyses of GW150914, GW190521\_074359, and GW200129\_065458~\cite{Chung:2025wbg}. The estimate is based on the posteriors for $\delta \hat{f}_{220}$ and remnant mass and spin from the \pseobnr analysis, reweighted to a prior uniform in the dCS coupling length. More robust constraints could be obtained from a Bayesian analysis that directly uses waveform predictions in dCS gravity~\cite{Silva:2022srr, Chung:2025wbg}.

\textit{Bounding post-Newtonian inspiral parameters---} The inspiral regime can
be treated perturbatively within the \ac{PN} framework~\cite{Blanchet:2014}, an
expansion in powers of $v/c$ where the $n$PN order corresponds to
$\mathcal{O}([v/c]^{2n})$. As the intrinsic parameters of the binary uniquely
determine the PN coefficients $\varphi_i$ in the \ac{GW} phase at each order, we
can construct a consistency test of \ac{GR} by introducing deformation
parameters at each PN
order~\cite{Blanchet:1994ez,Blanchet:1994ex,Arun:2006hn,Yunes:2009ke,Mishra:2010tp,Li:2011cg,Agathos:2013upa}. 
We only consider variations in the individual \ac{PN} coefficients independently, treating them as free coefficients that constrain the
degree to which deviations from \ac{GR} agree with the data.
The inspiral deviations are constructed so as to represent a shift to the non-spinning 
PN coefficient, i.e., $\varphi_i \rightarrow (1 + \delta \hat{\varphi}_i )
\varphi_i^{\rm NS} + \varphi_i^{\rm S}$, where $\varphi_i^{\rm NS}$ denotes the
non-spinning coefficient and $\varphi_i^{\rm S}$ is the spin-dependent part of the PN
coefficient. 
In \ac{GR}, the coefficients at $-1$PN and $0.5$PN are explicitly
zero and should be interpreted as absolute deviations, while the other
coefficients are expressed as fractional deviations. 

As in the \mbox{\ac{GWTC-4.0}} analysis~\cite{TGR-GWTC4}, we use two independent pipelines: \ac{FTI}~\cite{LIGOScientific:2018dkp,Mehta:2022pcn} and \ac{TIGER}~\cite{AgathosEtAl:2014,Meidam:2017dgf,Roy:2025gzv}. 
The pipelines have several methodological differences, including the cut-off frequency at which corrections are turned off and choice of waveform models used. 
The \ac{FTI} pipeline employs the \seobnrhmrom model~\cite{Pompili:2023tna}, which is restricted to aligned-spin configurations, whereas \ac{TIGER} utilizes the precessing-spin waveform model \phenomxphm~\cite{Pratten:2020ceb,Colleoni:2024knd}. Due to technical changes in the pipelines, only events first reported in \mbox{\ac{GWTC-4.0}} that meet the \ac{FTI} or \ac{TIGER} selection criteria are analyzed~\cite{TGR-GWTC4}. 
For \ac{FTI}, we use $\nFTIEventsOFourA$ events and for \ac{TIGER}, we use $\nTIGEREventsOFourA$ events.

\begin{figure}
    \includegraphics[width=\columnwidth]{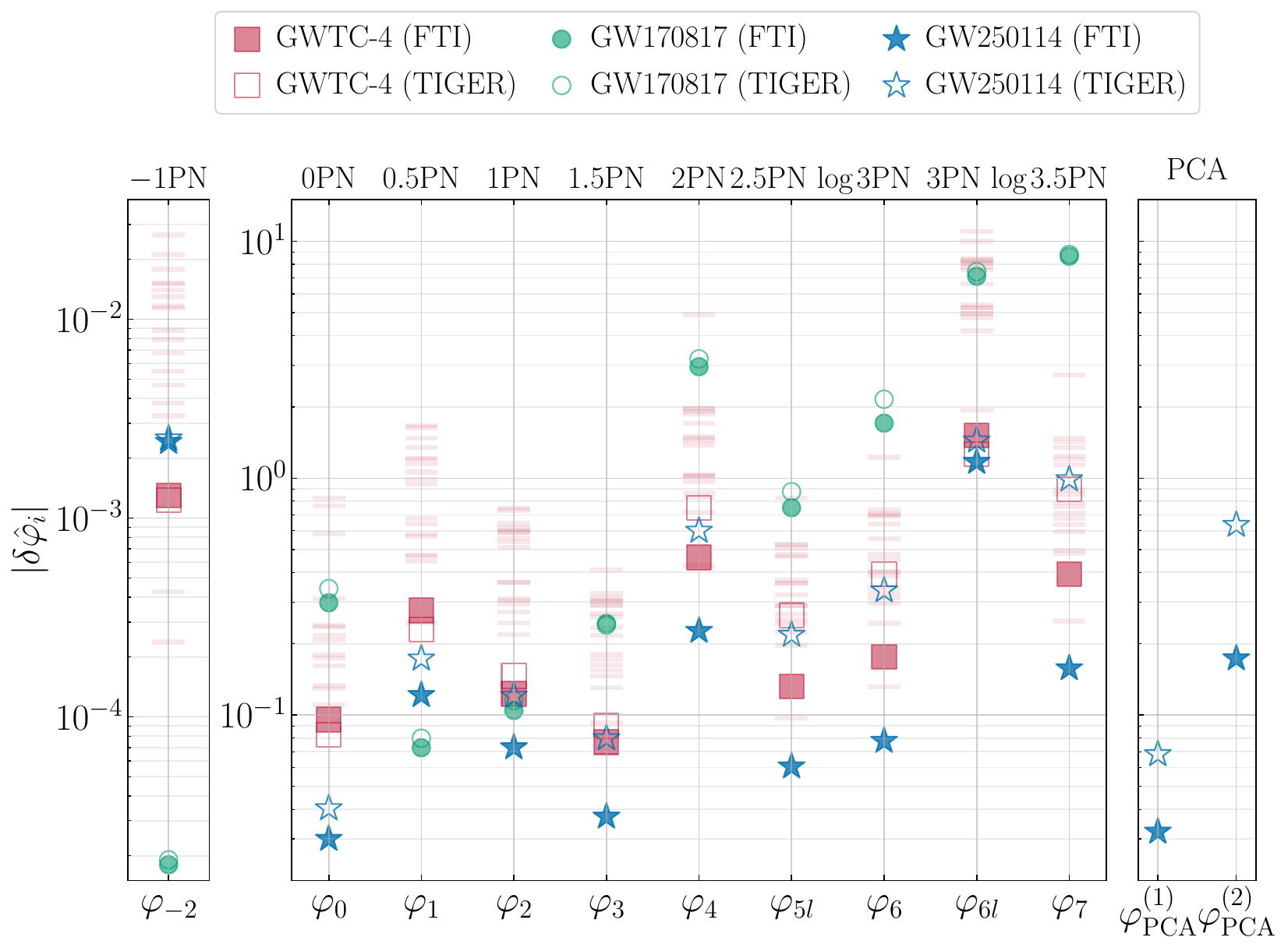}
    \caption{\label{fig:fti_tiger} Constraints on the inspiral phase from \eventname. The $90 \%$ upper bounds on the magnitude $| \delta \hat{\varphi}_i |$ of the \ac{PN} inspiral deviation
    coefficients from $-1$PN to $3.5$PN and the first two leading \ac{PCA} parameters. 
    For \eventname, blue filled (unfilled) stars are for \ac{FTI} (\ac{TIGER}). The horizontal
    red stripes mark the results from individual events from \mbox{\ac{GWTC-4.0}} using \ac{FTI}.
    Bounds obtained by hierarchically combining the results from \mbox{\ac{GWTC-4.0}} are shown in the filled
    (unfilled) red squares for \ac{FTI} (\ac{TIGER})~\cite{TGR-GWTC4}.  
    The right panel shows constraints on the two leading \ac{PCA} parameters
    that capture the dominant modes of deviation across the $1.5$--$3.5$ PN parameter space.}
\end{figure}

The results are summarized in Fig.~\ref{fig:fti_tiger}. 
Due to the significantly higher \ac{SNR} of \eventname during the inspiral phase compared to the rest of the observed \ac{BBH} population, the \ac{FTI} analysis of \eventname provides constraints on a subset of deviations that are 2--3 times more stringent than the joint constraints derived from a hierarchical analysis~\cite{Isi:2019asy,Zhong:2024pwb} of the \mbox{\ac{GWTC-4.0}} results~\cite{TGR-GWTC4}. 
The bounding fractional deviations to the leading-order \ac{PN} and 1.5\ac{PN} coefficients are $\delta \hat{\varphi}_{0} = \GWEventFTIDphiZero$ and $\delta \hat{\varphi}_{3} = \GWEventFTIDphiThree$ for \eventname compared to $\delta \hat{\varphi}_{0} = \GWTCFourFTIDphiZero$ and $\delta \hat{\varphi}_{3} = \GWTCFourFTIDphiThree$ for \mbox{\ac{GWTC-4.0}}. 
\eventname being a shorter signal, we do not place competitive constraints on the dipole $\varphi_{-2}$ compared to those obtained from GW170817~\cite{LIGOScientific:2018dkp}. 
Overall, the \ac{TIGER} pipeline yields constraints that are less stringent than the combined \mbox{\ac{GWTC-4.0}} results~\cite{TGR-GWTC4}, likely due to the differing treatment of the cutoff frequency and transition to merger--ringdown between the pipelines~\cite{TGR-GWTC4}, and the inclusion of spin-precession in the \ac{TIGER} analysis. 

To address the limitation of varying individual \ac{PN} coefficients independently, 
we perform a \acf{PCA} to probe correlated deviations across multiple \ac{PN} orders~\cite{Pai:2012mv,Shoom:2021mdj,Saleem:2021nsb,Mahapatra:2025cwk}, focusing on the 1.5 to 3PN coefficients~\cite{Saleem:2021nsb,Mahapatra:2025cwk}. 

The \ac{PCA} analysis identifies the principal directions of parameter covariance, with the leading component
corresponding to the linear combination of PN coefficients that is best
constrained by the data. The leading \ac{PCA} component can therefore yield
tighter bounds than the individual \ac{PN} coefficients, while the sub-leading
component, being orthogonal to this optimal direction, can have weaker constraints
than the best measured \ac{PN} coefficients. We find that the \ac{FTI} pipeline constrains the leading \ac{PCA} 
component to $\delta\hat{\varphi}_{\mathrm{PCA}}^{(1)} = \GWPCAoneStatsFTI$ (slightly better constrained than $\delta \hat{\varphi}_{3} = \GWEventFTIDphiThree$) and
the sub-leading component to $\delta\hat{\varphi}_{\mathrm{PCA}}^{(2)} =
\GWPCAtwoStatsFTI$, consistent with \ac{GR} (see Fig.~\ref{fig:fti_tiger}). A comparison of the PCA results to \mbox{\ac{GWTC-4.0}} is technically challenging as the PCA components correspond to different linear combinations of \ac{PN} parameters for each event. This means that hierarchical inference requires modelling the joint distribution across all six \ac{PN} coefficients~\cite{Mahapatra:2025cwk}. Further details are presented in the Supplemental Material.

\textit{Signal consistency tests---} 
We now construct an analysis complementary to the other tests,
focusing on the consistency between different portions of the signal,
by employing dimensionless deviation parameters that quantify the
fractional difference between the remnant mass $M_{\mathrm{f}}$ and spin $\chi_{\mathrm{f}}$ 
inferred from the low- ($f < f^{\rm IMR}_{\rm c}$) and high-frequency ($f > f^{\rm
  IMR}_{\rm c}$) portions of the GW signal~\cite{Hughes:2004vw,
  Ghosh:2016qgn, Ghosh:2017gfp}. The cutoff $f^{\rm IMR}_{\rm c}$ is taken to 
be the GW frequency of the $(2, 2)$-mode at the innermost circular orbit of the remnant
Kerr BH~\cite{Bardeen:1972fi}. The remnant properties are calculated using
\ac{NR}-calibrated fits for the final-state~\cite{Hofmann:2016yih,
  Healy:2016lce, PhysRevD.95.064024} applied to the median values of
the redshifted component masses, spin magnitudes, and spin angles as
inferred from the full \ac{IMR} analysis~\cite{LIGOScientific:2019fpa,
  LIGOScientific:2020tif,LIGOScientific:2021sio}.  Inference is
performed in each of the frequency regimes using
\phenomxphm~\cite{Pratten:2020ceb,Colleoni:2024knd}, with priors that
are uniform in component mass and spin magnitude, and isotropic in the
spin orientation.  This choice of priors leads to highly non-uniform
priors on the deviation parameters, and we therefore
reweight the posteriors to impose uniform priors on
them~\cite{LIGOScientific:2020tif, LIGOScientific:2021sio}.
  
\begin{figure}
    \includegraphics[width=\linewidth]{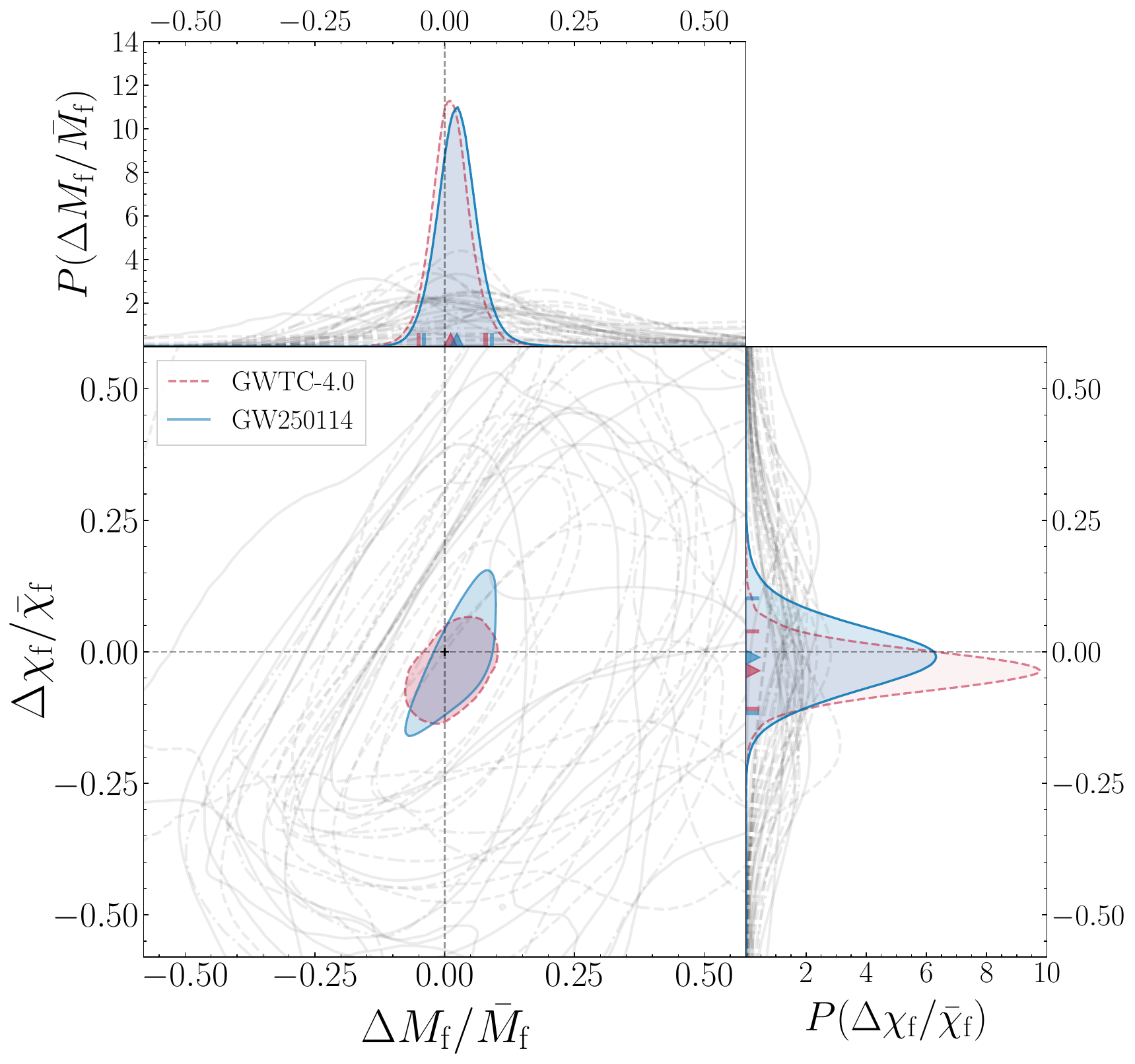}
    \caption{\label{fig:IMRCT_2DPlot} Consistency of \eventname with a \ac{BBH} in \ac{GR} using data from the inspiral and merger--ringdown. The 90\% credible regions of the two-dimensional posteriors on $\Delta M_{\rm f} / \bar{M}_{\rm f}$ and $\Delta \chi_{\rm f} / \bar{\chi}_{\rm f}$ for \eventname (filled blue), with (0, 0) being the expected value for \ac{GR}. The side panels show the marginalized posteriors. The gray two-dimensional posteriors show the results from individual events in \mbox{GWTC-4.0}~\cite{TGR-GWTC4}, while the red shaded posteriors denote constraints derived from a hierarchical inference on these events. Triangles mark the median values and vertical bars the symmetric $90$\% credible interval.}
\end{figure}

If \ac{GR} is valid, and our waveform models are sufficiently accurate, the analysis in each of
these regimes should yield consistent results.  Thus, we have $\Delta
M_{\rm f}/\bar{M}_{\rm f}= 2 (M^{\rm insp}_{\rm f} - M^{\rm
  postinsp}_{\rm f})/(M^{\rm insp}_{\rm f} + M^{\rm postinsp}_{\rm
  f})$, and $\Delta \chi_{\rm f}/\bar{\chi}_{\rm f} = 2 (\chi^{\rm
  insp}_{\rm f} - \chi^{\rm postinsp}_{\rm f})/(\chi^{\rm insp}_{\rm
  f}+ \chi^{\rm postinsp}_{\rm f})$, such that the \ac{GR} limit is
given by $\Delta M_{\rm f} / \bar{M}_{\rm f} = \Delta \chi_{\rm f} /
\bar{\chi}_{\rm f} = 0$. The core results are summarized in 
Fig.~\ref{fig:IMRCT_2DPlot}. We infer $\Delta M_{\rm f}
/ \bar{M}_{\rm f} = \GWTwoFiveZeroOneOneFourIMRCTdMf$
and $\Delta \chi_{\mathrm{f}} / \bar{\chi}_{\mathrm{f}} =
\GWTwoFiveZeroOneOneFourIMRCTdchif$ from \eventname.
We compare these results with the ones obtained by hierarchically
combining $\nIMRCTEventsGWTCFour$ events, of which $\nIMRCTEventsOFourA$ are first reported in
\mbox{GWTC-4.0}~\cite{TGR-GWTC4}, which yields $\Delta M_{\rm f} /
\bar{M}_{\rm f} = \GWTCFourIMRCTdMf$ and $\Delta
\chi_{\mathrm{f}} / \bar{\chi}_{\mathrm{f}} =
\GWTCFourIMRCTdchif$. 
Remarkably, the constraints derived from \eventname alone yield a consistency test of comparable
stringency to the combined analysis of
\mbox{\ac{GWTC-4.0}}~\cite{TGR-GWTC4}. 
The \ac{GR} quantile for the two-dimensional posteriors from \eventname is
$\GWTwoFiveZeroOneOneFourIMRCTGRQuantile$, compared
to $\GWTCFourIMRCTGRQuantile$ for the
\mbox{\ac{GWTC-4.0}} analysis~\cite{TGR-GWTC4}.

We also exploit the results of the consistency test to determine at
what statistical significance the Hawking area
theorem~\cite{Hawking:1971tu}, a fundamental consequence of the second
law of \ac{BH} mechanics, holds.  This theorem states that the horizon
area of a \ac{BH} cannot decrease over time.  
Our analysis yields a credibility of $\IMRCTAreaSigmaGauss \sigma_{\rm IMRCT}$, representing the statistical significance in standard deviations of the difference between the mean total area of the initial \ac{BH}s and the mean final-\ac{BH}'s area. This test differs from the more agnostic 
strategy followed in \citet{GW250114}, where the initial and final areas of the objects are computed excluding the GW data around the 
merger signal. In addition, the test performed here splits the data in the frequency-domain, which is not equivalent to the time-domain analysis in \citet{GW250114}. See the Supplemental Material for the main results in Fig.~\ref{fig:IMRCT_AreaLaw} and further details.

Finally, in the Supplemental Material we also report the results of a residuals test~\cite{Cornish:2011ys, Vallisneri:2012qq,LIGOScientific:2016lio}, which looks 
for excess coherent power in the detector network after the maximum-likelihood waveform has been subtracted from the data. The upper limit for the residual network \ac{SNR} is $\ResidualsSNR$ at 90\% credibility ($p$-value $\ResidualsPValue$, see Supplemental Material). Thus, we do not find any statistically significant coherent power beyond what is expected from the noise background.

\textit{Conclusion---} The outstanding improvement of the LIGO detectors in the last
  decade~\cite{membersoftheLIGOScientific:2024elc,Ganapathy:2022hgu,Capote:2024rmo}
  has enabled unprecedented observations~\cite{LIGOScientific:2025slb,LIGOScientific:2025pvj}. In particular, \eventname
was observed with the largest \ac{SNR} to date ($\SNRInspiralNRSur$ up
to merger and $\SNRPostInspiralNRSur$ post-merger)~\cite{GW250114},
approximately three times that of the similar event
GW150914~\cite{LIGOScientific:2016aoc}. We have performed the most
  strigent suite of tests of \ac{GR} and the Kerr nature for a
  \ac{BBH} coalescence to date. Probing the inspiral, merger and
ringdown stages, we have set constraints comparable, and in some cases 
2--3 times more stringent, than the ones obtained
combining tens loudest events of
\mbox{\ac{GWTC-4.0}}~\cite{TGR-GWTC4}. At least three \acp{QNM} have been identified or constrained with several methods and models: the quadrupolar 220 fundamental and first overtone 221, and the hexadecapolar 440 mode. We have found that their spectroscopic pattern~\cite{Detweiler:1980gk,Nollert:1999ji,Kokkotas:1999bd,Dreyer:2003bv, Berti:2005ys, Gossan:2011ha} aligns with the Kerr metric prediction, and their amplitudes are consistent with those measured in a \ac{NR} simulation of \eventname-like 
systems in \ac{GR}. In summary, the single, loud event GW250114 
has yielded the scientific return of dozens of previous detections, offering a preview 
of the unprecedented science that upcoming LIGO--Virgo--KAGRA observing runs~\cite{Aasi:2013wya} will unlock.

Strain data from the LIGO detectors for
\eventname are available from the Gravitational Wave
Open Science Center \cite{GWOSC_page}. All the material required for reproducing the figures, including scripts and posterior distributions from the analyses, is available
in the data release \cite{zenodo_data_release}.

\textit{Acknowledgements---} 
This work made use of the following software, listed in alphabetical order:
\arviz \cite{arviz_2019}, 
\Asimov \cite{Williams:2022pgn},
\astropy \cite{astropy:2013,astropy:2018,astropy:2022}, 
\BayesWave \cite{Cornish:2020dwh},
\bilby \cite{Ashton:2018jfp, Romero-Shaw:2020owr}, 
\bilbyTGR \cite{ashton_2025_15676285},
\cpnest \cite{cpnest},
\dynesty \cite{Speagle:2019ivv},
\gwpy \cite{gwpy, duncan_macleod_2024_12734623}, 
\hfivepy \cite{collette_python_hdf5_2014}, 
\phenomxphm \cite{Pratten:2020ceb, Colleoni:2024knd}, 
\jupyter and \ipython \cite{ipython_2007,jupyter_2016,jupyter_2021}, 
\lalsuite \cite{lalsuite, swiglal}, 
\matplotlib \cite{matplotlib_3_7_3_zenodo,Hunter:2007}, 
\nrsur \cite{Varma:2019csw},
\numpy \cite{numpy}, 
\pandas \cite{mckinney-proc-scipy-2010, pandas_13819579}, 
\pesummary \cite{Hoy:2020vys}, 
\pyring \cite{pyRing, Carullo:2019flw}, 
\pyseobnr \cite{Mihaylov:2023bkc},
\python \cite{python},
\qnmcode \cite{Stein:2019mop},
\ringdown \cite{ringdown,Isi:2021iql}, 
\scipy \cite{2020SciPy-NMeth, scipy_10543017}, 
\seaborn \cite{Waskom2021}, 
\seobnrphm and \seobnrhmrom \cite{Pompili:2023tna, Khalil:2023kep, vandeMeent:2023ols, Ramos-Buades:2023ehm}, 
\sxs \cite{boyle_2025_16277847},
\tqdm \cite{tqdm_14002015}, 

Late in the preparation of this manuscript, an error in the likelihood function used in the \bilby inference code was discovered, leading to an overly constrained likelihood~\cite{Talbot:2025vth}. The impact of this error is discussed in detail in the GWTC-4.0 methods paper~\cite{LIGOScientific:2025yae}. Following from investigations presented therein, we expect differences in the inferred posterior distributions due to this error to be small. A full correction of the results was not possible for the current version of the manuscript. Results for \pseobnr, \ac{FTI}, and \ac{TIGER} have been corrected. For the IMR consistency test, only events analyzed before \mbox{GWTC-4.0} remain to be corrected. Updated results for \ac{PCA} and the \ac{IMR} consistency test using a corrected likelihood function will be provided in a future revision, before publication. Other analyses are unaffected by this error.

%

This material is based upon work supported by NSF's LIGO Laboratory, which is a
major facility fully funded by the National Science Foundation.
The authors also gratefully acknowledge the support of
the Science and Technology Facilities Council (STFC) of the
United Kingdom, the Max-Planck-Society (MPS), and the State of
Niedersachsen/Germany for support of the construction of Advanced LIGO 
and construction and operation of the GEO\,600 detector. 
Additional support for Advanced LIGO was provided by the Australian Research Council.
The authors gratefully acknowledge the Italian Istituto Nazionale di Fisica Nucleare (INFN),  
the French Centre National de la Recherche Scientifique (CNRS) and
the Netherlands Organization for Scientific Research (NWO) 
for the construction and operation of the Virgo detector
and the creation and support  of the EGO consortium. 
The authors also gratefully acknowledge research support from these agencies as well as by 
the Council of Scientific and Industrial Research of India, 
the Department of Science and Technology, India,
the Science \& Engineering Research Board (SERB), India,
the Ministry of Human Resource Development, India,
the Spanish Agencia Estatal de Investigaci\'on (AEI),
the Spanish Ministerio de Ciencia, Innovaci\'on y Universidades,
the European Union NextGenerationEU/PRTR (PRTR-C17.I1),
the ICSC - CentroNazionale di Ricerca in High Performance Computing, Big Data
and Quantum Computing, funded by the European Union NextGenerationEU,
the Comunitat Auton\`oma de les Illes Balears through the Conselleria d'Educaci\'o i Universitats,
the Conselleria d'Innovaci\'o, Universitats, Ci\`encia i Societat Digital de la Generalitat Valenciana and
the CERCA Programme Generalitat de Catalunya, Spain,
the Polish National Agency for Academic Exchange,
the National Science Centre of Poland and the European Union - European Regional
Development Fund;
the Foundation for Polish Science (FNP),
the Polish Ministry of Science and Higher Education,
the Swiss National Science Foundation (SNSF),
the Russian Science Foundation,
the European Commission,
the European Social Funds (ESF),
the European Regional Development Funds (ERDF),
the Royal Society, 
the Scottish Funding Council, 
the Scottish Universities Physics Alliance, 
the Hungarian Scientific Research Fund (OTKA),
the French Lyon Institute of Origins (LIO),
the Belgian Fonds de la Recherche Scientifique (FRS-FNRS), 
Actions de Recherche Concert\'ees (ARC) and
Fonds Wetenschappelijk Onderzoek - Vlaanderen (FWO), Belgium,
the Paris \^{I}le-de-France Region, 
the National Research, Development and Innovation Office of Hungary (NKFIH), 
the National Research Foundation of Korea,
the Natural Sciences and Engineering Research Council of Canada (NSERC),
the Canadian Foundation for Innovation (CFI),
the Brazilian Ministry of Science, Technology, and Innovations,
the International Center for Theoretical Physics South American Institute for Fundamental Research (ICTP-SAIFR), 
the Research Grants Council of Hong Kong,
the National Natural Science Foundation of China (NSFC),
the Israel Science Foundation (ISF),
the US-Israel Binational Science Fund (BSF),
the Leverhulme Trust, 
the Research Corporation,
the National Science and Technology Council (NSTC), Taiwan,
the United States Department of Energy,
and
the Kavli Foundation.
The authors gratefully acknowledge the support of the NSF, STFC, INFN and CNRS for provision of computational resources.

This work was supported by MEXT,
the JSPS Leading-edge Research Infrastructure Program,
JSPS Grant-in-Aid for Specially Promoted Research 26000005,
JSPS Grant-in-Aid for Scientific Research on Innovative Areas 2402: 24103006,
24103005, and 2905: JP17H06358, JP17H06361 and JP17H06364,
JSPS Core-to-Core Program A.\ Advanced Research Networks,
JSPS Grants-in-Aid for Scientific Research (S) 17H06133 and 20H05639,
JSPS Grant-in-Aid for Transformative Research Areas (A) 20A203: JP20H05854,
the joint research program of the Institute for Cosmic Ray Research,
University of Tokyo,
the National Research Foundation (NRF),
the Computing Infrastructure Project of the Global Science experimental Data hub
Center (GSDC) at KISTI,
the Korea Astronomy and Space Science Institute (KASI),
the Ministry of Science and ICT (MSIT) in Korea,
Academia Sinica (AS),
the AS Grid Center (ASGC) and the National Science and Technology Council (NSTC)
in Taiwan under grants including the Science Vanguard Research Program,
the Advanced Technology Center (ATC) of NAOJ,
and the Mechanical Engineering Center of KEK.

For the purpose of open access, the authors have applied a Creative Commons Attribution (CC BY)
license to any Author Accepted Manuscript version arising.
We request that citations to this article use 'A. G. Abac {\it et al.} (LIGO-Virgo-KAGRA Collaboration), ...' or similar phrasing, depending on journal convention.




\clearpage

\section*{Supplemental  material}

\subsection{Significance estimation for a non-negative  \ac{QNM} amplitude}
\label{sec:amplest}

In the main text we quote the significance with which the amplitude $A$ of a
given \ac{QNM} is shown to be $>0$.  We use the one-dimensional \ac{HPD}
credible regions based on samples drawn from the posterior density. The one
dimensional $p$ \ac{HPD} is the shortest interval that contains a fraction $p$
of the samples. It can be computed by considering the $\left\lfloor \left( 1 -
p\right) N\right\rfloor$ intervals in the sorted samples that each include
$\left\lceil pN \right\rceil$ samples, where $N$ is the total number of samples
available, and choosing the shortest one (i.e.~the one with the highest
estimated density). Using sample-based \ac{HPD} intervals in this way can
produce significant differences between the true and estimated interval when $p
\simeq 1/N$~\cite{GW250114}, but in all cases in this paper, we have sufficient
samples so that $p N \gg 1$, and this is not an issue.

To convert a $p$ \ac{HPD} into a significance level, $x \sigma$, we relate these
quantities through the Gaussian distribution, via 
\begin{equation}
	p = \int_{-x}^{x} \mathrm{d} y \, \phi(y),
\end{equation}
where $\phi(y)$ is the standard normal density.  For example, a $3\sigma$
interval contains a fraction $p = 0.9973$ of the samples.

\subsection{Validity regimes of QNM models} 
\label{sec:modelvalidation}

Analyses of \ac{NR} simulations of \ac{BBH} mergers, like \eventname, with Eq.~\eqref{eq:h-from-modes} show that the time interval in which fits at consecutive start times yield exponentially consistent amplitudes varies based on the mode content~\cite{Baibhav:2023clw,Cheung:2023vki,Mitman:2025hgy}. To determine when Eq.~\eqref{eq:h-from-modes} is valid for the models we consider, we study the equal-mass, non-spinning \ac{NR} \ac{BBH} simulation \sxsThreeSixOneSeven~\cite{SXS,Scheel:2025jct}, whose intrinsic parameters lie within the 90\% credible region of \eventname. For each QNM model, we perform a linear least-squares fit to the $(2,2)$ mode with the start times $t_>\in[0,40]t_{M_{\rm f}}$ in steps of $0.1t_{M_{\rm f}}$ and measure the complex amplitude of each \ac{QNM} at $t_{\rm peak}$~\cite{Cook:2020otn}. We define a stability window of size $10t_{M_{\rm f}}$ as a region in which the \acp{QNM} in the model have amplitudes whose fractional variation is comparable to our 50\% credible level measurement uncertainties in \eventname ($\QNMModelOneStabilityTolerance\%$ for the 220 at $t_>=\QNMModelOneTime t_{M_{\rm f}}$ with the 220 model, $\QNMModelTwoStabilityTolerance\%$ for the 221 at $t_>=\QNMModelTwoTime t_{M_{\rm f}}$ with the 220+221 model, and $\QNMModelThreeStabilityTolerance\%$ for the 222 at $t_>=\QNMModelThreeTime t_{M_{\rm f}}$ with the 220+221+222 model; these times are chosen based on the value used in~\citet{GW250114} and the times identified in Fig.~\ref{fig:221}) and whose values agree with the most stable value to the same uncertainty. We define the stability regime as the union of all such windows. For the 220, 220+221, 220+221+222 models, we find these regimes to be approximately $[\QNMModelOneStabilityRegimeLeft,\QNMModelOneStabilityRegimeRight]t_{M_{\rm f}}$, $[\QNMModelTwoStabilityRegimeLeft,\QNMModelTwoStabilityRegimeRight]t_{M_{\rm f}}$, and $[\QNMModelThreeStabilityRegimeLeft,\QNMModelThreeStabilityRegimeRight]t_{M_{\rm f}}$.

\subsection{Comparison to \ac{NR}-informed predictions} 
\label{sec:NRcomparison}

\begin{figure}
	\includegraphics[width=\linewidth]{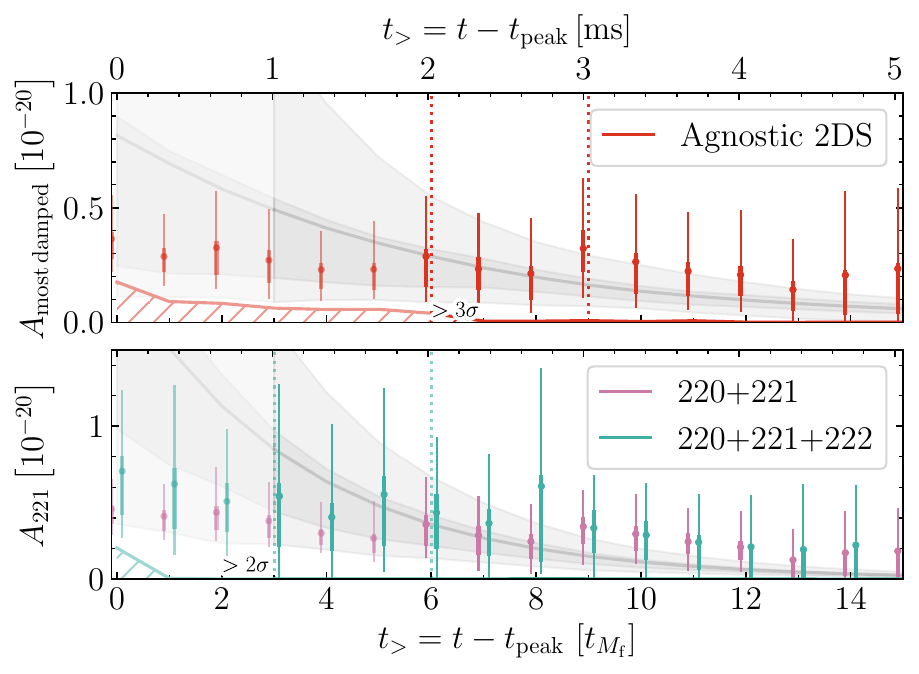}
	\caption{\label{fig:221} Consistency of a \ac{NR}-simulation's post-merger data with two \acp{QNM}.
		Identical to Fig.~\ref{fig:221_over_time}, but instead for a simulated signal using the equal-mass, non-spinning \ac{NR} \ac{BBH} simulation \sxsThreeSixOneSeven with Gaussian noise; fits are performed with \ringdown.
	}
\end{figure}

In Fig.~\ref{fig:221_over_time}, we have found that at least two agnostic damped 
sinusoids are required to explain the post-merger signal, and  when assuming 
a Kerr remnant, we have observed that the 221 amplitudes remain
  non-zero at $> 3 \sigma$ up to $9t_{M_{\rm f}}$ after the peak. 
Here, we find similar results when employing a simulated signal of the \ac{NR} simulation
\sxsThreeSixOneSeven~\cite{SXS,Scheel:2025jct}. The
signal is injected into Gaussian noise generated using the power
spectral density estimated from \eventname, with extrinsic parameters
set to the preliminary-reference maximum-likelihood values inferred
using \nrsur~\cite{GW250114}. We generate 10 different noise
realizations and choose the one such that the analysis of the
simulated signal most closely resembles that obtained for \eventname.
Like in Fig.~\ref{fig:221_over_time}, we find (see upper panel in Fig.~\ref{fig:221}) that  
at around $t_> \gtrsim 9t_{M_{\rm f}}$ the amplitude of the more rapidly decaying damped sinusoid becomes
consistent with zero at the 90\% level. Furthermore, at around $t_>
\lesssim 6 t_{M_{\rm f}}$ (lower panel), the amplitude of the 221 mode in the
220+221 model also starts to deviate from its extrapolated values;
but, by adding the 222 mode to the fit one can again recover
amplitude consistency at early times around $t_> \gtrsim 3 t_{M_{\rm f}}$.

We perform another test to verify that the relative amplitudes and phases of the 220 and 221 \ac{QNM}s are consistent with predictions from \ac{NR}. This requires mapping the detector \ac{QNM} amplitudes in Eq.~\eqref{eq:h-from-modes} to the remnant-frame \ac{QNM} amplitudes, with which the strain over the two-sphere in the remnant frame can be written as
\begin{align}
	\label{eq:h-over-sphere}
	h&\equiv  h_+ - ih_\times\nonumber\\
	&= \sum_{\substack{\ell \geq 2 \\ -\ell\leq m \leq \ell \\ n \geq 0}} C_{\ell m n} e^{-i\omega_{\ell m n}t}\,\phantom{}_{-2}S_{\ell m n}(M_{\rm f}\chi_{\rm f}\omega_{\ell m n},\theta_{JN}, \varphi),
\end{align}
where $\omega_{\ell m n}\equiv 2\pi f_{\ell m n}-i/\tau_{\ell m n}$ and $\phantom{}_{-2}S_{\ell m n}(M_{\rm f}\chi_{\rm f}\omega_{\ell m n}, \theta_{JN}, \varphi)$ is the spin-weight $-2$ $\ell m n$ spheroidal harmonic with oblateness $M_{\rm f}\chi_{\rm f}\omega_{\ell m n}$ and evaluation point on the two-sphere $(\theta_{JN},\varphi)$. The complex amplitudes $C_{\ell m n}$ in Eq.~\eqref{eq:h-over-sphere} are related to the complex amplitudes in Eq.~\eqref{eq:h-from-modes} via~\cite{Isi:2021iql}
\begin{subequations}
\begin{align}
	C_{\ell +|m|n}\,\phantom{}_{-2}S_{\ell +|m|n}(\theta_{JN},\varphi)&=A_{\ell m n}^{\rm R},\\
	C_{\ell -|m|n}\,\phantom{}_{-2}S_{\ell -|m|n}(\theta_{JN},\varphi)&=A_{\ell m n}^{\rm L}.
\end{align}
\end{subequations}
However, because $A_{220}^{\rm R}$ is measured to be larger than $A_{220}^{\rm L}$ for this event, from here on we focus exclusively on the $\ell|m| n$ \acp{QNM}. Using the \ac{QNM} amplitudes over the two-sphere, we can compare the relative amplitudes and phases of the 220 and 221 \acp{QNM} via
\begin{subequations}
\begin{align}
	\label{eq:A_ratio}
	\overline{A}_{221}&= \Big |\frac{C_{221}}{C_{220}} \Big |,\\
	\label{eq:d_phi}
	\Delta\phi_{221}&=\arg\left[\frac{C_{221}}{C_{220}}\right].
\end{align}
\end{subequations}

\begin{figure}
	\includegraphics[width=\linewidth]{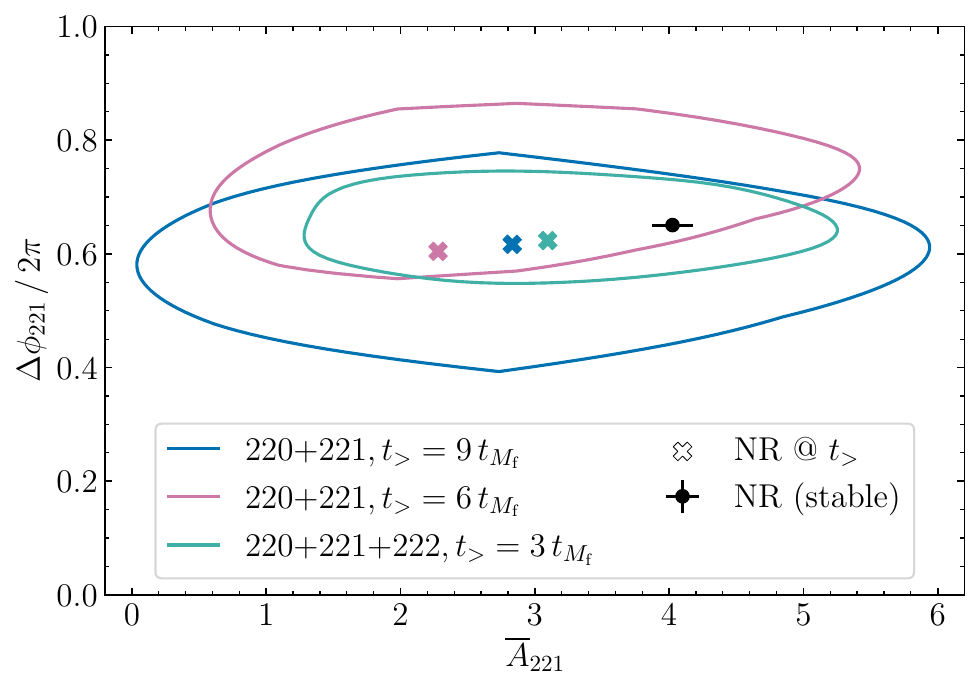}
	\caption{\label{fig:A_phi_vs_NR} Consistency of the 220 and 221 \acp{QNM} from \eventname with NR predictions. The 50\% credible regions of the two-dimensional posteriors on $\Delta\phi_{221}$ and $\overline{A}_{221}$ for \eventname, at varying fit start times, as measured by \ringdown. Colored crosses represent fits to the equal-mass, non-spinning \ac{NR} \ac{BBH} simulation \sxsThreeSixOneSeven. For $t_> = t - t_{\rm peak}[t_{M_\mathrm{f}}] \in\{6,9\}t_{M_{\mathrm{f}}}$ the 220+221 model is used, while for $t_>=3t_{M_{\mathrm{f}}}$, the 220+221+222 model is used. The black marker represents a stable value of these quantities over 
a window ranging from $t_>\in\left[\QNMModelThreeStabilityRegimeRightMTen,\QNMModelThreeStabilityRegimeRight\right]t_{M_{\mathrm{f}}}$, with the data point the mean and the error bars the $1\sigma$ variation over said window.}
\end{figure}

\noindent The quantities $\overline{A}_{\ell m n}$ and
$\Delta\phi_{\ell m n}$ should be interpreted as the amplitude ratio and phase
difference between the 220 and 221 \ac{QNM}s on the two-sphere extrapolated
to $t_{\mathrm{peak}}$. Each contour in Fig.~\ref{fig:A_phi_vs_NR} shows the
50\% credible region for the fit performed at the start time indicated in the
legend. For $t_>\in\{6,9\}t_{M_{\mathrm{f}}}$ the 220+221 model is used,
while for $t_>=3t_{M_{\mathrm{f}}}$, the 220+221+222 model is used.
The corresponding crosses represent the values extracted from fits to the NR
simulation \sxsThreeSixOneSeven performed using a linear least-squares fit to the $(2,2)$ mode~\cite{Cook:2020otn}. The marker with
error bars is obtained by fitting the 220+221+222 model over a $10t_{M_{\rm f}}$ window in which these quantities have stabilized. That is, we fit the NR waveform at times $t_{>}\in\left[\QNMModelThreeStabilityRegimeRightMTen,\QNMModelThreeStabilityRegimeRight\right]t_{M_{\mathrm{f}}}$---the latest $10t_{M_{f}}$ contained in the regime of validity for the 220+221+222 model, measure the mean and standard deviation of $C_{220}$ and $C_{221}$ over said $10t_{M_{\mathrm{f}}}$ window, and then compute $\overline{A}_{221}$ and $\Delta\phi_{221}$ accordingly. As can be seen, from $t_{>}\in\left[3,9\right]t_{M_{\mathrm{f}}}$ the amplitudes and
phases extracted from the data are broadly consistent with the values predicted. More specifically, they are consistent with the most stable value to $\geq\NRConfidence\%$ credibility. Their values evolve because the amplitudes of the exponentially damped-sinusoid model that are used are insufficient to recover particularly stable
overtone amplitudes for these $t_>$ values at effectively infinite \ac{SNR}.

Overall, these results indicate that the 221 amplitudes and phases measured
from \eventname when using different ringdown models at different starting times are consistent with expectations from a \ac{BBH} in \ac{GR}.

\subsection{Final mass and spin consistency}

\begin{figure}
	\includegraphics[width=\linewidth]{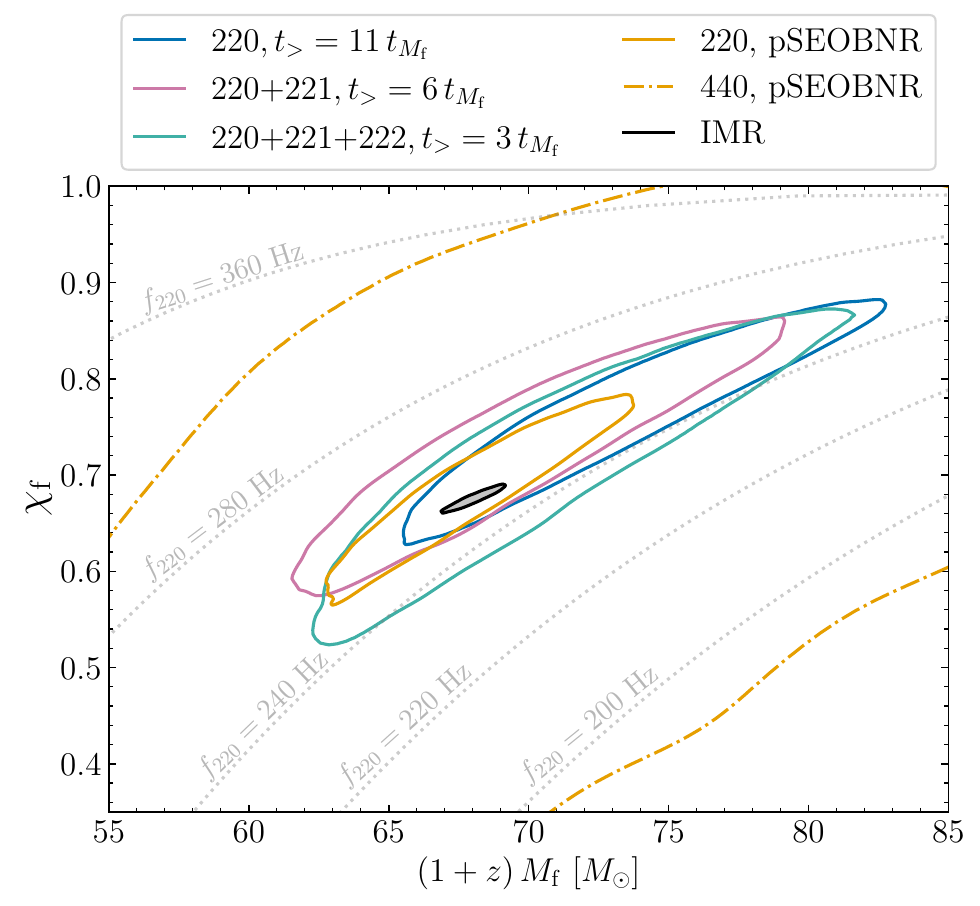}
	\caption{\label{fig:M_chi_over_time} Final mass and spin consistency. The $90\%$ credible regions of the remnant mass and spin inferred
		from a series of ringdown fits with \pyring:
		using the 220 model at $t_{>} = 11 \, t_{M_{\mathrm{f}}}$, 
		the 220+221 model at $t_{>} = 6\, t_{M_{\mathrm{f}}}$, 
		and the 220+221+222 model at $t_{>} = 3 \, t_{M_{\mathrm{f}}}$. 
		We also show results for the 220 and $440$ modes with the \pseobnr analysis,
		as well as the \ac{IMR} results of the full signal.
	}
\end{figure}

In Fig.~\ref{fig:M_chi_over_time} we show the mass and spin posteriors obtained
from the 220, 220+221, and 220+221+222 model fits at $t_{>}=11 \,
t_{M_{\mathrm{f}}}$, $t_{>}=6 \, t_{M_{\mathrm{f}}}$, and $t_{>}=3 \,
t_{M_{\mathrm{f}}}$, and from the 220 and 440 \pseobnr model. 
For all these fits, the inferred mass and spin are
consistent with that from the full \ac{IMR} analysis at 90\% credibility.

\subsection{Detectability of quadrupolar first overtone via model selection}

\begin{figure}
	\includegraphics[width=\linewidth]{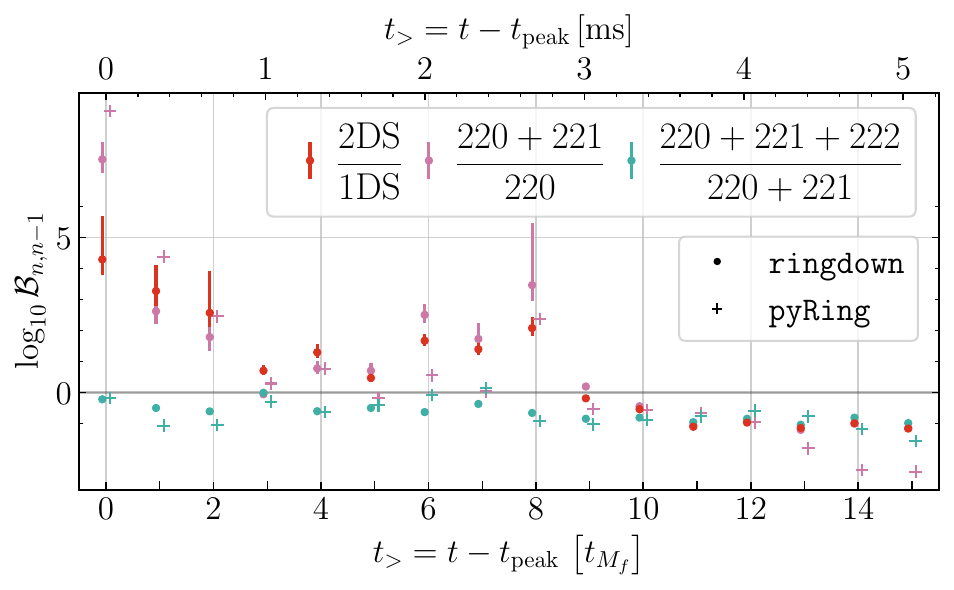}
	\caption{\label{fig:BF_over_time} Consistency of post-merger data with two \acp{QNM} via model selection. 
		The Bayes factors comparing the analyzed \ac{QNM} model to a nested
		model with one fewer mode. Dots indicate the estimated \ringdown Bayes
		factor and plus the \pyring Bayes factor.}
\end{figure}

In the main text, we have assessed \ac{QNM} detectability based on the
posterior amplitude support away from zero. A complementary criterion is represented
by the Bayes factor, which quantifies the ratio of the evidences of competing models,
with each evidence defined as the integral of the likelihood weighted by the prior~\cite{Morey2016}.
In Fig.~\ref{fig:BF_over_time}, we show the $\log_{10}$ Bayes factors of the two-mode
model compared to the one-mode model over time. The results from \pyring are obtained from the \cpnest
nested sampler used in inference~\cite{cpnest}, while the ones from \ringdown are estimated from the
221 amplitudes using the Savage--Dickey ratio~\cite{Dickey1971}. For nested models, these two approaches
are equivalent~\cite{Zimmerman:2019wzo}. For Savage--Dickey, we derive the amplitude maximum prior from
the \pyring maximum priors on the right- and left-handed polarized contributions of the modes
(between $[0,5 \times 10^{-20}]$) combined with the spin-weighted spherical harmonics~\cite{GW250114}.
The reported values correspond to the median and 90\% credible interval of 1000 Bayes factor estimates
with bootstrap resamples of the amplitude samples. The \pyring uncertainties are not displayed since
they are comparable to the marker size; they represent the half-width of the 90\% credible interval
estimated from the four nested sampling chains.

We find positive evidence for the presence of the 221 as late as $8 t_{M_\mathrm{f}}$ ($9 t_{M_\mathrm{f}}$)
from \pyring (\ringdown), with $\log_{10}$ Bayes factors of $\pyRingSixM$ ($\ringdownSixM$) at $6 t_{M_\mathrm{f}}$,
$\pyRingEightM$ ($\ringdownEightM$) at $8 t_{M_\mathrm{f}}$ and $\pyRingNineM$ ($\ringdownNineM$) at $9 t_{M_\mathrm{f}}$.
A similar trend is observed using agnostic sinusoids. The presence of a second overtone 222 is not significantly preferred at any time. These results are in agreement with the \ac{QNMRF} detection statistics in Fig.~\ref{fig:qnmrf}, as well as with the non-zero amplitude consistency in Fig.~\ref{fig:221_over_time}, as discussed in the main text. They are also in accord with 
predictions from analysis of similarly loud simulated signals~\cite{Akyuz:2025seg}.

The difference between the two pipelines is due to different prior volumes and data length used. In fact,
we are able to obtain the same values within the Bayes factor uncertainty when analyzing with \pyring $0.6~\mathrm{s}$
of data with similar priors to \ringdown, and when running \ringdown on the same \pyring settings.
For a complete description of the differences in settings between the codes, see Supplemental Material in~\citet{GW250114}.

\subsection{\ac{QNM} rational filter}

The \ac{QNMRF} analysis~\cite{Ma:2022wpv} is designed to isolate and remove specific complex-valued \acp{QNM} from the ringdown signal in the frequency domain. The resulting residual is then compared to pure colored Gaussian noise~\cite{Ma:2023cwe,Ma:2023vvr}. The filter is constructed for a given set of \acp{QNM}, corresponding to specific \ac{BH} masses and spins. 

\ac{QNMRF} computes the mode detection statistic, $\mathcal{D}$, defined as a comparison between
two ringdown model hypotheses: $\mathcal{H}$, which includes an additional mode,
and $\mathcal{H'}$, which excludes it. $\mathcal{D}$ is analogous to a
logarithmic Bayes factor, yet differs from the Bayes factors used in other
time-domain ringdown analyses~\cite{Lu:2025mwp}. To assess statistical
significance, we then take a frequentist approach to estimate \acp{FAP} due to the background noise, and determine a threshold,
$\mathcal{D}_{1\%}$, corresponding to a $1\%$ \ac{FAP}. Given the data $d$, the
statistic is expressed as:
\begin{equation}
\mathcal{D}(\mathcal{H}:\mathcal{H}') = \log_{10} \frac{\mathcal{Z}(d | \mathcal{H})}{\mathcal{Z}(d | \mathcal{H'})},
\end{equation}
where $\mathcal{Z}(d|\mathcal{H})$ and $\mathcal{Z}(d|\mathcal{H'})$ denote the evidences under hypotheses $\mathcal{H}$ and $\mathcal{H'}$.

In Fig.~\ref{fig:qnmrf}, the pink crosses represent the offset of $\mathcal{D}[\mathcal{H}(220+221):\mathcal{H}'(220)]$ relative to the 1\% \ac{FAP} threshold evaluated on background noise, while the green plus indicate the offset of $\mathcal{D}[\mathcal{H}(220+221+222):\mathcal{H}'(220+221)]$ relative to its corresponding threshold. 

For each ringdown model hypothesis, we also compute the joint posterior quantile $p(M_{\rm f}^{\rm IMR},\chi_{\rm f}^{\rm IMR})$ of the remnant \ac{BH} mass and spin, inferred from the full \ac{IMR} analysis, using \nrsur. This credible contour represents the region on which the inferred parameters $(M_{\rm f}^{\rm IMR},\chi_{\rm f}^{\rm IMR})$ lie, serving as a consistency check with the \ac{IMR} results~\cite{Lu:2025mwp}. A lower $p$ value indicates a better match between the \ac{IMR} analysis and the given \ac{QNM} hypothesis. Among all possible additional modes added to the nested model, one at a time, the mode that (i) yields a detection statistic $\mathcal{D}$ above the threshold and simultaneously (ii) results in the greatest reduction in $p(M_{\rm f}^{\rm IMR},\chi_{\rm f}^{\rm IMR})$ is considered confidently identified for a given starting time.
In Fig.~\ref{fig:qnmrf}, when the two-mode model $\mathcal{H}(220+221)$ is favored over the single-mode model $\mathcal{H}'(220)$, i.e. when $\mathcal{D}[\mathcal{H}(220+221):\mathcal{H}'(220)] > \mathcal{D}_{1\%}$, we find that the corresponding joint posterior quantile $p(M_{\rm f}^{\rm IMR},\chi_{\rm f}^{\rm IMR})$ also decreases across starting times from $4 \, t_{M_{\mathrm{f}}}$ to $10 \, t_{M_{\mathrm{f}}}$, indicating improved consistency with the \ac{IMR}-inferred remnant parameters. Similarly, when the three-mode model $\mathcal{H}(220+221+222)$ is preferred over $\mathcal{H}'(220+221)$ with $\mathcal{D} > \mathcal{D}_{1\%}$ from starting times $1 \, t_{M_{\mathrm{f}}}$ to $5 \, t_{M_{\mathrm{f}}}$, we observe a reduction in $p(M_{\rm f}^{\rm IMR}, \chi_{\rm f}^{\rm IMR})$. These results suggest that these additional modes enhance the fit to the data and also improve the \ac{IMR} consistency.

\begin{figure}
	\includegraphics[width=\linewidth]{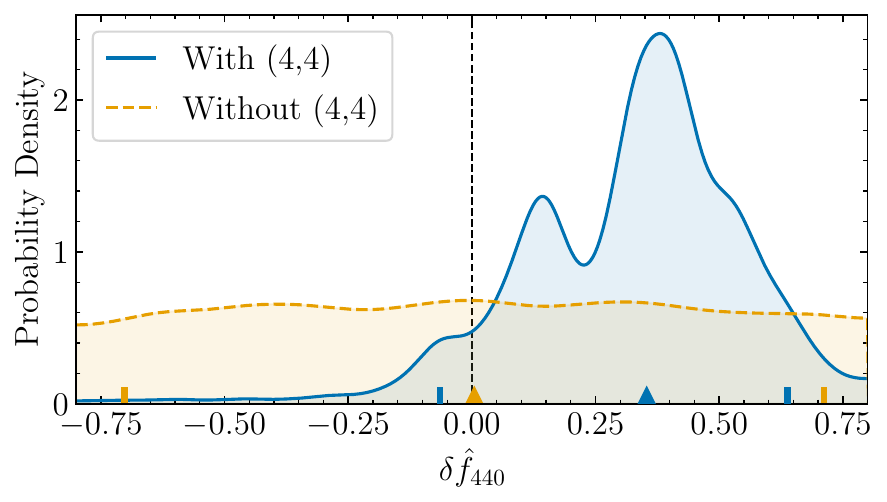}
	\caption{\label{fig:ridgeline_domega_dtau_injection} Validation of the 440-mode constraint using simulated signals.
		Posterior distributions for the fractional deviation in the $440$ \ac{QNM} frequency, $\delta \hat{f}_{440}$, obtained from simulated \ac{NR} signals consistent with \eventname. The blue curve shows results for a simulated signal that includes the $(\ell,|m|) = (4,4)$ multipoles, while the orange curve corresponds to a simulated signal with the modes removed. Triangles mark the median values and vertical bars the symmetric $90$\% credible interval.}
\end{figure}

\subsection{Validation of the $(4,4)$ fundamental-mode frequency constraint} 

To validate the constraint on the $440$ \ac{QNM} frequency, we perform targeted simulated-signal studies, recovering the signals using the \pseobnr model.
We analyze a synthetic signal generated using the equal-mass, non-spinning \ac{NR} simulation \sxsThreeSixOneSeven~\cite{SXS,Scheel:2025jct,boyle_2025_16277847}, with extrinsic parameters compatible with \eventname, and injected into Gaussian noise.
Several features observed in the real \eventname data, particularly those associated with the $440$ \ac{QNM}, can be qualitatively reproduced in this Gaussian noise injection. 
As shown in Fig.~\ref{fig:ridgeline_domega_dtau_injection}, the $440$ \ac{QNM} frequency is recovered with comparable precision to the real-signal analysis, while the damping time remains largely unconstrained.

In both the real event and Gaussian-noise injections, the posterior for $\delta \hat{f}_{440}$ can exhibit a multimodal structure, especially under a wide, uninformative prior for $\delta \hat{\tau}_{440}$. By contrast, an analogous injection into zero-noise data yields posteriors peaked at $\delta \hat{f}_{440}=0$ with no significant substructure.
The observed multimodality in $\delta \hat{f}_{440}$ is primarily driven by samples with large values of $\delta \hat{\tau}_{440}$, corresponding to long-lived, nearly sinusoidal modes. Indeed, for a purely sinusoidal signal, the likelihood is expected to exhibit secondary maxima with regular spacing in frequency~\cite{Xin:2025voy}.
When using an extended prior allowing for $\delta \hat{\tau}_{440}$ values up to 4, multimodality in $\delta \hat{f}_{440}$ is clearly associated with large $\delta \hat{\tau}_{440}$ samples. 
The maximum-likelihood parameters lie near the GR-consistent central mode, indicating that the secondary peaks are not favored by the data.
Given the low \ac{SNR} in the 440 mode, we do not expect to constrain deviations to its damping time. 
For the main results shown in Fig.~\ref{fig:ridgeline_domega_dtau}, we adopt a uniform prior on $\delta \hat{\tau}_{440}\in[-0.8, 0.8]$ consistent with that used for the ringdown constraints on the 221 mode. 

Finally, we repeat the Gaussian-noise injection, using the same NR simulation and parameters, but with the $(\ell, |m|) = (4,4)$ multipoles explicitly removed from the signal. 
In this case, the resulting constraint on $\delta \hat{f}_{440}$ is uninformative (orange curve in Fig.~\ref{fig:ridgeline_domega_dtau_injection}) suggesting that the posteriors recovered in the full injections are driven by the presence of the $(4,4)$ multipoles in the data. 
These findings reinforce the interpretation that the constraint on $\delta \hat{f}_{440}$ obtained from \eventname reflects a genuine physical feature of the signal rather than an artifact of the analysis or noise.

\subsection{Principal component analysis}
A limitation of the \ac{FTI} and \ac{TIGER} results presented above is that
individual PN deformation parameters are varied one at a time with all other
parameters being fixed to the \ac{GR} baseline. Whilst
robust~\cite{Meidam:2017dgf}, single-parameter tests do not probe
correlations across multiple PN orders, potentially missing more complex
departures from \ac{GR}. An alternative scheme was proposed
in~\cite{Saleem:2021nsb,Mahapatra:2025cwk}, in which six PN deformation parameters are
simultaneously varied, taken to be the $1.5$PN to the $3.5$PN parameters. The
$-1$PN, $0$PN, $0.5$PN, and $1$PN terms are fixed to their \ac{GR} values. The
approach is to estimate the joint posterior for the standard binary parameters
plus the 6 PN deformation parameters. Then, one marginalizes over the \ac{GR}
parameters to yield a six-dimensional posterior for the PN deformations that
captures correlated deviations~\cite{Pai:2012mv,Shoom:2021mdj,Saleem:2021nsb,Mahapatra:2025cwk},
though strong parameter correlations can often render the posteriors
uninformative or weakly constrained. Priors on the deformation parameters are
taken to be uniform, such that $\delta \hat{\varphi}^{(i)}_{\rm prior} \sim
\mathcal{U}(-20,20)$.

To mitigate against this potential shortcoming, we apply a \ac{PCA} to the six-dimensional
posteriors, diagonalize the covariance matrix and 
identify the linear combination of PN deformation parameters that are
best constrained by the data~\cite{Saleem:2021nsb,Mahapatra:2025cwk}.  The new basis,
$\delta \hat{\varphi}_{\rm PCA}^{(i)}$, provides orthogonal directions
that minimize posterior widths.  The \ac{PCA} analysis is applied to
the \ac{TIGER} framework, using
\phenomxphm~\cite{Pratten:2020ceb,Colleoni:2024knd}, and the \ac{FTI}
framework, using \seobnrhmrom~\cite{Pompili:2023tna}.  We find that
the leading two \ac{PCA} parameters are informative using the
\ac{TIGER} pipeline, and the leading three when using \ac{FTI}, with
all tests being consistent with zero.
The $90\%$ credible bounds on the leading \ac{PCA} parameter $\delta\hat{\varphi}_{\mathrm{PCA}}^{(1)}$ for \eventname are $\GWPCAoneStatsFTI$ (\ac{FTI}) and $\GWPCAoneStatsTIGER$ (\ac{TIGER}) respectively.
The leading and sub-leading \ac{PCA} parameters can be re-expressed as weighted combinations of the PN deformation coefficients. From the \ac{FTI} analysis of \eventname, we find
\begin{subequations}
\begin{align}
\delta\hat{\varphi}_{\mathrm{PCA}}^{(1)} &= 0.7482\,\delta\hat{\varphi}_3 - 0.1337\,\delta\hat{\varphi}_4 + 0.4910\,\delta\hat{\varphi}_{5l} \nonumber \\ &\qquad - 0.3797\,\delta\hat{\varphi}_6 + 0.0240\,\delta\hat{\varphi}_{6l} + 0.1908\,\delta\hat{\varphi}_7,\\
\delta\hat{\varphi}_{\mathrm{PCA}}^{(2)} &= 0.6387\,\delta\hat{\varphi}_3 + 0.0139\,\delta\hat{\varphi}_4 - 0.4136\,\delta\hat{\varphi}_{5l} \nonumber \\ &\qquad + 0.5397\,\delta\hat{\varphi}_6 + 0.0235\,\delta\hat{\varphi}_{6l} - 0.3592\,\delta\hat{\varphi}_7.
\end{align}
\end{subequations}
The \ac{PCA} coefficients are dominated by the $1.5$PN, $2.5$PN log, and $3$PN terms, in broad agreement with the individual PN coefficient analysis, as seen in Fig~\ref{fig:fti_tiger}. This analysis demonstrates that even when allowing for correlated deviations across multiple \ac{PN} orders, the deviations away from \ac{GR} inferred from \eventname alone are constrained to be negligible.

\subsection{Bounds on the black hole area theorem}
A key outcome of~\citet{GW250114} was a precision constraint on Hawking's area theorem~\cite{Hawking:1971tu}, a fundamental consequence of the second law of \ac{BH} 
mechanics stating that the horizon area of a \ac{BH} cannot decrease over time. 
In practice, this implies that for \ac{BH} mergers, the area of the final remnant must exceed the combined area of the two progenitor \acp{BH}~\cite{Hawking:1971tu}. 
Analogously to~\citet{GW250114}, we test this prediction by independently estimating the initial and final \ac{BH} areas using different portions of the signal; however, 
differently from~\citet{GW250114}, we employ the entire signal, whereas in the other analysis the data around merger are excluded.
Our approach closely follows the \ac{IMR} consistency test. We constrain the masses and spins of the \ac{BH}s in the inspiral and post-inspiral phases, which we directly map to the initial and final areas. 

\begin{figure}
    \includegraphics[width=\linewidth]{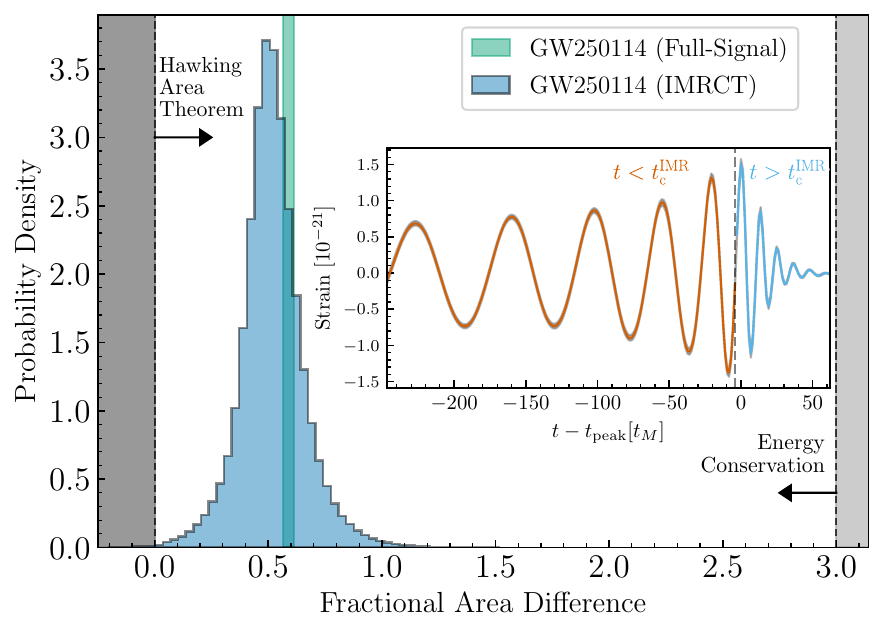}
    \caption{\label{fig:IMRCT_AreaLaw} 
    \ac{BH} area-law test using the entirety of \eventname. The fractional difference between the area of the final \ac{BH} $\mathcal{A}_{\rm f}$ and the total area of the initial \ac{BH}'s $\mathcal{A}_{\rm i}$ as calculated using the \ac{IMR} consistency test on \eventname. The grey shaded region on the left marks the region in which the area theorem is violated, $(\mathcal{A}_{\rm f} - \mathcal{A}_{\rm i}) / \mathcal{A}_{\rm i} < 0$. The grey shaded region on the right highlights the region excluded by energy conservation $M_\mathrm{f} \leq m_1 + m_2$. The vertical green band is the $90$\% credible interval inferred from the full-signal analysis in~\citet{GW250114}. The inset schematically shows a reconstructed (grey) signal in LIGO Livingston using the full-signal analysis~\cite{GW250114}, along with the inspiral (orange) and post-inspiral (light blue) regions used in the \ac{IMR} consistency test, such that the time $t_{\rm c}^{\rm IMR}$ corresponds to the transition frequency $f^{\rm IMR}_{\rm c}$, noting that the data is split in the frequency-domain and not the time-domain. 
    }
\end{figure}

The areas are calculated using the Kerr formula~\cite{Bardeen:1973gs}
\begin{align}
\mathcal{A} (m , \chi) &= 8 \pi \left( \frac{G m}{c^2} \right)^2 \left( 1 +  \sqrt{ 1 - \chi^2} \right)\,,
\end{align}
where $m$ and $\chi$ are the \ac{BH} mass and dimensionless spin. For the initial area $\mathcal{A}_{\rm i}$, we infer the individual \ac{BH} masses and spins from the inspiral, and the total area is calculated as $\mathcal{A}_{\rm i} = \mathcal{A}_1 + \mathcal{A}_2$. For the final area $\mathcal{A}_{\rm f}$, we employ \ac{NR} calibrated fits to estimate the remnant \ac{BH} mass and spin from the progenitor parameters~\cite{Jimenez-Forteza:2016oae,Hofmann:2016yih}, emphasizing that the initial \ac{BH} source properties used in this calculation are inferred exclusively from post-inspiral data. 
In Fig.~\ref{fig:IMRCT_AreaLaw}, we show the fractional difference between the final and initial areas, $(\mathcal{A}_{\rm f} - \mathcal{A}_{\rm i}) / \mathcal{A}_{\rm i}$. 
We find that \eventname is consistent with the area theorem at the $\IMRCTAreaSigmaGauss \sigma_{\rm IMRCT}$ credibility level. Here, the significance $X\sigma_{\rm IMRCT}$ is calculated from the difference in areas and defined as the ratio of the difference in means to the standard deviation of the differences~\cite{GW250114},
\begin{align}
\label{eq:area_significance}
X = \frac{\mu_{\rm f} - \mu_{\rm i}}{\sqrt{\sigma^2_{\rm f} + \sigma^2_{\rm i}}},
\end{align}
which expresses how many standard deviations the observed mean deviates from zero. Here $\mu_{\rm i}$ and $\mu_{\rm f}$ denote the means of the initial and final areas respectively, while $\sigma_{\rm i}$ and $\sigma_{\rm f}$ are their corresponding standard deviations. As discussed in~\citet{GW250114}, this estimate is less sensitive to sampling errors in the distribution tails since it relies only on the first two cumulants.
This bound is slightly more stringent than that presented in~\citet{GW250114}, due to stronger \ac{GR} assumptions and use of the complete signal. 
Moreover, the test performed here splits the data in the frequency-domain, which is not equivalent to the time-domain analysis done in \citet{GW250114}. 
Using the fractional difference to calculate the significance, we find $\IMRCTdAreaSigmaGauss \sigma_{\rm IMRCT}$, with differences being driven by uncertainty in the initial area normalization. 
In Fig.~\ref{fig:IMRCT_AreaLaw}, we also show the $90$\% credible interval from the full-signal analysis using \nrsur~\cite{GW250114}, which coherently describe the complete signal assuming both \ac{GR} and the area theorem. It yields the most stringent bound because it employs the full \ac{SNR} of \eventname, instead of using a smaller portion 
associated either to the inspiral or the post-inspiral phases.

\subsection{Residuals test}

The residuals test~\cite{LIGOScientific:2016lio} is a statistical analysis that checks for the presence of excess coherent power remaining in the detector network after subtracting the best-fit waveform from the data~\cite{Cornish:2011ys, Vallisneri:2012qq}. Significant residual power could indicate the presence of additional physical effects that are not captured by current \ac{BBH} models, modeling systematics, or unaccounted instrumental noise artifacts.

We perform the residual data by subtracting from the original data the maximum-likelihood \nrsur waveform model. If the model adequately captures the \ac{GW} signal, the resulting residuals should be consistent with stationary Gaussian noise. The residual data is then analyzed using \BayesWave~\cite{Cornish:2020dwh}, and the 90\% credible upper limit on the network \ac{SNR} $\rho_{90}$ is calculated. To compare this $\rho_{90}$ with its expected distribution, segments of data around the signal (with no injected signal) are also analyzed and the probability of obtaining an $\rho_{90}$ higher than that of the residual data is calculated and reported as the \textit{p}-value $= P(\rho_{\mathrm{90}}^{n} \geq \rho_\mathrm{90})$, where $\rho_{\mathrm{90}}^{n}$ is the 90\% credible upper limit on the coherent \ac{SNR} of the background, noise-only segments. 

A higher $p$-value suggests that the residual power is more consistent with instrumental noise, indicating insufficient evidence to reject the null hypothesis that the residual power originates from noise. For a single event, we also expect the $p$-value to be a random draw from a uniform distribution on the interval (0,1]. Furthermore, the goodness of fit of the \ac{GR} template for the signal in the data can be quantified by calculating the 90\% credible lower limit on the fitting factor (FF), given by:
\begin{equation}
\mathrm{FF}_{\mathrm{90}} = \frac{\rho_{\mathrm{GR}}}{\sqrt{\rho^2_{\mathrm{GR}} + \rho^2_{\mathrm{90}}}}\,,
\end{equation}
where $\rho_{\mathrm{GR}}$ is the optimal network \ac{SNR} for the maximum-likelihood waveform \cite{TGR-GWTC4}.

For \eventname, we find that $\rho_{90} = \ResidualsSNR$ with a $p$-value of $\ResidualsPValue$. The calculated $\mathrm{FF}_{90}$ is $\ResidualsFF$. Based on this, we do not find any significant coherent power beyond what is expected from noise.


\bibliography{bibliography}

\begin{thebibliography}{230}%
\makeatletter
\providecommand \@ifxundefined [1]{%
 \@ifx{#1\undefined}
}%
\providecommand \@ifnum [1]{%
 \ifnum #1\expandafter \@firstoftwo
 \else \expandafter \@secondoftwo
 \fi
}%
\providecommand \@ifx [1]{%
 \ifx #1\expandafter \@firstoftwo
 \else \expandafter \@secondoftwo
 \fi
}%
\providecommand \natexlab [1]{#1}%
\providecommand \enquote  [1]{``#1''}%
\providecommand \bibnamefont  [1]{#1}%
\providecommand \bibfnamefont [1]{#1}%
\providecommand \citenamefont [1]{#1}%
\providecommand \href@noop [0]{\@secondoftwo}%
\providecommand \href [0]{\begingroup \@sanitize@url \@href}%
\providecommand \@href[1]{\@@startlink{#1}\@@href}%
\providecommand \@@href[1]{\endgroup#1\@@endlink}%
\providecommand \@sanitize@url [0]{\catcode `\\12\catcode `\$12\catcode `\&12\catcode `\#12\catcode `\^12\catcode `\_12\catcode `\%12\relax}%
\providecommand \@@startlink[1]{}%
\providecommand \@@endlink[0]{}%
\providecommand \url  [0]{\begingroup\@sanitize@url \@url }%
\providecommand \@url [1]{\endgroup\@href {#1}{\urlprefix }}%
\providecommand \urlprefix  [0]{URL }%
\providecommand \Eprint [0]{\href }%
\providecommand \doibase [0]{https://doi.org/}%
\providecommand \selectlanguage [0]{\@gobble}%
\providecommand \bibinfo  [0]{\@secondoftwo}%
\providecommand \bibfield  [0]{\@secondoftwo}%
\providecommand \translation [1]{[#1]}%
\providecommand \BibitemOpen [0]{}%
\providecommand \bibitemStop [0]{}%
\providecommand \bibitemNoStop [0]{.\EOS\space}%
\providecommand \EOS [0]{\spacefactor3000\relax}%
\providecommand \BibitemShut  [1]{\csname bibitem#1\endcsname}%
\let\auto@bib@innerbib\@empty
\bibitem [{\citenamefont {Aasi}\ \emph {et~al.}(2015)\citenamefont {Aasi} \emph {et~al.}}]{TheLIGOScientific:2014jea}%
  \BibitemOpen
  \bibfield  {author} {\bibinfo {author} {\bibfnamefont {J.}~\bibnamefont {Aasi}} \emph {et~al.} (\bibinfo {collaboration} {LIGO Scientific Collaboration}),\ }\bibfield  {title} {\bibinfo {title} {{Advanced LIGO}},\ }\href {https://doi.org/10.1088/0264-9381/32/7/074001} {\bibfield  {journal} {\bibinfo  {journal} {Class. Quantum Grav.}\ }\textbf {\bibinfo {volume} {32}},\ \bibinfo {pages} {074001} (\bibinfo {year} {2015})},\ \Eprint {https://arxiv.org/abs/1411.4547} {arXiv:1411.4547 [gr-qc]} \BibitemShut {NoStop}%
\bibitem [{\citenamefont {Abac}\ \emph {et~al.}(2025{\natexlab{a}})\citenamefont {Abac} \emph {et~al.}}]{GW250114}%
  \BibitemOpen
  \bibfield  {author} {\bibinfo {author} {\bibfnamefont {A.~G.}\ \bibnamefont {Abac}} \emph {et~al.},\ }\bibfield  {title} {\bibinfo {title} {{GW250114: testing Hawking's area law and the Kerr nature of black holes}},\ }\Eprint {https://arxiv.org/abs/2509.0000} {arXiv:2509.0000 [gr-qc]}  (\bibinfo {year} {2025}{\natexlab{a}})\BibitemShut {NoStop}%
\bibitem [{\citenamefont {Acernese}\ \emph {et~al.}(2015)\citenamefont {Acernese} \emph {et~al.}}]{TheVirgo:2014hva}%
  \BibitemOpen
  \bibfield  {author} {\bibinfo {author} {\bibfnamefont {F.}~\bibnamefont {Acernese}} \emph {et~al.} (\bibinfo {collaboration} {Virgo Collaboration}),\ }\bibfield  {title} {\bibinfo {title} {{Advanced Virgo: a second-generation interferometric gravitational wave detector}},\ }\href {https://doi.org/10.1088/0264-9381/32/2/024001} {\bibfield  {journal} {\bibinfo  {journal} {Class. Quantum Grav.}\ }\textbf {\bibinfo {volume} {32}},\ \bibinfo {pages} {024001} (\bibinfo {year} {2015})},\ \Eprint {https://arxiv.org/abs/1408.3978} {arXiv:1408.3978 [gr-qc]} \BibitemShut {NoStop}%
\bibitem [{\citenamefont {Akutsu}\ \emph {et~al.}(2021)\citenamefont {Akutsu} \emph {et~al.}}]{KAGRA:2020tym}%
  \BibitemOpen
  \bibfield  {author} {\bibinfo {author} {\bibfnamefont {T.}~\bibnamefont {Akutsu}} \emph {et~al.} (\bibinfo {collaboration} {KAGRA}),\ }\bibfield  {title} {\bibinfo {title} {{Overview of KAGRA: Detector design and construction history}},\ }\href {https://doi.org/10.1093/ptep/ptaa125} {\bibfield  {journal} {\bibinfo  {journal} {PTEP}\ }\textbf {\bibinfo {volume} {2021}},\ \bibinfo {pages} {05A101} (\bibinfo {year} {2021})},\ \Eprint {https://arxiv.org/abs/2005.05574} {arXiv:2005.05574 [physics.ins-det]} \BibitemShut {NoStop}%
\bibitem [{\citenamefont {Einstein}(1915)}]{Einstein:1915}%
  \BibitemOpen
  \bibfield  {author} {\bibinfo {author} {\bibfnamefont {A.}~\bibnamefont {Einstein}},\ }\bibfield  {title} {\bibinfo {title} {{On the General Thoery of Relativity}},\ }\href@noop {} {\bibfield  {journal} {\bibinfo  {journal} {Sitzungsber.\ K.\ Preuss.\ Akad.\ Wiss.}\ }\textbf {\bibinfo {volume} {49}},\ \bibinfo {pages} {778} (\bibinfo {year} {1915})}\BibitemShut {NoStop}%
\bibitem [{\citenamefont {Kerr}(1963)}]{Kerr:1963ud}%
  \BibitemOpen
  \bibfield  {author} {\bibinfo {author} {\bibfnamefont {R.~P.}\ \bibnamefont {Kerr}},\ }\bibfield  {title} {\bibinfo {title} {{Gravitational field of a spinning mass as an example of algebraically special metrics}},\ }\href {https://doi.org/10.1103/PhysRevLett.11.237} {\bibfield  {journal} {\bibinfo  {journal} {Phys. Rev. Lett.}\ }\textbf {\bibinfo {volume} {11}},\ \bibinfo {pages} {237} (\bibinfo {year} {1963})}\BibitemShut {NoStop}%
\bibitem [{\citenamefont {Arnowitt}\ \emph {et~al.}(2008)\citenamefont {Arnowitt}, \citenamefont {Deser},\ and\ \citenamefont {Misner}}]{Arnowitt:1962hi}%
  \BibitemOpen
  \bibfield  {author} {\bibinfo {author} {\bibfnamefont {R.~L.}\ \bibnamefont {Arnowitt}}, \bibinfo {author} {\bibfnamefont {S.}~\bibnamefont {Deser}},\ and\ \bibinfo {author} {\bibfnamefont {C.~W.}\ \bibnamefont {Misner}},\ }\bibfield  {title} {\bibinfo {title} {{The Dynamics of general relativity}},\ }\href {https://doi.org/10.1007/s10714-008-0661-1} {\bibfield  {journal} {\bibinfo  {journal} {Gen. Rel. Grav.}\ }\textbf {\bibinfo {volume} {40}},\ \bibinfo {pages} {1997} (\bibinfo {year} {2008})},\ \Eprint {https://arxiv.org/abs/gr-qc/0405109} {arXiv:gr-qc/0405109} \BibitemShut {NoStop}%
\bibitem [{\citenamefont {Foures-Bruhat}(1952)}]{Foures-Bruhat:1952grw}%
  \BibitemOpen
  \bibfield  {author} {\bibinfo {author} {\bibfnamefont {Y.}~\bibnamefont {Foures-Bruhat}},\ }\bibfield  {title} {\bibinfo {title} {{Theoreme d'existence pour certains systemes derivees partielles non lineaires}},\ }\href {https://doi.org/10.1007/BF02392131} {\bibfield  {journal} {\bibinfo  {journal} {Acta Mat.}\ }\textbf {\bibinfo {volume} {88}},\ \bibinfo {pages} {141} (\bibinfo {year} {1952})}\BibitemShut {NoStop}%
\bibitem [{\citenamefont {Schwarzschild}(1916)}]{Schwarzschild:1916uq}%
  \BibitemOpen
  \bibfield  {author} {\bibinfo {author} {\bibfnamefont {K.}~\bibnamefont {Schwarzschild}},\ }\bibfield  {title} {\bibinfo {title} {{On the gravitational field of a mass point according to Einstein's theory}},\ }\href@noop {} {\bibfield  {journal} {\bibinfo  {journal} {Sitzungsber. Preuss. Akad. Wiss. Berlin (Math. Phys. )}\ }\textbf {\bibinfo {volume} {1916}},\ \bibinfo {pages} {189} (\bibinfo {year} {1916})},\ \Eprint {https://arxiv.org/abs/physics/9905030} {arXiv:physics/9905030} \BibitemShut {NoStop}%
\bibitem [{\citenamefont {Newman}\ \emph {et~al.}(1965)\citenamefont {Newman}, \citenamefont {Couch}, \citenamefont {Chinnapared}, \citenamefont {Exton}, \citenamefont {Prakash},\ and\ \citenamefont {Torrence}}]{Newman:1965my}%
  \BibitemOpen
  \bibfield  {author} {\bibinfo {author} {\bibfnamefont {E.~T.}\ \bibnamefont {Newman}}, \bibinfo {author} {\bibfnamefont {E.}~\bibnamefont {Couch}}, \bibinfo {author} {\bibfnamefont {K.}~\bibnamefont {Chinnapared}}, \bibinfo {author} {\bibfnamefont {A.}~\bibnamefont {Exton}}, \bibinfo {author} {\bibfnamefont {A.}~\bibnamefont {Prakash}},\ and\ \bibinfo {author} {\bibfnamefont {R.}~\bibnamefont {Torrence}},\ }\bibfield  {title} {\bibinfo {title} {{Metric of a Rotating, Charged Mass}},\ }\href {https://doi.org/10.1063/1.1704351} {\bibfield  {journal} {\bibinfo  {journal} {J. Math. Phys.}\ }\textbf {\bibinfo {volume} {6}},\ \bibinfo {pages} {918} (\bibinfo {year} {1965})}\BibitemShut {NoStop}%
\bibitem [{\citenamefont {Israel}(1968)}]{Israel:1967za}%
  \BibitemOpen
  \bibfield  {author} {\bibinfo {author} {\bibfnamefont {W.}~\bibnamefont {Israel}},\ }\bibfield  {title} {\bibinfo {title} {{Event horizons in static electrovac space-times}},\ }\href {https://doi.org/10.1007/BF01645859} {\bibfield  {journal} {\bibinfo  {journal} {Commun. Math. Phys.}\ }\textbf {\bibinfo {volume} {8}},\ \bibinfo {pages} {245} (\bibinfo {year} {1968})}\BibitemShut {NoStop}%
\bibitem [{\citenamefont {Carter}(1971)}]{Carter:1971zc}%
  \BibitemOpen
  \bibfield  {author} {\bibinfo {author} {\bibfnamefont {B.}~\bibnamefont {Carter}},\ }\bibfield  {title} {\bibinfo {title} {{Axisymmetric Black Hole Has Only Two Degrees of Freedom}},\ }\href {https://doi.org/10.1103/PhysRevLett.26.331} {\bibfield  {journal} {\bibinfo  {journal} {Phys. Rev. Lett.}\ }\textbf {\bibinfo {volume} {26}},\ \bibinfo {pages} {331} (\bibinfo {year} {1971})}\BibitemShut {NoStop}%
\bibitem [{\citenamefont {Hawking}(1972)}]{Hawking:1971vc}%
  \BibitemOpen
  \bibfield  {author} {\bibinfo {author} {\bibfnamefont {S.~W.}\ \bibnamefont {Hawking}},\ }\bibfield  {title} {\bibinfo {title} {{Black holes in general relativity}},\ }\href {https://doi.org/10.1007/BF01877517} {\bibfield  {journal} {\bibinfo  {journal} {Commun. Math. Phys.}\ }\textbf {\bibinfo {volume} {25}},\ \bibinfo {pages} {152} (\bibinfo {year} {1972})}\BibitemShut {NoStop}%
\bibitem [{\citenamefont {Robinson}(1975)}]{Robinson:1975bv}%
  \BibitemOpen
  \bibfield  {author} {\bibinfo {author} {\bibfnamefont {D.~C.}\ \bibnamefont {Robinson}},\ }\bibfield  {title} {\bibinfo {title} {Uniqueness of the {Kerr} black hole},\ }\href {https://doi.org/10.1103/PhysRevLett.34.905} {\bibfield  {journal} {\bibinfo  {journal} {Phys. Rev. Lett.}\ }\textbf {\bibinfo {volume} {34}},\ \bibinfo {pages} {905} (\bibinfo {year} {1975})}\BibitemShut {NoStop}%
\bibitem [{\citenamefont {Mazur}(1982)}]{Mazur:1982db}%
  \BibitemOpen
  \bibfield  {author} {\bibinfo {author} {\bibfnamefont {P.~O.}\ \bibnamefont {Mazur}},\ }\bibfield  {title} {\bibinfo {title} {Proof of uniqueness of the {Kerr-Newman} black hole solution},\ }\href {https://doi.org/10.1088/0305-4470/15/10/021} {\bibfield  {journal} {\bibinfo  {journal} {J. Phys. A}\ }\textbf {\bibinfo {volume} {15}},\ \bibinfo {pages} {3173} (\bibinfo {year} {1982})}\BibitemShut {NoStop}%
\bibitem [{\citenamefont {Carter}(1968)}]{Carter:1968rr}%
  \BibitemOpen
  \bibfield  {author} {\bibinfo {author} {\bibfnamefont {B.}~\bibnamefont {Carter}},\ }\bibfield  {title} {\bibinfo {title} {{Global structure of the Kerr family of gravitational fields}},\ }\href {https://doi.org/10.1103/PhysRev.174.1559} {\bibfield  {journal} {\bibinfo  {journal} {Phys. Rev.}\ }\textbf {\bibinfo {volume} {174}},\ \bibinfo {pages} {1559} (\bibinfo {year} {1968})}\BibitemShut {NoStop}%
\bibitem [{\citenamefont {Penrose}(1969)}]{Penrose:1969pc}%
  \BibitemOpen
  \bibfield  {author} {\bibinfo {author} {\bibfnamefont {R.}~\bibnamefont {Penrose}},\ }\bibfield  {title} {\bibinfo {title} {{Gravitational collapse: The role of general relativity}},\ }\href@noop {} {\bibfield  {journal} {\bibinfo  {journal} {Riv. Nuovo Cim.}\ }\textbf {\bibinfo {volume} {1}},\ \bibinfo {pages} {252} (\bibinfo {year} {1969})}\BibitemShut {NoStop}%
\bibitem [{\citenamefont {Penrose}\ and\ \citenamefont {Floyd}(1971)}]{Penrose:1971uk}%
  \BibitemOpen
  \bibfield  {author} {\bibinfo {author} {\bibfnamefont {R.}~\bibnamefont {Penrose}}\ and\ \bibinfo {author} {\bibfnamefont {R.~M.}\ \bibnamefont {Floyd}},\ }\bibfield  {title} {\bibinfo {title} {{Extraction of rotational energy from a black hole}},\ }\href {https://doi.org/10.1038/physci229177a0} {\bibfield  {journal} {\bibinfo  {journal} {Nature}\ }\textbf {\bibinfo {volume} {229}},\ \bibinfo {pages} {177} (\bibinfo {year} {1971})}\BibitemShut {NoStop}%
\bibitem [{\citenamefont {Bardeen}\ \emph {et~al.}(1973)\citenamefont {Bardeen}, \citenamefont {Carter},\ and\ \citenamefont {Hawking}}]{Bardeen:1973gs}%
  \BibitemOpen
  \bibfield  {author} {\bibinfo {author} {\bibfnamefont {J.~M.}\ \bibnamefont {Bardeen}}, \bibinfo {author} {\bibfnamefont {B.}~\bibnamefont {Carter}},\ and\ \bibinfo {author} {\bibfnamefont {S.~W.}\ \bibnamefont {Hawking}},\ }\bibfield  {title} {\bibinfo {title} {{The Four laws of black hole mechanics}},\ }\href {https://doi.org/10.1007/BF01645742} {\bibfield  {journal} {\bibinfo  {journal} {Commun. Math. Phys.}\ }\textbf {\bibinfo {volume} {31}},\ \bibinfo {pages} {161} (\bibinfo {year} {1973})}\BibitemShut {NoStop}%
\bibitem [{\citenamefont {Schmidt}(1963)}]{Schmidt:1963wkp}%
  \BibitemOpen
  \bibfield  {author} {\bibinfo {author} {\bibfnamefont {M.}~\bibnamefont {Schmidt}},\ }\bibfield  {title} {\bibinfo {title} {{3C 273 : A Star-Like Object with Large Red-Shift}},\ }\href {https://doi.org/10.1038/1971040a0} {\bibfield  {journal} {\bibinfo  {journal} {Nature}\ }\textbf {\bibinfo {volume} {197}},\ \bibinfo {pages} {1040} (\bibinfo {year} {1963})}\BibitemShut {NoStop}%
\bibitem [{\citenamefont {Dafermos}\ \emph {et~al.}(2019)\citenamefont {Dafermos}, \citenamefont {Holzegel},\ and\ \citenamefont {Rodnianski}}]{Dafermos:2016uzj}%
  \BibitemOpen
  \bibfield  {author} {\bibinfo {author} {\bibfnamefont {M.}~\bibnamefont {Dafermos}}, \bibinfo {author} {\bibfnamefont {G.}~\bibnamefont {Holzegel}},\ and\ \bibinfo {author} {\bibfnamefont {I.}~\bibnamefont {Rodnianski}},\ }\bibfield  {title} {\bibinfo {title} {{The linear stability of the Schwarzschild solution to gravitational perturbations}},\ }\href {https://doi.org/10.4310/acta.2019.v222.n1.a1} {\bibfield  {journal} {\bibinfo  {journal} {Acta Mat.}\ }\textbf {\bibinfo {volume} {222}},\ \bibinfo {pages} {1} (\bibinfo {year} {2019})},\ \Eprint {https://arxiv.org/abs/1601.06467} {arXiv:1601.06467 [gr-qc]} \BibitemShut {NoStop}%
\bibitem [{\citenamefont {Teixeira~da Costa}(2020)}]{TeixeiradaCosta:2019skg}%
  \BibitemOpen
  \bibfield  {author} {\bibinfo {author} {\bibfnamefont {R.}~\bibnamefont {Teixeira~da Costa}},\ }\bibfield  {title} {\bibinfo {title} {{Mode stability for the Teukolsky equation on extremal and subextremal Kerr spacetimes}},\ }\href {https://doi.org/10.1007/s00220-020-03796-z} {\bibfield  {journal} {\bibinfo  {journal} {Commun. Math. Phys.}\ }\textbf {\bibinfo {volume} {378}},\ \bibinfo {pages} {705} (\bibinfo {year} {2020})},\ \Eprint {https://arxiv.org/abs/1910.02854} {arXiv:1910.02854 [gr-qc]} \BibitemShut {NoStop}%
\bibitem [{\citenamefont {Klainerman}\ and\ \citenamefont {Szeftel}(2023)}]{Klainerman:2021qzy}%
  \BibitemOpen
  \bibfield  {author} {\bibinfo {author} {\bibfnamefont {S.}~\bibnamefont {Klainerman}}\ and\ \bibinfo {author} {\bibfnamefont {J.}~\bibnamefont {Szeftel}},\ }\bibfield  {title} {\bibinfo {title} {{Kerr stability for small angular momentum}},\ }\href {https://doi.org/10.4310/PAMQ.2023.v19.n3.a1} {\bibfield  {journal} {\bibinfo  {journal} {Pure Appl. Math. Quart.}\ }\textbf {\bibinfo {volume} {19}},\ \bibinfo {pages} {791} (\bibinfo {year} {2023})},\ \Eprint {https://arxiv.org/abs/2104.11857} {arXiv:2104.11857 [math.AP]} \BibitemShut {NoStop}%
\bibitem [{\citenamefont {Dafermos}\ \emph {et~al.}(2021)\citenamefont {Dafermos}, \citenamefont {Holzegel}, \citenamefont {Rodnianski},\ and\ \citenamefont {Taylor}}]{Dafermos:2021cbw}%
  \BibitemOpen
  \bibfield  {author} {\bibinfo {author} {\bibfnamefont {M.}~\bibnamefont {Dafermos}}, \bibinfo {author} {\bibfnamefont {G.}~\bibnamefont {Holzegel}}, \bibinfo {author} {\bibfnamefont {I.}~\bibnamefont {Rodnianski}},\ and\ \bibinfo {author} {\bibfnamefont {M.}~\bibnamefont {Taylor}},\ }\bibfield  {title} {\bibinfo {title} {{The non-linear stability of the Schwarzschild family of black holes}},\ }\Eprint {https://arxiv.org/abs/2104.08222} {arXiv:2104.08222 [gr-qc]}  (\bibinfo {year} {2021})\BibitemShut {NoStop}%
\bibitem [{\citenamefont {Hawking}\ and\ \citenamefont {Penrose}(1970)}]{Hawking:1970zqf}%
  \BibitemOpen
  \bibfield  {author} {\bibinfo {author} {\bibfnamefont {S.~W.}\ \bibnamefont {Hawking}}\ and\ \bibinfo {author} {\bibfnamefont {R.}~\bibnamefont {Penrose}},\ }\bibfield  {title} {\bibinfo {title} {{The Singularities of gravitational collapse and cosmology}},\ }\href {https://doi.org/10.1098/rspa.1970.0021} {\bibfield  {journal} {\bibinfo  {journal} {Proc. Roy. Soc. Lond. A}\ }\textbf {\bibinfo {volume} {314}},\ \bibinfo {pages} {529} (\bibinfo {year} {1970})}\BibitemShut {NoStop}%
\bibitem [{\citenamefont {Almheiri}\ \emph {et~al.}(2021)\citenamefont {Almheiri}, \citenamefont {Hartman}, \citenamefont {Maldacena}, \citenamefont {Shaghoulian},\ and\ \citenamefont {Tajdini}}]{Almheiri:2020cfm}%
  \BibitemOpen
  \bibfield  {author} {\bibinfo {author} {\bibfnamefont {A.}~\bibnamefont {Almheiri}}, \bibinfo {author} {\bibfnamefont {T.}~\bibnamefont {Hartman}}, \bibinfo {author} {\bibfnamefont {J.}~\bibnamefont {Maldacena}}, \bibinfo {author} {\bibfnamefont {E.}~\bibnamefont {Shaghoulian}},\ and\ \bibinfo {author} {\bibfnamefont {A.}~\bibnamefont {Tajdini}},\ }\bibfield  {title} {\bibinfo {title} {{The entropy of Hawking radiation}},\ }\href {https://doi.org/10.1103/RevModPhys.93.035002} {\bibfield  {journal} {\bibinfo  {journal} {Rev. Mod. Phys.}\ }\textbf {\bibinfo {volume} {93}},\ \bibinfo {pages} {035002} (\bibinfo {year} {2021})},\ \Eprint {https://arxiv.org/abs/2006.06872} {arXiv:2006.06872 [hep-th]} \BibitemShut {NoStop}%
\bibitem [{\citenamefont {Raju}(2022)}]{Raju:2020smc}%
  \BibitemOpen
  \bibfield  {author} {\bibinfo {author} {\bibfnamefont {S.}~\bibnamefont {Raju}},\ }\bibfield  {title} {\bibinfo {title} {{Lessons from the information paradox}},\ }\href {https://doi.org/10.1016/j.physrep.2021.10.001} {\bibfield  {journal} {\bibinfo  {journal} {Phys. Rept.}\ }\textbf {\bibinfo {volume} {943}},\ \bibinfo {pages} {1} (\bibinfo {year} {2022})},\ \Eprint {https://arxiv.org/abs/2012.05770} {arXiv:2012.05770 [hep-th]} \BibitemShut {NoStop}%
\bibitem [{\citenamefont {Will}(2014)}]{Will:2014kxa}%
  \BibitemOpen
  \bibfield  {author} {\bibinfo {author} {\bibfnamefont {C.~M.}\ \bibnamefont {Will}},\ }\bibfield  {title} {\bibinfo {title} {{The Confrontation between General Relativity and Experiment}},\ }\href {https://doi.org/10.12942/lrr-2014-4} {\bibfield  {journal} {\bibinfo  {journal} {Living Rev. Relativity}\ }\textbf {\bibinfo {volume} {17}},\ \bibinfo {pages} {4} (\bibinfo {year} {2014})},\ \Eprint {https://arxiv.org/abs/1403.7377} {arXiv:1403.7377 [gr-qc]} \BibitemShut {NoStop}%
\bibitem [{\citenamefont {Berti}\ \emph {et~al.}(2015)\citenamefont {Berti} \emph {et~al.}}]{Berti:2015itd}%
  \BibitemOpen
  \bibfield  {author} {\bibinfo {author} {\bibfnamefont {E.}~\bibnamefont {Berti}} \emph {et~al.},\ }\bibfield  {title} {\bibinfo {title} {{Testing General Relativity with Present and Future Astrophysical Observations}},\ }\href {https://doi.org/10.1088/0264-9381/32/24/243001} {\bibfield  {journal} {\bibinfo  {journal} {Class. Quantum Grav.}\ }\textbf {\bibinfo {volume} {32}},\ \bibinfo {pages} {243001} (\bibinfo {year} {2015})},\ \Eprint {https://arxiv.org/abs/1501.07274} {arXiv:1501.07274 [gr-qc]} \BibitemShut {NoStop}%
\bibitem [{\citenamefont {Yunes}\ \emph {et~al.}(2025)\citenamefont {Yunes}, \citenamefont {Siemens},\ and\ \citenamefont {Yagi}}]{Yunes:2024lzm}%
  \BibitemOpen
  \bibfield  {author} {\bibinfo {author} {\bibfnamefont {N.}~\bibnamefont {Yunes}}, \bibinfo {author} {\bibfnamefont {X.}~\bibnamefont {Siemens}},\ and\ \bibinfo {author} {\bibfnamefont {K.}~\bibnamefont {Yagi}},\ }\bibfield  {title} {\bibinfo {title} {{Gravitational-Wave Tests of General Relativity with Ground-Based Detectors and Pulsar-Timing Arrays}},\ }\href {https://doi.org/10.1007/s41114-024-00054-9} {\bibfield  {journal} {\bibinfo  {journal} {Living Rev. Relativity}\ }\textbf {\bibinfo {volume} {28}},\ \bibinfo {pages} {3} (\bibinfo {year} {2025})},\ \Eprint {https://arxiv.org/abs/2408.05240} {arXiv:2408.05240 [gr-qc]} \BibitemShut {NoStop}%
\bibitem [{\citenamefont {Abac}\ \emph {et~al.}(2025{\natexlab{b}})\citenamefont {Abac} \emph {et~al.}}]{LIGOScientific:2025hdt}%
  \BibitemOpen
  \bibfield  {author} {\bibinfo {author} {\bibfnamefont {A.~G.}\ \bibnamefont {Abac}} \emph {et~al.} (\bibinfo {collaboration} {LIGO Scientific, VIRGO, KAGRA}),\ }\bibfield  {title} {\bibinfo {title} {{GWTC-4.0: An Introduction to Version 4.0 of the Gravitational-Wave Transient Catalog}},\ }\href@noop {} {\  (\bibinfo {year} {2025}{\natexlab{b}})},\ \Eprint {https://arxiv.org/abs/2508.18080} {arXiv:2508.18080 [gr-qc]} \BibitemShut {NoStop}%
\bibitem [{LIG(2025{\natexlab{a}})}]{LIGOScientific:2025slb}%
  \BibitemOpen
  \bibfield  {title} {\bibinfo {title} {{GWTC-4.0: Updating the Gravitational-Wave Transient Catalog with Observations from the First Part of the Fourth LIGO-Virgo-KAGRA Observing Run}},\ }\href@noop {} {\  (\bibinfo {year} {2025}{\natexlab{a}})},\ \Eprint {https://arxiv.org/abs/2508.18082} {arXiv:2508.18082 [gr-qc]} \BibitemShut {NoStop}%
\bibitem [{\citenamefont {{Blanchet}}\ and\ \citenamefont {{Sathyaprakash}}(1995)}]{1995PhRvL..74.1067B}%
  \BibitemOpen
  \bibfield  {author} {\bibinfo {author} {\bibfnamefont {L.}~\bibnamefont {{Blanchet}}}\ and\ \bibinfo {author} {\bibfnamefont {B.~S.}\ \bibnamefont {{Sathyaprakash}}},\ }\bibfield  {title} {\bibinfo {title} {{Detecting a Tail Effect in Gravitational-Wave Experiments}},\ }\href {https://doi.org/10.1103/PhysRevLett.74.1067} {\bibfield  {journal} {\bibinfo  {journal} {Physical Review Letters}\ }\textbf {\bibinfo {volume} {74}},\ \bibinfo {pages} {1067} (\bibinfo {year} {1995})}\BibitemShut {NoStop}%
\bibitem [{\citenamefont {Abbott}\ \emph {et~al.}(2016{\natexlab{a}})\citenamefont {Abbott} \emph {et~al.}}]{GW150914_paper}%
  \BibitemOpen
  \bibfield  {author} {\bibinfo {author} {\bibfnamefont {B.~P.}\ \bibnamefont {Abbott}} \emph {et~al.} (\bibinfo {collaboration} {LIGO Scientific Collaboration, Virgo Collaboration}),\ }\bibfield  {title} {\bibinfo {title} {{Observation of Gravitational Waves from a Binary Black Hole Merger}},\ }\href {https://doi.org/10.1103/PhysRevLett.116.061102} {\bibfield  {journal} {\bibinfo  {journal} {Phys. Rev. Lett.}\ }\textbf {\bibinfo {volume} {116}},\ \bibinfo {eid} {061102} (\bibinfo {year} {2016}{\natexlab{a}})},\ \Eprint {https://arxiv.org/abs/1602.03837} {arXiv:1602.03837 [gr-qc]} \BibitemShut {NoStop}%
\bibitem [{\citenamefont {Abbott}\ \emph {et~al.}(2016{\natexlab{b}})\citenamefont {Abbott} \emph {et~al.}}]{LIGOScientific:2016lio}%
  \BibitemOpen
  \bibfield  {author} {\bibinfo {author} {\bibfnamefont {B.~P.}\ \bibnamefont {Abbott}} \emph {et~al.} (\bibinfo {collaboration} {LIGO Scientific, Virgo}),\ }\bibfield  {title} {\bibinfo {title} {{Tests of general relativity with GW150914}},\ }\href {https://doi.org/10.1103/PhysRevLett.116.221101} {\bibfield  {journal} {\bibinfo  {journal} {Phys. Rev. Lett.}\ }\textbf {\bibinfo {volume} {116}},\ \bibinfo {pages} {221101} (\bibinfo {year} {2016}{\natexlab{b}})},\ \bibinfo {note} {[Erratum: Phys.Rev.Lett. 121, 129902 (2018)]},\ \Eprint {https://arxiv.org/abs/1602.03841} {arXiv:1602.03841 [gr-qc]} \BibitemShut {NoStop}%
\bibitem [{\citenamefont {Abbott}\ \emph {et~al.}(2019{\natexlab{a}})\citenamefont {Abbott} \emph {et~al.}}]{GWTC1}%
  \BibitemOpen
  \bibfield  {author} {\bibinfo {author} {\bibfnamefont {B.~P.}\ \bibnamefont {Abbott}} \emph {et~al.} (\bibinfo {collaboration} {LIGO Scientific Collaboration, Virgo Collaboration}),\ }\bibfield  {title} {\bibinfo {title} {{GWTC-1: A Gravitational-Wave Transient Catalog of Compact Binary Mergers Observed by LIGO and Virgo during the First and Second Observing Runs}},\ }\href {https://doi.org/10.1103/PhysRevX.9.031040} {\bibfield  {journal} {\bibinfo  {journal} {Phys. Rev. X}\ }\textbf {\bibinfo {volume} {9}},\ \bibinfo {pages} {031040} (\bibinfo {year} {2019}{\natexlab{a}})},\ \Eprint {https://arxiv.org/abs/1811.12907} {arXiv:1811.12907 [astro-ph.HE]} \BibitemShut {NoStop}%
\bibitem [{\citenamefont {Abbott}\ \emph {et~al.}(2021{\natexlab{a}})\citenamefont {Abbott} \emph {et~al.}}]{GWTC2}%
  \BibitemOpen
  \bibfield  {author} {\bibinfo {author} {\bibfnamefont {R.}~\bibnamefont {Abbott}} \emph {et~al.} (\bibinfo {collaboration} {LIGO Scientific, Virgo}),\ }\bibfield  {title} {\bibinfo {title} {{GWTC-2: Compact Binary Coalescences Observed by LIGO and Virgo During the First Half of the Third Observing Run}},\ }\href {https://doi.org/10.1103/PhysRevX.11.021053} {\bibfield  {journal} {\bibinfo  {journal} {Phys. Rev. X}\ }\textbf {\bibinfo {volume} {11}},\ \bibinfo {pages} {021053} (\bibinfo {year} {2021}{\natexlab{a}})},\ \Eprint {https://arxiv.org/abs/2010.14527} {arXiv:2010.14527 [gr-qc]} \BibitemShut {NoStop}%
\bibitem [{\citenamefont {Abbott}\ \emph {et~al.}(2024)\citenamefont {Abbott} \emph {et~al.}}]{GWTC2p1}%
  \BibitemOpen
  \bibfield  {author} {\bibinfo {author} {\bibfnamefont {R.}~\bibnamefont {Abbott}} \emph {et~al.} (\bibinfo {collaboration} {LIGO Scientific, Virgo}),\ }\bibfield  {title} {\bibinfo {title} {{GWTC-2.1: Deep extended catalog of compact binary coalescences observed by LIGO and Virgo during the first half of the third observing run}},\ }\href {https://doi.org/10.1103/PhysRevD.109.022001} {\bibfield  {journal} {\bibinfo  {journal} {Phys. Rev. D}\ }\textbf {\bibinfo {volume} {109}},\ \bibinfo {pages} {022001} (\bibinfo {year} {2024})},\ \Eprint {https://arxiv.org/abs/2108.01045} {arXiv:2108.01045 [gr-qc]} \BibitemShut {NoStop}%
\bibitem [{\citenamefont {Abbott}\ \emph {et~al.}(2023)\citenamefont {Abbott} \emph {et~al.}}]{GWTC3}%
  \BibitemOpen
  \bibfield  {author} {\bibinfo {author} {\bibfnamefont {R.}~\bibnamefont {Abbott}} \emph {et~al.} (\bibinfo {collaboration} {KAGRA, Virgo, LIGO Scientific}),\ }\bibfield  {title} {\bibinfo {title} {{GWTC-3: Compact Binary Coalescences Observed by LIGO and Virgo during the Second Part of the Third Observing Run}},\ }\href {https://doi.org/10.1103/PhysRevX.13.041039} {\bibfield  {journal} {\bibinfo  {journal} {Phys. Rev. X}\ }\textbf {\bibinfo {volume} {13}},\ \bibinfo {pages} {041039} (\bibinfo {year} {2023})},\ \Eprint {https://arxiv.org/abs/2111.03606} {arXiv:2111.03606 [gr-qc]} \BibitemShut {NoStop}%
\bibitem [{\citenamefont {Abbott}\ \emph {et~al.}(2019{\natexlab{b}})\citenamefont {Abbott} \emph {et~al.}}]{LIGOScientific:2018dkp}%
  \BibitemOpen
  \bibfield  {author} {\bibinfo {author} {\bibfnamefont {B.~P.}\ \bibnamefont {Abbott}} \emph {et~al.} (\bibinfo {collaboration} {LIGO Scientific, Virgo}),\ }\bibfield  {title} {\bibinfo {title} {{Tests of General Relativity with GW170817}},\ }\href {https://doi.org/10.1103/PhysRevLett.123.011102} {\bibfield  {journal} {\bibinfo  {journal} {Phys. Rev. Lett.}\ }\textbf {\bibinfo {volume} {123}},\ \bibinfo {pages} {011102} (\bibinfo {year} {2019}{\natexlab{b}})},\ \Eprint {https://arxiv.org/abs/1811.00364} {arXiv:1811.00364 [gr-qc]} \BibitemShut {NoStop}%
\bibitem [{\citenamefont {Abbott}\ \emph {et~al.}(2019{\natexlab{c}})\citenamefont {Abbott} \emph {et~al.}}]{LIGOScientific:2019fpa}%
  \BibitemOpen
  \bibfield  {author} {\bibinfo {author} {\bibfnamefont {B.~P.}\ \bibnamefont {Abbott}} \emph {et~al.} (\bibinfo {collaboration} {LIGO Scientific Collaboration, Virgo Collaboration}),\ }\bibfield  {title} {\bibinfo {title} {{Tests of General Relativity with the Binary Black Hole Signals from the LIGO-Virgo Catalog GWTC-1}},\ }\href {https://doi.org/10.1103/PhysRevD.100.104036} {\bibfield  {journal} {\bibinfo  {journal} {Phys. Rev. D}\ }\textbf {\bibinfo {volume} {100}},\ \bibinfo {pages} {104036} (\bibinfo {year} {2019}{\natexlab{c}})},\ \Eprint {https://arxiv.org/abs/1903.04467} {arXiv:1903.04467 [gr-qc]} \BibitemShut {NoStop}%
\bibitem [{\citenamefont {Abbott}\ \emph {et~al.}(2021{\natexlab{b}})\citenamefont {Abbott} \emph {et~al.}}]{LIGOScientific:2020tif}%
  \BibitemOpen
  \bibfield  {author} {\bibinfo {author} {\bibfnamefont {R.}~\bibnamefont {Abbott}} \emph {et~al.} (\bibinfo {collaboration} {LIGO Scientific Collaboration, Virgo Collaboration}),\ }\bibfield  {title} {\bibinfo {title} {{Tests of general relativity with binary black holes from the second LIGO-Virgo gravitational-wave transient catalog}},\ }\href {https://doi.org/10.1103/PhysRevD.103.122002} {\bibfield  {journal} {\bibinfo  {journal} {Phys. Rev. D}\ }\textbf {\bibinfo {volume} {103}},\ \bibinfo {pages} {122002} (\bibinfo {year} {2021}{\natexlab{b}})},\ \Eprint {https://arxiv.org/abs/2010.14529} {arXiv:2010.14529 [gr-qc]} \BibitemShut {NoStop}%
\bibitem [{\citenamefont {Abbott}\ \emph {et~al.}(2021{\natexlab{c}})\citenamefont {Abbott} \emph {et~al.}}]{LIGOScientific:2021sio}%
  \BibitemOpen
  \bibfield  {author} {\bibinfo {author} {\bibfnamefont {R.}~\bibnamefont {Abbott}} \emph {et~al.} (\bibinfo {collaboration} {LIGO Scientific, Virgo, KAGRA}),\ }\bibfield  {title} {\bibinfo {title} {{Tests of General Relativity with GWTC-3}},\ }\Eprint {https://arxiv.org/abs/2112.06861} {arXiv:2112.06861 [gr-qc]}  (\bibinfo {year} {2021}{\natexlab{c}})\BibitemShut {NoStop}%
\bibitem [{\citenamefont {Freire}\ \emph {et~al.}(2012)\citenamefont {Freire}, \citenamefont {Wex}, \citenamefont {Esposito-Farese}, \citenamefont {Verbiest}, \citenamefont {Bailes}, \citenamefont {Jacoby}, \citenamefont {Kramer}, \citenamefont {Stairs}, \citenamefont {Antoniadis},\ and\ \citenamefont {Janssen}}]{Freire:2012mg}%
  \BibitemOpen
  \bibfield  {author} {\bibinfo {author} {\bibfnamefont {P.~C.~C.}\ \bibnamefont {Freire}}, \bibinfo {author} {\bibfnamefont {N.}~\bibnamefont {Wex}}, \bibinfo {author} {\bibfnamefont {G.}~\bibnamefont {Esposito-Farese}}, \bibinfo {author} {\bibfnamefont {J.~P.~W.}\ \bibnamefont {Verbiest}}, \bibinfo {author} {\bibfnamefont {M.}~\bibnamefont {Bailes}}, \bibinfo {author} {\bibfnamefont {B.~A.}\ \bibnamefont {Jacoby}}, \bibinfo {author} {\bibfnamefont {M.}~\bibnamefont {Kramer}}, \bibinfo {author} {\bibfnamefont {I.~H.}\ \bibnamefont {Stairs}}, \bibinfo {author} {\bibfnamefont {J.}~\bibnamefont {Antoniadis}},\ and\ \bibinfo {author} {\bibfnamefont {G.~H.}\ \bibnamefont {Janssen}},\ }\bibfield  {title} {\bibinfo {title} {{The relativistic pulsar-white dwarf binary PSR J1738+0333 II. The most stringent test of scalar-tensor gravity}},\ }\href {https://doi.org/10.1111/j.1365-2966.2012.21253.x} {\bibfield  {journal} {\bibinfo  {journal} {Mon. Not. Roy. Astron. Soc.}\ }\textbf {\bibinfo {volume} {423}},\ \bibinfo {pages} {3328} (\bibinfo {year} {2012})},\ \Eprint {https://arxiv.org/abs/1205.1450} {arXiv:1205.1450 [astro-ph.GA]} \BibitemShut {NoStop}%
\bibitem [{\citenamefont {Kramer}\ \emph {et~al.}(2021)\citenamefont {Kramer} \emph {et~al.}}]{Kramer:2021jcw}%
  \BibitemOpen
  \bibfield  {author} {\bibinfo {author} {\bibfnamefont {M.}~\bibnamefont {Kramer}} \emph {et~al.},\ }\bibfield  {title} {\bibinfo {title} {{Strong-Field Gravity Tests with the Double Pulsar}},\ }\href {https://doi.org/10.1103/PhysRevX.11.041050} {\bibfield  {journal} {\bibinfo  {journal} {Phys. Rev. X}\ }\textbf {\bibinfo {volume} {11}},\ \bibinfo {pages} {041050} (\bibinfo {year} {2021})},\ \Eprint {https://arxiv.org/abs/2112.06795} {arXiv:2112.06795 [astro-ph.HE]} \BibitemShut {NoStop}%
\bibitem [{\citenamefont {Abuter}\ \emph {et~al.}(2018)\citenamefont {Abuter} \emph {et~al.}}]{GRAVITY:2018ofz}%
  \BibitemOpen
  \bibfield  {author} {\bibinfo {author} {\bibfnamefont {R.}~\bibnamefont {Abuter}} \emph {et~al.} (\bibinfo {collaboration} {GRAVITY}),\ }\bibfield  {title} {\bibinfo {title} {{Detection of the gravitational redshift in the orbit of the star S2 near the Galactic centre massive black hole}},\ }\href {https://doi.org/10.1051/0004-6361/201833718} {\bibfield  {journal} {\bibinfo  {journal} {Astron. Astrophys.}\ }\textbf {\bibinfo {volume} {615}},\ \bibinfo {pages} {L15} (\bibinfo {year} {2018})},\ \Eprint {https://arxiv.org/abs/1807.09409} {arXiv:1807.09409 [astro-ph.GA]} \BibitemShut {NoStop}%
\bibitem [{\citenamefont {Do}\ \emph {et~al.}(2019)\citenamefont {Do} \emph {et~al.}}]{Do:2019txf}%
  \BibitemOpen
  \bibfield  {author} {\bibinfo {author} {\bibfnamefont {T.}~\bibnamefont {Do}} \emph {et~al.},\ }\bibfield  {title} {\bibinfo {title} {{Relativistic redshift of the star S0-2 orbiting the Galactic center supermassive black hole}},\ }\href {https://doi.org/10.1126/science.aav8137} {\bibfield  {journal} {\bibinfo  {journal} {Science}\ }\textbf {\bibinfo {volume} {365}},\ \bibinfo {pages} {664} (\bibinfo {year} {2019})},\ \Eprint {https://arxiv.org/abs/1907.10731} {arXiv:1907.10731 [astro-ph.GA]} \BibitemShut {NoStop}%
\bibitem [{\citenamefont {Akiyama}\ \emph {et~al.}(2019)\citenamefont {Akiyama} \emph {et~al.}}]{EventHorizonTelescope:2019dse}%
  \BibitemOpen
  \bibfield  {author} {\bibinfo {author} {\bibfnamefont {K.}~\bibnamefont {Akiyama}} \emph {et~al.} (\bibinfo {collaboration} {Event Horizon Telescope}),\ }\bibfield  {title} {\bibinfo {title} {{First M87 Event Horizon Telescope Results. I. The Shadow of the Supermassive Black Hole}},\ }\href {https://doi.org/10.3847/2041-8213/ab0ec7} {\bibfield  {journal} {\bibinfo  {journal} {Astrophys. J. Lett.}\ }\textbf {\bibinfo {volume} {875}},\ \bibinfo {pages} {L1} (\bibinfo {year} {2019})},\ \Eprint {https://arxiv.org/abs/1906.11238} {arXiv:1906.11238 [astro-ph.GA]} \BibitemShut {NoStop}%
\bibitem [{\citenamefont {Clifton}\ \emph {et~al.}(2012)\citenamefont {Clifton}, \citenamefont {Ferreira}, \citenamefont {Padilla},\ and\ \citenamefont {Skordis}}]{Clifton:2011jh}%
  \BibitemOpen
  \bibfield  {author} {\bibinfo {author} {\bibfnamefont {T.}~\bibnamefont {Clifton}}, \bibinfo {author} {\bibfnamefont {P.~G.}\ \bibnamefont {Ferreira}}, \bibinfo {author} {\bibfnamefont {A.}~\bibnamefont {Padilla}},\ and\ \bibinfo {author} {\bibfnamefont {C.}~\bibnamefont {Skordis}},\ }\bibfield  {title} {\bibinfo {title} {{Modified Gravity and Cosmology}},\ }\href {https://doi.org/10.1016/j.physrep.2012.01.001} {\bibfield  {journal} {\bibinfo  {journal} {Phys. Rep.}\ }\textbf {\bibinfo {volume} {513}},\ \bibinfo {pages} {1} (\bibinfo {year} {2012})},\ \Eprint {https://arxiv.org/abs/1106.2476} {arXiv:1106.2476 [astro-ph.CO]} \BibitemShut {NoStop}%
\bibitem [{\citenamefont {Einstein}(1918)}]{Einstein:1918btx}%
  \BibitemOpen
  \bibfield  {author} {\bibinfo {author} {\bibfnamefont {A.}~\bibnamefont {Einstein}},\ }\bibfield  {title} {\bibinfo {title} {{{\"U}ber Gravitationswellen}},\ }\href@noop {} {\bibfield  {journal} {\bibinfo  {journal} {Sitzungsber. Preuss. Akad. Wiss. Berlin (Math. Phys. )}\ }\textbf {\bibinfo {volume} {1918}},\ \bibinfo {pages} {154} (\bibinfo {year} {1918})}\BibitemShut {NoStop}%
\bibitem [{\citenamefont {Pretorius}(2005)}]{Pretorius:2005gq}%
  \BibitemOpen
  \bibfield  {author} {\bibinfo {author} {\bibfnamefont {F.}~\bibnamefont {Pretorius}},\ }\bibfield  {title} {\bibinfo {title} {{Evolution of binary black hole spacetimes}},\ }\href {https://doi.org/10.1103/PhysRevLett.95.121101} {\bibfield  {journal} {\bibinfo  {journal} {\prl}\ }\textbf {\bibinfo {volume} {95}},\ \bibinfo {pages} {121101} (\bibinfo {year} {2005})},\ \Eprint {https://arxiv.org/abs/0507014} {arXiv:0507014 [gr-qc]} \BibitemShut {NoStop}%
\bibitem [{\citenamefont {Campanelli}\ \emph {et~al.}(2007)\citenamefont {Campanelli}, \citenamefont {Lousto}, \citenamefont {Zlochower}, \citenamefont {Krishnan},\ and\ \citenamefont {Merritt}}]{Campanelli:2006fy}%
  \BibitemOpen
  \bibfield  {author} {\bibinfo {author} {\bibfnamefont {M.}~\bibnamefont {Campanelli}}, \bibinfo {author} {\bibfnamefont {C.~O.}\ \bibnamefont {Lousto}}, \bibinfo {author} {\bibfnamefont {Y.}~\bibnamefont {Zlochower}}, \bibinfo {author} {\bibfnamefont {B.}~\bibnamefont {Krishnan}},\ and\ \bibinfo {author} {\bibfnamefont {D.}~\bibnamefont {Merritt}},\ }\bibfield  {title} {\bibinfo {title} {{Spin Flips and Precession in Black-Hole-Binary Mergers}},\ }\href {https://doi.org/10.1103/PhysRevD.75.064030} {\bibfield  {journal} {\bibinfo  {journal} {\prd}\ }\textbf {\bibinfo {volume} {75}},\ \bibinfo {pages} {064030} (\bibinfo {year} {2007})},\ \Eprint {https://arxiv.org/abs/gr-qc/0612076} {arXiv:gr-qc/0612076} \BibitemShut {NoStop}%
\bibitem [{\citenamefont {Baker}\ \emph {et~al.}(2006)\citenamefont {Baker}, \citenamefont {Centrella}, \citenamefont {Choi}, \citenamefont {Koppitz},\ and\ \citenamefont {van Meter}}]{Baker:2005vv}%
  \BibitemOpen
  \bibfield  {author} {\bibinfo {author} {\bibfnamefont {J.~G.}\ \bibnamefont {Baker}}, \bibinfo {author} {\bibfnamefont {J.}~\bibnamefont {Centrella}}, \bibinfo {author} {\bibfnamefont {D.-I.}\ \bibnamefont {Choi}}, \bibinfo {author} {\bibfnamefont {M.}~\bibnamefont {Koppitz}},\ and\ \bibinfo {author} {\bibfnamefont {J.}~\bibnamefont {van Meter}},\ }\bibfield  {title} {\bibinfo {title} {{Gravitational wave extraction from an inspiraling configuration of merging black holes}},\ }\href {https://doi.org/10.1103/PhysRevLett.96.111102} {\bibfield  {journal} {\bibinfo  {journal} {\prl}\ }\textbf {\bibinfo {volume} {96}},\ \bibinfo {pages} {111102} (\bibinfo {year} {2006})},\ \Eprint {https://arxiv.org/abs/gr-qc/0511103} {arXiv:gr-qc/0511103} \BibitemShut {NoStop}%
\bibitem [{\citenamefont {Vishveshwara}(1970)}]{Vishveshwara:1970zz}%
  \BibitemOpen
  \bibfield  {author} {\bibinfo {author} {\bibfnamefont {C.~V.}\ \bibnamefont {Vishveshwara}},\ }\bibfield  {title} {\bibinfo {title} {{Scattering of Gravitational Radiation by a Schwarzschild Black-hole}},\ }\href {https://doi.org/10.1038/227936a0} {\bibfield  {journal} {\bibinfo  {journal} {Nature}\ }\textbf {\bibinfo {volume} {227}},\ \bibinfo {pages} {936} (\bibinfo {year} {1970})}\BibitemShut {NoStop}%
\bibitem [{\citenamefont {Press}(1971)}]{Press1971}%
  \BibitemOpen
  \bibfield  {author} {\bibinfo {author} {\bibfnamefont {W.~H.}\ \bibnamefont {Press}},\ }\bibfield  {title} {\bibinfo {title} {{Long Wave Trains of Gravitational Waves from a Vibrating Black Hole}},\ }\href {https://doi.org/10.1086/180849} {\bibfield  {journal} {\bibinfo  {journal} {Astrophys. J.}\ }\textbf {\bibinfo {volume} {170}},\ \bibinfo {pages} {L105} (\bibinfo {year} {1971})}\BibitemShut {NoStop}%
\bibitem [{\citenamefont {Regge}(1957)}]{Regge:1957td}%
  \BibitemOpen
  \bibfield  {author} {\bibinfo {author} {\bibfnamefont {T.}~\bibnamefont {Regge}},\ }\bibfield  {title} {\bibinfo {title} {{Stability of a Schwarzschild singularity}},\ }\href {https://doi.org/10.1103/PhysRev.108.1063} {\bibfield  {journal} {\bibinfo  {journal} {Phys. Rev.}\ }\textbf {\bibinfo {volume} {108}},\ \bibinfo {pages} {1063} (\bibinfo {year} {1957})}\BibitemShut {NoStop}%
\bibitem [{\citenamefont {Zerilli}(1970)}]{Zerilli:1970se}%
  \BibitemOpen
  \bibfield  {author} {\bibinfo {author} {\bibfnamefont {F.~J.}\ \bibnamefont {Zerilli}},\ }\bibfield  {title} {\bibinfo {title} {{Effective potential for even parity Regge-Wheeler gravitational perturbation equations}},\ }\href {https://doi.org/10.1103/PhysRevLett.24.737} {\bibfield  {journal} {\bibinfo  {journal} {Phys. Rev. Lett.}\ }\textbf {\bibinfo {volume} {24}},\ \bibinfo {pages} {737} (\bibinfo {year} {1970})}\BibitemShut {NoStop}%
\bibitem [{\citenamefont {Teukolsky}(1973)}]{Teukolsky:1973ha}%
  \BibitemOpen
  \bibfield  {author} {\bibinfo {author} {\bibfnamefont {S.~A.}\ \bibnamefont {Teukolsky}},\ }\bibfield  {title} {\bibinfo {title} {{Perturbations of a rotating black hole. 1. Fundamental equations for gravitational electromagnetic and neutrino field perturbations}},\ }\href {https://doi.org/10.1086/152444} {\bibfield  {journal} {\bibinfo  {journal} {Astrophys. J.}\ }\textbf {\bibinfo {volume} {185}},\ \bibinfo {pages} {635} (\bibinfo {year} {1973})}\BibitemShut {NoStop}%
\bibitem [{\citenamefont {Chandrasekhar}\ and\ \citenamefont {Detweiler}(1975)}]{Chandrasekhar:1975zza}%
  \BibitemOpen
  \bibfield  {author} {\bibinfo {author} {\bibfnamefont {S.}~\bibnamefont {Chandrasekhar}}\ and\ \bibinfo {author} {\bibfnamefont {S.~L.}\ \bibnamefont {Detweiler}},\ }\bibfield  {title} {\bibinfo {title} {{The quasi-normal modes of the Schwarzschild black hole}},\ }\href {https://doi.org/10.1098/rspa.1975.0112} {\bibfield  {journal} {\bibinfo  {journal} {Proc. R. Soc. A}\ }\textbf {\bibinfo {volume} {344}},\ \bibinfo {pages} {441} (\bibinfo {year} {1975})}\BibitemShut {NoStop}%
\bibitem [{\citenamefont {Detweiler}(1977)}]{Detweiler:1977gy}%
  \BibitemOpen
  \bibfield  {author} {\bibinfo {author} {\bibfnamefont {S.~L.}\ \bibnamefont {Detweiler}},\ }\bibfield  {title} {\bibinfo {title} {{Resonant oscillations of a rapidly rotating black hole}},\ }\href {https://doi.org/10.1098/rspa.1977.0005} {\bibfield  {journal} {\bibinfo  {journal} {Proc. Roy. Soc. Lond. A}\ }\textbf {\bibinfo {volume} {352}},\ \bibinfo {pages} {381} (\bibinfo {year} {1977})}\BibitemShut {NoStop}%
\bibitem [{\citenamefont {Detweiler}(1980)}]{Detweiler:1980gk}%
  \BibitemOpen
  \bibfield  {author} {\bibinfo {author} {\bibfnamefont {S.~L.}\ \bibnamefont {Detweiler}},\ }\bibfield  {title} {\bibinfo {title} {{Black holes and gravitational waves. III. The resonant frequencies of rotating holes}},\ }\href {https://doi.org/10.1086/158109} {\bibfield  {journal} {\bibinfo  {journal} {Astrophys. J.}\ }\textbf {\bibinfo {volume} {239}},\ \bibinfo {pages} {292} (\bibinfo {year} {1980})}\BibitemShut {NoStop}%
\bibitem [{\citenamefont {Leaver}(1985)}]{Leaver:1985}%
  \BibitemOpen
  \bibfield  {author} {\bibinfo {author} {\bibfnamefont {E.~W.}\ \bibnamefont {Leaver}},\ }\href@noop {} {\bibfield  {journal} {\bibinfo  {journal} {Proc. R. Soc. London Ser. A}\ }\textbf {\bibinfo {volume} {402}},\ \bibinfo {pages} {285} (\bibinfo {year} {1985})}\BibitemShut {NoStop}%
\bibitem [{\citenamefont {Price}(1972{\natexlab{a}})}]{Price:1971fb}%
  \BibitemOpen
  \bibfield  {author} {\bibinfo {author} {\bibfnamefont {R.~H.}\ \bibnamefont {Price}},\ }\bibfield  {title} {\bibinfo {title} {{Nonspherical perturbations of relativistic gravitational collapse. 1. Scalar and gravitational perturbations}},\ }\href {https://doi.org/10.1103/PhysRevD.5.2419} {\bibfield  {journal} {\bibinfo  {journal} {Phys. Rev. D}\ }\textbf {\bibinfo {volume} {5}},\ \bibinfo {pages} {2419} (\bibinfo {year} {1972}{\natexlab{a}})}\BibitemShut {NoStop}%
\bibitem [{\citenamefont {Price}(1972{\natexlab{b}})}]{Price:1972pw}%
  \BibitemOpen
  \bibfield  {author} {\bibinfo {author} {\bibfnamefont {R.~H.}\ \bibnamefont {Price}},\ }\bibfield  {title} {\bibinfo {title} {{Nonspherical Perturbations of Relativistic Gravitational Collapse. II. Integer-Spin, Zero-Rest-Mass Fields}},\ }\href {https://doi.org/10.1103/PhysRevD.5.2439} {\bibfield  {journal} {\bibinfo  {journal} {Phys. Rev. D}\ }\textbf {\bibinfo {volume} {5}},\ \bibinfo {pages} {2439} (\bibinfo {year} {1972}{\natexlab{b}})}\BibitemShut {NoStop}%
\bibitem [{\citenamefont {Nollert}(1999)}]{Nollert:1999ji}%
  \BibitemOpen
  \bibfield  {author} {\bibinfo {author} {\bibfnamefont {H.-P.}\ \bibnamefont {Nollert}},\ }\bibfield  {title} {\bibinfo {title} {{TOPICAL REVIEW: Quasinormal modes: the characteristic `sound' of black holes and neutron stars}},\ }\href {https://doi.org/10.1088/0264-9381/16/12/201} {\bibfield  {journal} {\bibinfo  {journal} {Class. Quant. Grav.}\ }\textbf {\bibinfo {volume} {16}},\ \bibinfo {pages} {R159} (\bibinfo {year} {1999})}\BibitemShut {NoStop}%
\bibitem [{\citenamefont {Kokkotas}\ and\ \citenamefont {Schmidt}(1999)}]{Kokkotas:1999bd}%
  \BibitemOpen
  \bibfield  {author} {\bibinfo {author} {\bibfnamefont {K.~D.}\ \bibnamefont {Kokkotas}}\ and\ \bibinfo {author} {\bibfnamefont {B.~G.}\ \bibnamefont {Schmidt}},\ }\bibfield  {title} {\bibinfo {title} {{Quasinormal modes of stars and black holes}},\ }\href {https://doi.org/10.12942/lrr-1999-2} {\bibfield  {journal} {\bibinfo  {journal} {Living Rev. Rel.}\ }\textbf {\bibinfo {volume} {2}},\ \bibinfo {pages} {2} (\bibinfo {year} {1999})},\ \Eprint {https://arxiv.org/abs/gr-qc/9909058} {arXiv:gr-qc/9909058} \BibitemShut {NoStop}%
\bibitem [{\citenamefont {Dreyer}\ \emph {et~al.}(2004)\citenamefont {Dreyer}, \citenamefont {Kelly}, \citenamefont {Krishnan}, \citenamefont {Finn}, \citenamefont {Garrison},\ and\ \citenamefont {Lopez-Aleman}}]{Dreyer:2003bv}%
  \BibitemOpen
  \bibfield  {author} {\bibinfo {author} {\bibfnamefont {O.}~\bibnamefont {Dreyer}}, \bibinfo {author} {\bibfnamefont {B.~J.}\ \bibnamefont {Kelly}}, \bibinfo {author} {\bibfnamefont {B.}~\bibnamefont {Krishnan}}, \bibinfo {author} {\bibfnamefont {L.~S.}\ \bibnamefont {Finn}}, \bibinfo {author} {\bibfnamefont {D.}~\bibnamefont {Garrison}},\ and\ \bibinfo {author} {\bibfnamefont {R.}~\bibnamefont {Lopez-Aleman}},\ }\bibfield  {title} {\bibinfo {title} {{Black hole spectroscopy: Testing general relativity through gravitational wave observations}},\ }\href {https://doi.org/10.1088/0264-9381/21/4/003} {\bibfield  {journal} {\bibinfo  {journal} {Class. Quantum Grav.}\ }\textbf {\bibinfo {volume} {21}},\ \bibinfo {pages} {787} (\bibinfo {year} {2004})},\ \Eprint {https://arxiv.org/abs/gr-qc/0309007} {arXiv:gr-qc/0309007} \BibitemShut {NoStop}%
\bibitem [{\citenamefont {Berti}\ \emph {et~al.}(2006)\citenamefont {Berti}, \citenamefont {Cardoso},\ and\ \citenamefont {Will}}]{Berti:2005ys}%
  \BibitemOpen
  \bibfield  {author} {\bibinfo {author} {\bibfnamefont {E.}~\bibnamefont {Berti}}, \bibinfo {author} {\bibfnamefont {V.}~\bibnamefont {Cardoso}},\ and\ \bibinfo {author} {\bibfnamefont {C.~M.}\ \bibnamefont {Will}},\ }\bibfield  {title} {\bibinfo {title} {{On gravitational-wave spectroscopy of massive black holes with the space interferometer LISA}},\ }\href {https://doi.org/10.1103/PhysRevD.73.064030} {\bibfield  {journal} {\bibinfo  {journal} {Phys. Rev. D}\ }\textbf {\bibinfo {volume} {73}},\ \bibinfo {pages} {064030} (\bibinfo {year} {2006})},\ \Eprint {https://arxiv.org/abs/gr-qc/0512160} {arXiv:gr-qc/0512160} \BibitemShut {NoStop}%
\bibitem [{\citenamefont {Gossan}\ \emph {et~al.}(2012)\citenamefont {Gossan}, \citenamefont {Veitch},\ and\ \citenamefont {Sathyaprakash}}]{Gossan:2011ha}%
  \BibitemOpen
  \bibfield  {author} {\bibinfo {author} {\bibfnamefont {S.}~\bibnamefont {Gossan}}, \bibinfo {author} {\bibfnamefont {J.}~\bibnamefont {Veitch}},\ and\ \bibinfo {author} {\bibfnamefont {B.~S.}\ \bibnamefont {Sathyaprakash}},\ }\bibfield  {title} {\bibinfo {title} {{Bayesian model selection for testing the no-hair theorem with black hole ringdowns}},\ }\href {https://doi.org/10.1103/PhysRevD.85.124056} {\bibfield  {journal} {\bibinfo  {journal} {Phys. Rev. D}\ }\textbf {\bibinfo {volume} {85}},\ \bibinfo {pages} {124056} (\bibinfo {year} {2012})},\ \Eprint {https://arxiv.org/abs/1111.5819} {arXiv:1111.5819 [gr-qc]} \BibitemShut {NoStop}%
\bibitem [{\citenamefont {Meidam}\ \emph {et~al.}(2014)\citenamefont {Meidam}, \citenamefont {Agathos}, \citenamefont {Van Den~Broeck}, \citenamefont {Veitch},\ and\ \citenamefont {Sathyaprakash}}]{Meidam:2014jpa}%
  \BibitemOpen
  \bibfield  {author} {\bibinfo {author} {\bibfnamefont {J.}~\bibnamefont {Meidam}}, \bibinfo {author} {\bibfnamefont {M.}~\bibnamefont {Agathos}}, \bibinfo {author} {\bibfnamefont {C.}~\bibnamefont {Van Den~Broeck}}, \bibinfo {author} {\bibfnamefont {J.}~\bibnamefont {Veitch}},\ and\ \bibinfo {author} {\bibfnamefont {B.~S.}\ \bibnamefont {Sathyaprakash}},\ }\bibfield  {title} {\bibinfo {title} {{Testing the no-hair theorem with black hole ringdowns using TIGER}},\ }\href {https://doi.org/10.1103/PhysRevD.90.064009} {\bibfield  {journal} {\bibinfo  {journal} {Phys. Rev. D}\ }\textbf {\bibinfo {volume} {90}},\ \bibinfo {pages} {064009} (\bibinfo {year} {2014})},\ \Eprint {https://arxiv.org/abs/1406.3201} {arXiv:1406.3201 [gr-qc]} \BibitemShut {NoStop}%
\bibitem [{\citenamefont {Berti}\ \emph {et~al.}(2025)\citenamefont {Berti}, \citenamefont {Cardoso}, \citenamefont {Carullo} \emph {et~al.}}]{Berti:2025hly}%
  \BibitemOpen
  \bibfield  {author} {\bibinfo {author} {\bibfnamefont {E.}~\bibnamefont {Berti}}, \bibinfo {author} {\bibfnamefont {V.}~\bibnamefont {Cardoso}}, \bibinfo {author} {\bibfnamefont {G.}~\bibnamefont {Carullo}}, \emph {et~al.},\ }\bibfield  {title} {\bibinfo {title} {{Black hole spectroscopy: from theory to experiment}},\ }\Eprint {https://arxiv.org/abs/2505.23895} {arXiv:2505.23895 [gr-qc]}  (\bibinfo {year} {2025})\BibitemShut {NoStop}%
\bibitem [{\citenamefont {Penrose}(2002)}]{Penrose:2002col}%
  \BibitemOpen
  \bibfield  {author} {\bibinfo {author} {\bibfnamefont {R.}~\bibnamefont {Penrose}},\ }\bibfield  {title} {\bibinfo {title} {{“Golden Oldie”: Gravitational Collapse: The Role of General Relativity}},\ }\href {https://doi.org/10.1023/A:1016578408204} {\bibfield  {journal} {\bibinfo  {journal} {Gen. Relativ. Gravit.}\ }\textbf {\bibinfo {volume} {34}},\ \bibinfo {pages} {1141} (\bibinfo {year} {2002})}\BibitemShut {NoStop}%
\bibitem [{\citenamefont {{Klainerman}}(2002)}]{2002nmgm.meet...28K}%
  \BibitemOpen
  \bibfield  {author} {\bibinfo {author} {\bibfnamefont {S.}~\bibnamefont {{Klainerman}}},\ }\bibfield  {title} {\bibinfo {title} {{Mathematical Challenges of General Relativity}},\ }in\ \href {https://doi.org/10.1142/9789812777386_0003} {\emph {\bibinfo {booktitle} {The Ninth Marcel Grossmann Meeting}}},\ \bibinfo {editor} {edited by\ \bibinfo {editor} {\bibfnamefont {V.~G.}\ \bibnamefont {{Gurzadyan}}}, \bibinfo {editor} {\bibfnamefont {R.~T.}\ \bibnamefont {{Jantzen}}},\ and\ \bibinfo {editor} {\bibfnamefont {R.}~\bibnamefont {{Ruffini}}}}\ (\bibinfo {year} {2002})\ pp.\ \bibinfo {pages} {28--43}\BibitemShut {NoStop}%
\bibitem [{\citenamefont {Chru{\'s}ciel}\ \emph {et~al.}(2012)\citenamefont {Chru{\'s}ciel}, \citenamefont {Lopes~Costa},\ and\ \citenamefont {Heusler}}]{Chrusciel:2012jk}%
  \BibitemOpen
  \bibfield  {author} {\bibinfo {author} {\bibfnamefont {P.~T.}\ \bibnamefont {Chru{\'s}ciel}}, \bibinfo {author} {\bibfnamefont {J.}~\bibnamefont {Lopes~Costa}},\ and\ \bibinfo {author} {\bibfnamefont {M.}~\bibnamefont {Heusler}},\ }\bibfield  {title} {\bibinfo {title} {{Stationary Black Holes: Uniqueness and Beyond}},\ }\href {https://doi.org/10.12942/lrr-2012-7} {\bibfield  {journal} {\bibinfo  {journal} {Living Rev. Relativity}\ }\textbf {\bibinfo {volume} {15}},\ \bibinfo {pages} {7} (\bibinfo {year} {2012})},\ \Eprint {https://arxiv.org/abs/1205.6112} {arXiv:1205.6112 [gr-qc]} \BibitemShut {NoStop}%
\bibitem [{\citenamefont {Gibbons}(1975)}]{Gibbons:1975kk}%
  \BibitemOpen
  \bibfield  {author} {\bibinfo {author} {\bibfnamefont {G.~W.}\ \bibnamefont {Gibbons}},\ }\bibfield  {title} {\bibinfo {title} {{Vacuum Polarization and the Spontaneous Loss of Charge by Black Holes}},\ }\href {https://doi.org/10.1007/BF01609829} {\bibfield  {journal} {\bibinfo  {journal} {Commun. Math. Phys.}\ }\textbf {\bibinfo {volume} {44}},\ \bibinfo {pages} {245} (\bibinfo {year} {1975})}\BibitemShut {NoStop}%
\bibitem [{\citenamefont {Blandford}\ and\ \citenamefont {Znajek}(1977)}]{Blandford:1977ds}%
  \BibitemOpen
  \bibfield  {author} {\bibinfo {author} {\bibfnamefont {R.~D.}\ \bibnamefont {Blandford}}\ and\ \bibinfo {author} {\bibfnamefont {R.~L.}\ \bibnamefont {Znajek}},\ }\bibfield  {title} {\bibinfo {title} {{Electromagnetic extractions of energy from Kerr black holes}},\ }\href {https://doi.org/10.1093/mnras/179.3.433} {\bibfield  {journal} {\bibinfo  {journal} {Mon. Not. R. Astron. Soc.}\ }\textbf {\bibinfo {volume} {179}},\ \bibinfo {pages} {433} (\bibinfo {year} {1977})}\BibitemShut {NoStop}%
\bibitem [{\citenamefont {{Hanni}}(1982)}]{1982PhRvD..25.2509H}%
  \BibitemOpen
  \bibfield  {author} {\bibinfo {author} {\bibfnamefont {R.~S.}\ \bibnamefont {{Hanni}}},\ }\bibfield  {title} {\bibinfo {title} {{Limits on the charge of a collapsed object}},\ }\href {https://doi.org/10.1103/PhysRevD.25.2509} {\bibfield  {journal} {\bibinfo  {journal} {Phys. Rev. D}\ }\textbf {\bibinfo {volume} {25}},\ \bibinfo {pages} {2509} (\bibinfo {year} {1982})}\BibitemShut {NoStop}%
\bibitem [{\citenamefont {Carullo}\ \emph {et~al.}(2022)\citenamefont {Carullo}, \citenamefont {Laghi}, \citenamefont {Johnson-McDaniel}, \citenamefont {Del~Pozzo}, \citenamefont {Dias}, \citenamefont {Godazgar},\ and\ \citenamefont {Santos}}]{Carullo:2021oxn}%
  \BibitemOpen
  \bibfield  {author} {\bibinfo {author} {\bibfnamefont {G.}~\bibnamefont {Carullo}}, \bibinfo {author} {\bibfnamefont {D.}~\bibnamefont {Laghi}}, \bibinfo {author} {\bibfnamefont {N.~K.}\ \bibnamefont {Johnson-McDaniel}}, \bibinfo {author} {\bibfnamefont {W.}~\bibnamefont {Del~Pozzo}}, \bibinfo {author} {\bibfnamefont {O.~J.~C.}\ \bibnamefont {Dias}}, \bibinfo {author} {\bibfnamefont {M.}~\bibnamefont {Godazgar}},\ and\ \bibinfo {author} {\bibfnamefont {J.~E.}\ \bibnamefont {Santos}},\ }\bibfield  {title} {\bibinfo {title} {{Constraints on Kerr-Newman black holes from merger-ringdown gravitational-wave observations}},\ }\href {https://doi.org/10.1103/PhysRevD.105.062009} {\bibfield  {journal} {\bibinfo  {journal} {Phys. Rev. D}\ }\textbf {\bibinfo {volume} {105}},\ \bibinfo {pages} {062009} (\bibinfo {year} {2022})},\ \Eprint {https://arxiv.org/abs/2109.13961} {arXiv:2109.13961 [gr-qc]} \BibitemShut {NoStop}%
\bibitem [{\citenamefont {Carullo}\ \emph {et~al.}(2019)\citenamefont {Carullo}, \citenamefont {Del~Pozzo},\ and\ \citenamefont {Veitch}}]{Carullo:2019flw}%
  \BibitemOpen
  \bibfield  {author} {\bibinfo {author} {\bibfnamefont {G.}~\bibnamefont {Carullo}}, \bibinfo {author} {\bibfnamefont {W.}~\bibnamefont {Del~Pozzo}},\ and\ \bibinfo {author} {\bibfnamefont {J.}~\bibnamefont {Veitch}},\ }\bibfield  {title} {\bibinfo {title} {{Observational Black Hole Spectroscopy: A time-domain multimode analysis of GW150914}},\ }\href {https://doi.org/10.1103/PhysRevD.99.123029} {\bibfield  {journal} {\bibinfo  {journal} {Phys. Rev. D}\ }\textbf {\bibinfo {volume} {99}},\ \bibinfo {pages} {123029} (\bibinfo {year} {2019})},\ \bibinfo {note} {[Erratum: \href{http://doi.org/10.1103/PhysRevD.100.089903}{Phys.\ Rev.\ D {\bf{100}}, 089903(E) (2019)}]},\ \Eprint {https://arxiv.org/abs/1902.07527} {arXiv:1902.07527 [gr-qc]} \BibitemShut {NoStop}%
\bibitem [{\citenamefont {Isi}\ \emph {et~al.}(2019{\natexlab{a}})\citenamefont {Isi}, \citenamefont {Giesler}, \citenamefont {Farr}, \citenamefont {Scheel},\ and\ \citenamefont {Teukolsky}}]{Isi:2019aib}%
  \BibitemOpen
  \bibfield  {author} {\bibinfo {author} {\bibfnamefont {M.}~\bibnamefont {Isi}}, \bibinfo {author} {\bibfnamefont {M.}~\bibnamefont {Giesler}}, \bibinfo {author} {\bibfnamefont {W.~M.}\ \bibnamefont {Farr}}, \bibinfo {author} {\bibfnamefont {M.~A.}\ \bibnamefont {Scheel}},\ and\ \bibinfo {author} {\bibfnamefont {S.~A.}\ \bibnamefont {Teukolsky}},\ }\bibfield  {title} {\bibinfo {title} {{Testing the no-hair theorem with GW150914}},\ }\href {https://doi.org/10.1103/PhysRevLett.123.111102} {\bibfield  {journal} {\bibinfo  {journal} {Phys. Rev. Lett.}\ }\textbf {\bibinfo {volume} {123}},\ \bibinfo {pages} {111102} (\bibinfo {year} {2019}{\natexlab{a}})},\ \Eprint {https://arxiv.org/abs/1905.00869} {arXiv:1905.00869 [gr-qc]} \BibitemShut {NoStop}%
\bibitem [{\citenamefont {Capano}\ \emph {et~al.}(2021)\citenamefont {Capano}, \citenamefont {Cabero}, \citenamefont {Westerweck}, \citenamefont {Abedi}, \citenamefont {Kastha}, \citenamefont {Nitz}, \citenamefont {Nielsen},\ and\ \citenamefont {Krishnan}}]{Capano:2021etf}%
  \BibitemOpen
  \bibfield  {author} {\bibinfo {author} {\bibfnamefont {C.~D.}\ \bibnamefont {Capano}}, \bibinfo {author} {\bibfnamefont {M.}~\bibnamefont {Cabero}}, \bibinfo {author} {\bibfnamefont {J.}~\bibnamefont {Westerweck}}, \bibinfo {author} {\bibfnamefont {J.}~\bibnamefont {Abedi}}, \bibinfo {author} {\bibfnamefont {S.}~\bibnamefont {Kastha}}, \bibinfo {author} {\bibfnamefont {A.~H.}\ \bibnamefont {Nitz}}, \bibinfo {author} {\bibfnamefont {A.~B.}\ \bibnamefont {Nielsen}},\ and\ \bibinfo {author} {\bibfnamefont {B.}~\bibnamefont {Krishnan}},\ }\bibfield  {title} {\bibinfo {title} {{Observation of a multimode quasi-normal spectrum from a perturbed black hole}},\ }\Eprint {https://arxiv.org/abs/2105.05238} {arXiv:2105.05238 [gr-qc]}  (\bibinfo {year} {2021})\BibitemShut {NoStop}%
\bibitem [{\citenamefont {Cotesta}\ \emph {et~al.}(2022)\citenamefont {Cotesta}, \citenamefont {Carullo}, \citenamefont {Berti},\ and\ \citenamefont {Cardoso}}]{Cotesta:2022pci}%
  \BibitemOpen
  \bibfield  {author} {\bibinfo {author} {\bibfnamefont {R.}~\bibnamefont {Cotesta}}, \bibinfo {author} {\bibfnamefont {G.}~\bibnamefont {Carullo}}, \bibinfo {author} {\bibfnamefont {E.}~\bibnamefont {Berti}},\ and\ \bibinfo {author} {\bibfnamefont {V.}~\bibnamefont {Cardoso}},\ }\bibfield  {title} {\bibinfo {title} {{Analysis of Ringdown Overtones in GW150914}},\ }\href {https://doi.org/10.1103/PhysRevLett.129.111102} {\bibfield  {journal} {\bibinfo  {journal} {Phys. Rev. Lett.}\ }\textbf {\bibinfo {volume} {129}},\ \bibinfo {pages} {111102} (\bibinfo {year} {2022})},\ \Eprint {https://arxiv.org/abs/2201.00822} {arXiv:2201.00822 [gr-qc]} \BibitemShut {NoStop}%
\bibitem [{\citenamefont {Siegel}\ \emph {et~al.}(2023)\citenamefont {Siegel}, \citenamefont {Isi},\ and\ \citenamefont {Farr}}]{Siegel:2023lxl}%
  \BibitemOpen
  \bibfield  {author} {\bibinfo {author} {\bibfnamefont {H.}~\bibnamefont {Siegel}}, \bibinfo {author} {\bibfnamefont {M.}~\bibnamefont {Isi}},\ and\ \bibinfo {author} {\bibfnamefont {W.~M.}\ \bibnamefont {Farr}},\ }\bibfield  {title} {\bibinfo {title} {{Ringdown of GW190521: Hints of multiple quasinormal modes with a precessional interpretation}},\ }\href {https://doi.org/10.1103/PhysRevD.108.064008} {\bibfield  {journal} {\bibinfo  {journal} {Phys. Rev. D}\ }\textbf {\bibinfo {volume} {108}},\ \bibinfo {pages} {064008} (\bibinfo {year} {2023})},\ \Eprint {https://arxiv.org/abs/2307.11975} {arXiv:2307.11975 [gr-qc]} \BibitemShut {NoStop}%
\bibitem [{\citenamefont {Gennari}\ \emph {et~al.}(2024)\citenamefont {Gennari}, \citenamefont {Carullo},\ and\ \citenamefont {Del~Pozzo}}]{Gennari:2023gmx}%
  \BibitemOpen
  \bibfield  {author} {\bibinfo {author} {\bibfnamefont {V.}~\bibnamefont {Gennari}}, \bibinfo {author} {\bibfnamefont {G.}~\bibnamefont {Carullo}},\ and\ \bibinfo {author} {\bibfnamefont {W.}~\bibnamefont {Del~Pozzo}},\ }\bibfield  {title} {\bibinfo {title} {{Searching for ringdown higher modes with a numerical relativity-informed post-merger model}},\ }\href {https://doi.org/10.1140/epjc/s10052-024-12550-x} {\bibfield  {journal} {\bibinfo  {journal} {Eur. Phys. J. C}\ }\textbf {\bibinfo {volume} {84}},\ \bibinfo {pages} {233} (\bibinfo {year} {2024})},\ \Eprint {https://arxiv.org/abs/2312.12515} {arXiv:2312.12515 [gr-qc]} \BibitemShut {NoStop}%
\bibitem [{\citenamefont {Varma}\ \emph {et~al.}(2019)\citenamefont {Varma}, \citenamefont {Field}, \citenamefont {Scheel}, \citenamefont {Blackman}, \citenamefont {Gerosa}, \citenamefont {Stein}, \citenamefont {Kidder},\ and\ \citenamefont {Pfeiffer}}]{Varma:2019csw}%
  \BibitemOpen
  \bibfield  {author} {\bibinfo {author} {\bibfnamefont {V.}~\bibnamefont {Varma}}, \bibinfo {author} {\bibfnamefont {S.~E.}\ \bibnamefont {Field}}, \bibinfo {author} {\bibfnamefont {M.~A.}\ \bibnamefont {Scheel}}, \bibinfo {author} {\bibfnamefont {J.}~\bibnamefont {Blackman}}, \bibinfo {author} {\bibfnamefont {D.}~\bibnamefont {Gerosa}}, \bibinfo {author} {\bibfnamefont {L.~C.}\ \bibnamefont {Stein}}, \bibinfo {author} {\bibfnamefont {L.~E.}\ \bibnamefont {Kidder}},\ and\ \bibinfo {author} {\bibfnamefont {H.~P.}\ \bibnamefont {Pfeiffer}},\ }\bibfield  {title} {\bibinfo {title} {{Surrogate models for precessing binary black hole simulations with unequal masses}},\ }\href {https://doi.org/10.1103/PhysRevResearch.1.033015} {\bibfield  {journal} {\bibinfo  {journal} {Phys. Rev. Research.}\ }\textbf {\bibinfo {volume} {1}},\ \bibinfo {pages} {033015} (\bibinfo {year} {2019})},\ \Eprint {https://arxiv.org/abs/1905.09300} {arXiv:1905.09300 [gr-qc]} \BibitemShut {NoStop}%
\bibitem [{\citenamefont {Gamboa}\ \emph {et~al.}(2025)\citenamefont {Gamboa} \emph {et~al.}}]{Gamboa:2024hli}%
  \BibitemOpen
  \bibfield  {author} {\bibinfo {author} {\bibfnamefont {A.}~\bibnamefont {Gamboa}} \emph {et~al.},\ }\bibfield  {title} {\bibinfo {title} {{Accurate waveforms for eccentric, aligned-spin binary black holes: The multipolar effective-one-body model seobnrv5ehm}},\ }\href {https://doi.org/10.1103/jxrc-z298} {\bibfield  {journal} {\bibinfo  {journal} {Phys. Rev. D}\ }\textbf {\bibinfo {volume} {112}},\ \bibinfo {pages} {044038} (\bibinfo {year} {2025})},\ \Eprint {https://arxiv.org/abs/2412.12823} {arXiv:2412.12823 [gr-qc]} \BibitemShut {NoStop}%
\bibitem [{\citenamefont {Nagar}\ \emph {et~al.}(2024)\citenamefont {Nagar}, \citenamefont {Gamba}, \citenamefont {Rettegno}, \citenamefont {Fantini},\ and\ \citenamefont {Bernuzzi}}]{Nagar:2024dzj}%
  \BibitemOpen
  \bibfield  {author} {\bibinfo {author} {\bibfnamefont {A.}~\bibnamefont {Nagar}}, \bibinfo {author} {\bibfnamefont {R.}~\bibnamefont {Gamba}}, \bibinfo {author} {\bibfnamefont {P.}~\bibnamefont {Rettegno}}, \bibinfo {author} {\bibfnamefont {V.}~\bibnamefont {Fantini}},\ and\ \bibinfo {author} {\bibfnamefont {S.}~\bibnamefont {Bernuzzi}},\ }\bibfield  {title} {\bibinfo {title} {{Effective-one-body waveform model for noncircularized, planar, coalescing black hole binaries: The importance of radiation reaction}},\ }\href {https://doi.org/10.1103/PhysRevD.110.084001} {\bibfield  {journal} {\bibinfo  {journal} {Phys. Rev. D}\ }\textbf {\bibinfo {volume} {110}},\ \bibinfo {pages} {084001} (\bibinfo {year} {2024})},\ \Eprint {https://arxiv.org/abs/2404.05288} {arXiv:2404.05288 [gr-qc]} \BibitemShut {NoStop}%
\bibitem [{\citenamefont {Leaver}(1986)}]{Leaver:1986gd}%
  \BibitemOpen
  \bibfield  {author} {\bibinfo {author} {\bibfnamefont {E.~W.}\ \bibnamefont {Leaver}},\ }\bibfield  {title} {\bibinfo {title} {{Spectral decomposition of the perturbation response of the Schwarzschild geometry}},\ }\href {https://doi.org/10.1103/PhysRevD.34.384} {\bibfield  {journal} {\bibinfo  {journal} {Phys. Rev. D}\ }\textbf {\bibinfo {volume} {34}},\ \bibinfo {pages} {384} (\bibinfo {year} {1986})}\BibitemShut {NoStop}%
\bibitem [{\citenamefont {Andersson}(1997)}]{Andersson:1996cm}%
  \BibitemOpen
  \bibfield  {author} {\bibinfo {author} {\bibfnamefont {N.}~\bibnamefont {Andersson}},\ }\bibfield  {title} {\bibinfo {title} {{Evolving test fields in a black hole geometry}},\ }\href {https://doi.org/10.1103/PhysRevD.55.468} {\bibfield  {journal} {\bibinfo  {journal} {Phys. Rev. D}\ }\textbf {\bibinfo {volume} {55}},\ \bibinfo {pages} {468} (\bibinfo {year} {1997})},\ \Eprint {https://arxiv.org/abs/gr-qc/9607064} {arXiv:gr-qc/9607064} \BibitemShut {NoStop}%
\bibitem [{\citenamefont {Zhu}\ \emph {et~al.}(2024)\citenamefont {Zhu} \emph {et~al.}}]{Zhu:2024dyl}%
  \BibitemOpen
  \bibfield  {author} {\bibinfo {author} {\bibfnamefont {H.}~\bibnamefont {Zhu}} \emph {et~al.},\ }\bibfield  {title} {\bibinfo {title} {{Imprints of changing mass and spin on black hole ringdown}},\ }\href {https://doi.org/10.1103/PhysRevD.110.124028} {\bibfield  {journal} {\bibinfo  {journal} {Phys. Rev. D}\ }\textbf {\bibinfo {volume} {110}},\ \bibinfo {pages} {124028} (\bibinfo {year} {2024})},\ \Eprint {https://arxiv.org/abs/2404.12424} {arXiv:2404.12424 [gr-qc]} \BibitemShut {NoStop}%
\bibitem [{\citenamefont {Chavda}\ \emph {et~al.}(2024)\citenamefont {Chavda}, \citenamefont {Lagos},\ and\ \citenamefont {Hui}}]{Chavda:2024awq}%
  \BibitemOpen
  \bibfield  {author} {\bibinfo {author} {\bibfnamefont {A.}~\bibnamefont {Chavda}}, \bibinfo {author} {\bibfnamefont {M.}~\bibnamefont {Lagos}},\ and\ \bibinfo {author} {\bibfnamefont {L.}~\bibnamefont {Hui}},\ }\bibfield  {title} {\bibinfo {title} {{The impact of initial conditions on quasi-normal modes}},\ }\href@noop {} {\  (\bibinfo {year} {2024})},\ \Eprint {https://arxiv.org/abs/2412.03435} {arXiv:2412.03435 [gr-qc]} \BibitemShut {NoStop}%
\bibitem [{\citenamefont {De~Amicis}\ \emph {et~al.}(2025)\citenamefont {De~Amicis}, \citenamefont {Cannizzaro}, \citenamefont {Carullo},\ and\ \citenamefont {Sberna}}]{DeAmicis:2025xuh}%
  \BibitemOpen
  \bibfield  {author} {\bibinfo {author} {\bibfnamefont {M.}~\bibnamefont {De~Amicis}}, \bibinfo {author} {\bibfnamefont {E.}~\bibnamefont {Cannizzaro}}, \bibinfo {author} {\bibfnamefont {G.}~\bibnamefont {Carullo}},\ and\ \bibinfo {author} {\bibfnamefont {L.}~\bibnamefont {Sberna}},\ }\bibfield  {title} {\bibinfo {title} {{Dynamical quasinormal mode excitation}},\ }\href@noop {} {\  (\bibinfo {year} {2025})},\ \Eprint {https://arxiv.org/abs/2506.21668} {arXiv:2506.21668 [gr-qc]} \BibitemShut {NoStop}%
\bibitem [{\citenamefont {London}\ \emph {et~al.}(2014)\citenamefont {London}, \citenamefont {Shoemaker},\ and\ \citenamefont {Healy}}]{London:2014cma}%
  \BibitemOpen
  \bibfield  {author} {\bibinfo {author} {\bibfnamefont {L.}~\bibnamefont {London}}, \bibinfo {author} {\bibfnamefont {D.}~\bibnamefont {Shoemaker}},\ and\ \bibinfo {author} {\bibfnamefont {J.}~\bibnamefont {Healy}},\ }\bibfield  {title} {\bibinfo {title} {{Modeling ringdown: Beyond the fundamental quasinormal modes}},\ }\href {https://doi.org/10.1103/PhysRevD.90.124032} {\bibfield  {journal} {\bibinfo  {journal} {Phys. Rev. D}\ }\textbf {\bibinfo {volume} {90}},\ \bibinfo {pages} {124032} (\bibinfo {year} {2014})},\ \bibinfo {note} {[Erratum: Phys.Rev.D 94, 069902 (2016)]},\ \Eprint {https://arxiv.org/abs/1404.3197} {arXiv:1404.3197 [gr-qc]} \BibitemShut {NoStop}%
\bibitem [{\citenamefont {Mitman}\ \emph {et~al.}(2023)\citenamefont {Mitman} \emph {et~al.}}]{Mitman:2022qdl}%
  \BibitemOpen
  \bibfield  {author} {\bibinfo {author} {\bibfnamefont {K.}~\bibnamefont {Mitman}} \emph {et~al.},\ }\bibfield  {title} {\bibinfo {title} {{Nonlinearities in Black Hole Ringdowns}},\ }\href {https://doi.org/10.1103/PhysRevLett.130.081402} {\bibfield  {journal} {\bibinfo  {journal} {Phys. Rev. Lett.}\ }\textbf {\bibinfo {volume} {130}},\ \bibinfo {pages} {081402} (\bibinfo {year} {2023})},\ \Eprint {https://arxiv.org/abs/2208.07380} {arXiv:2208.07380 [gr-qc]} \BibitemShut {NoStop}%
\bibitem [{\citenamefont {Cheung}\ \emph {et~al.}(2023)\citenamefont {Cheung} \emph {et~al.}}]{Cheung:2022rbm}%
  \BibitemOpen
  \bibfield  {author} {\bibinfo {author} {\bibfnamefont {M.~H.-Y.}\ \bibnamefont {Cheung}} \emph {et~al.},\ }\bibfield  {title} {\bibinfo {title} {{Nonlinear Effects in Black Hole Ringdown}},\ }\href {https://doi.org/10.1103/PhysRevLett.130.081401} {\bibfield  {journal} {\bibinfo  {journal} {Phys. Rev. Lett.}\ }\textbf {\bibinfo {volume} {130}},\ \bibinfo {pages} {081401} (\bibinfo {year} {2023})},\ \Eprint {https://arxiv.org/abs/2208.07374} {arXiv:2208.07374 [gr-qc]} \BibitemShut {NoStop}%
\bibitem [{\citenamefont {Ma}\ and\ \citenamefont {Yang}(2024)}]{Ma:2024qcv}%
  \BibitemOpen
  \bibfield  {author} {\bibinfo {author} {\bibfnamefont {S.}~\bibnamefont {Ma}}\ and\ \bibinfo {author} {\bibfnamefont {H.}~\bibnamefont {Yang}},\ }\bibfield  {title} {\bibinfo {title} {{Excitation of quadratic quasinormal modes for Kerr black holes}},\ }\href {https://doi.org/10.1103/PhysRevD.109.104070} {\bibfield  {journal} {\bibinfo  {journal} {Phys. Rev. D}\ }\textbf {\bibinfo {volume} {109}},\ \bibinfo {pages} {104070} (\bibinfo {year} {2024})},\ \Eprint {https://arxiv.org/abs/2401.15516} {arXiv:2401.15516 [gr-qc]} \BibitemShut {NoStop}%
\bibitem [{\citenamefont {Buonanno}\ \emph {et~al.}(2007)\citenamefont {Buonanno}, \citenamefont {Cook},\ and\ \citenamefont {Pretorius}}]{Buonanno:2006ui}%
  \BibitemOpen
  \bibfield  {author} {\bibinfo {author} {\bibfnamefont {A.}~\bibnamefont {Buonanno}}, \bibinfo {author} {\bibfnamefont {G.~B.}\ \bibnamefont {Cook}},\ and\ \bibinfo {author} {\bibfnamefont {F.}~\bibnamefont {Pretorius}},\ }\bibfield  {title} {\bibinfo {title} {{Inspiral, merger and ring-down of equal-mass black-hole binaries}},\ }\href {https://doi.org/10.1103/PhysRevD.75.124018} {\bibfield  {journal} {\bibinfo  {journal} {Phys. Rev. D}\ }\textbf {\bibinfo {volume} {75}},\ \bibinfo {pages} {124018} (\bibinfo {year} {2007})},\ \Eprint {https://arxiv.org/abs/gr-qc/0610122} {arXiv:gr-qc/0610122} \BibitemShut {NoStop}%
\bibitem [{\citenamefont {Berti}\ \emph {et~al.}(2007)\citenamefont {Berti} \emph {et~al.}}]{Berti:2007fi}%
  \BibitemOpen
  \bibfield  {author} {\bibinfo {author} {\bibfnamefont {E.}~\bibnamefont {Berti}} \emph {et~al.},\ }\bibfield  {title} {\bibinfo {title} {{Inspiral, merger and ringdown of unequal mass black hole binaries: {A} multipolar analysis}},\ }\href {https://doi.org/10.1103/PhysRevD.76.064034} {\bibfield  {journal} {\bibinfo  {journal} {\prd}\ }\textbf {\bibinfo {volume} {76}},\ \bibinfo {pages} {064034} (\bibinfo {year} {2007})},\ \Eprint {https://arxiv.org/abs/gr-qc/0703053} {arXiv:gr-qc/0703053} \BibitemShut {NoStop}%
\bibitem [{\citenamefont {Ching}\ \emph {et~al.}(1995)\citenamefont {Ching}, \citenamefont {Leung}, \citenamefont {Suen},\ and\ \citenamefont {Young}}]{Ching:1995tj}%
  \BibitemOpen
  \bibfield  {author} {\bibinfo {author} {\bibfnamefont {E.~S.~C.}\ \bibnamefont {Ching}}, \bibinfo {author} {\bibfnamefont {P.~T.}\ \bibnamefont {Leung}}, \bibinfo {author} {\bibfnamefont {W.~M.}\ \bibnamefont {Suen}},\ and\ \bibinfo {author} {\bibfnamefont {K.}~\bibnamefont {Young}},\ }\bibfield  {title} {\bibinfo {title} {{Wave propagation in gravitational systems: Late time behavior}},\ }\href {https://doi.org/10.1103/PhysRevD.52.2118} {\bibfield  {journal} {\bibinfo  {journal} {Phys. Rev. D}\ }\textbf {\bibinfo {volume} {52}},\ \bibinfo {pages} {2118} (\bibinfo {year} {1995})},\ \Eprint {https://arxiv.org/abs/gr-qc/9507035} {arXiv:gr-qc/9507035} \BibitemShut {NoStop}%
\bibitem [{\citenamefont {Ma}\ \emph {et~al.}(2025)\citenamefont {Ma}, \citenamefont {Scheel}, \citenamefont {Moxon}, \citenamefont {Nelli}, \citenamefont {Deppe}, \citenamefont {Kidder}, \citenamefont {Throwe},\ and\ \citenamefont {Vu}}]{Ma:2024hzq}%
  \BibitemOpen
  \bibfield  {author} {\bibinfo {author} {\bibfnamefont {S.}~\bibnamefont {Ma}}, \bibinfo {author} {\bibfnamefont {M.~A.}\ \bibnamefont {Scheel}}, \bibinfo {author} {\bibfnamefont {J.}~\bibnamefont {Moxon}}, \bibinfo {author} {\bibfnamefont {K.~C.}\ \bibnamefont {Nelli}}, \bibinfo {author} {\bibfnamefont {N.}~\bibnamefont {Deppe}}, \bibinfo {author} {\bibfnamefont {L.~E.}\ \bibnamefont {Kidder}}, \bibinfo {author} {\bibfnamefont {W.}~\bibnamefont {Throwe}},\ and\ \bibinfo {author} {\bibfnamefont {N.~L.}\ \bibnamefont {Vu}},\ }\bibfield  {title} {\bibinfo {title} {{Merging black holes with Cauchy-characteristic matching: Computation of late-time tails}},\ }\href {https://doi.org/10.1103/jd26-8q5w} {\bibfield  {journal} {\bibinfo  {journal} {Phys. Rev. D}\ }\textbf {\bibinfo {volume} {112}},\ \bibinfo {pages} {024003} (\bibinfo {year} {2025})},\ \Eprint {https://arxiv.org/abs/2412.06906} {arXiv:2412.06906 [gr-qc]} \BibitemShut {NoStop}%
\bibitem [{\citenamefont {De~Amicis}\ \emph {et~al.}(2024{\natexlab{a}})\citenamefont {De~Amicis}, \citenamefont {Albanesi},\ and\ \citenamefont {Carullo}}]{DeAmicis:2024not}%
  \BibitemOpen
  \bibfield  {author} {\bibinfo {author} {\bibfnamefont {M.}~\bibnamefont {De~Amicis}}, \bibinfo {author} {\bibfnamefont {S.}~\bibnamefont {Albanesi}},\ and\ \bibinfo {author} {\bibfnamefont {G.}~\bibnamefont {Carullo}},\ }\bibfield  {title} {\bibinfo {title} {{Inspiral-inherited ringdown tails}},\ }\href {https://doi.org/10.1103/PhysRevD.110.104005} {\bibfield  {journal} {\bibinfo  {journal} {Phys. Rev. D}\ }\textbf {\bibinfo {volume} {110}},\ \bibinfo {pages} {104005} (\bibinfo {year} {2024}{\natexlab{a}})},\ \Eprint {https://arxiv.org/abs/2406.17018} {arXiv:2406.17018 [gr-qc]} \BibitemShut {NoStop}%
\bibitem [{\citenamefont {De~Amicis}\ \emph {et~al.}(2024{\natexlab{b}})\citenamefont {De~Amicis} \emph {et~al.}}]{DeAmicis:2024eoy}%
  \BibitemOpen
  \bibfield  {author} {\bibinfo {author} {\bibfnamefont {M.}~\bibnamefont {De~Amicis}} \emph {et~al.},\ }\bibfield  {title} {\bibinfo {title} {{Late-time tails in nonlinear evolutions of merging black holes}},\ }\href@noop {} {\  (\bibinfo {year} {2024}{\natexlab{b}})},\ \Eprint {https://arxiv.org/abs/2412.06887} {arXiv:2412.06887 [gr-qc]} \BibitemShut {NoStop}%
\bibitem [{\citenamefont {Kamaretsos}\ \emph {et~al.}(2012)\citenamefont {Kamaretsos}, \citenamefont {Hannam},\ and\ \citenamefont {Sathyaprakash}}]{Kamaretsos:2012bs}%
  \BibitemOpen
  \bibfield  {author} {\bibinfo {author} {\bibfnamefont {I.}~\bibnamefont {Kamaretsos}}, \bibinfo {author} {\bibfnamefont {M.}~\bibnamefont {Hannam}},\ and\ \bibinfo {author} {\bibfnamefont {B.}~\bibnamefont {Sathyaprakash}},\ }\bibfield  {title} {\bibinfo {title} {{Is black-hole ringdown a memory of its progenitor?}},\ }\href {https://doi.org/10.1103/PhysRevLett.109.141102} {\bibfield  {journal} {\bibinfo  {journal} {Phys. Rev. Lett.}\ }\textbf {\bibinfo {volume} {109}},\ \bibinfo {pages} {141102} (\bibinfo {year} {2012})},\ \Eprint {https://arxiv.org/abs/1207.0399} {arXiv:1207.0399 [gr-qc]} \BibitemShut {NoStop}%
\bibitem [{\citenamefont {London}(2020)}]{London:2018gaq}%
  \BibitemOpen
  \bibfield  {author} {\bibinfo {author} {\bibfnamefont {L.~T.}\ \bibnamefont {London}},\ }\bibfield  {title} {\bibinfo {title} {{Modeling ringdown. II. Aligned-spin binary black holes, implications for data analysis and fundamental theory}},\ }\href {https://doi.org/10.1103/PhysRevD.102.084052} {\bibfield  {journal} {\bibinfo  {journal} {Phys. Rev. D}\ }\textbf {\bibinfo {volume} {102}},\ \bibinfo {pages} {084052} (\bibinfo {year} {2020})},\ \Eprint {https://arxiv.org/abs/1801.08208} {arXiv:1801.08208 [gr-qc]} \BibitemShut {NoStop}%
\bibitem [{\citenamefont {Jim\'enez~Forteza}\ \emph {et~al.}(2020)\citenamefont {Jim\'enez~Forteza}, \citenamefont {Bhagwat}, \citenamefont {Pani},\ and\ \citenamefont {Ferrari}}]{JimenezForteza:2020cve}%
  \BibitemOpen
  \bibfield  {author} {\bibinfo {author} {\bibfnamefont {X.}~\bibnamefont {Jim\'enez~Forteza}}, \bibinfo {author} {\bibfnamefont {S.}~\bibnamefont {Bhagwat}}, \bibinfo {author} {\bibfnamefont {P.}~\bibnamefont {Pani}},\ and\ \bibinfo {author} {\bibfnamefont {V.}~\bibnamefont {Ferrari}},\ }\bibfield  {title} {\bibinfo {title} {{Spectroscopy of binary black hole ringdown using overtones and angular modes}},\ }\href {https://doi.org/10.1103/PhysRevD.102.044053} {\bibfield  {journal} {\bibinfo  {journal} {Phys. Rev. D}\ }\textbf {\bibinfo {volume} {102}},\ \bibinfo {pages} {044053} (\bibinfo {year} {2020})},\ \Eprint {https://arxiv.org/abs/2005.03260} {arXiv:2005.03260 [gr-qc]} \BibitemShut {NoStop}%
\bibitem [{\citenamefont {Cheung}\ \emph {et~al.}(2024)\citenamefont {Cheung}, \citenamefont {Berti}, \citenamefont {Baibhav},\ and\ \citenamefont {Cotesta}}]{Cheung:2023vki}%
  \BibitemOpen
  \bibfield  {author} {\bibinfo {author} {\bibfnamefont {M.~H.-Y.}\ \bibnamefont {Cheung}}, \bibinfo {author} {\bibfnamefont {E.}~\bibnamefont {Berti}}, \bibinfo {author} {\bibfnamefont {V.}~\bibnamefont {Baibhav}},\ and\ \bibinfo {author} {\bibfnamefont {R.}~\bibnamefont {Cotesta}},\ }\bibfield  {title} {\bibinfo {title} {{Extracting linear and nonlinear quasinormal modes from black hole merger simulations}},\ }\href {https://doi.org/10.1103/PhysRevD.109.044069} {\bibfield  {journal} {\bibinfo  {journal} {Phys. Rev. D}\ }\textbf {\bibinfo {volume} {109}},\ \bibinfo {pages} {044069} (\bibinfo {year} {2024})},\ \bibinfo {note} {[Erratum: Phys.Rev.D 110, 049902 (2024)]},\ \Eprint {https://arxiv.org/abs/2310.04489} {arXiv:2310.04489 [gr-qc]} \BibitemShut {NoStop}%
\bibitem [{\citenamefont {Maga{\~n}a~Zertuche}\ \emph {et~al.}(2024)\citenamefont {Maga{\~n}a~Zertuche} \emph {et~al.}}]{MaganaZertuche:2024ajz}%
  \BibitemOpen
  \bibfield  {author} {\bibinfo {author} {\bibfnamefont {L.}~\bibnamefont {Maga{\~n}a~Zertuche}} \emph {et~al.},\ }\bibfield  {title} {\bibinfo {title} {{High-Precision Ringdown Surrogate Model for Non-Precessing Binary Black Holes}},\ }\Eprint {https://arxiv.org/abs/2408.05300} {arXiv:2408.05300 [gr-qc]}  (\bibinfo {year} {2024})\BibitemShut {NoStop}%
\bibitem [{\citenamefont {Nobili}\ \emph {et~al.}(2025)\citenamefont {Nobili}, \citenamefont {Bhagwat}, \citenamefont {Pacilio},\ and\ \citenamefont {Gerosa}}]{Nobili:2025ydt}%
  \BibitemOpen
  \bibfield  {author} {\bibinfo {author} {\bibfnamefont {F.}~\bibnamefont {Nobili}}, \bibinfo {author} {\bibfnamefont {S.}~\bibnamefont {Bhagwat}}, \bibinfo {author} {\bibfnamefont {C.}~\bibnamefont {Pacilio}},\ and\ \bibinfo {author} {\bibfnamefont {D.}~\bibnamefont {Gerosa}},\ }\bibfield  {title} {\bibinfo {title} {{Ringdown mode amplitudes of precessing binary black holes}},\ }\Eprint {https://arxiv.org/abs/2504.17021} {arXiv:2504.17021 [gr-qc]}  (\bibinfo {year} {2025})\BibitemShut {NoStop}%
\bibitem [{\citenamefont {Isi}\ and\ \citenamefont {Farr}(2021)}]{Isi:2021iql}%
  \BibitemOpen
  \bibfield  {author} {\bibinfo {author} {\bibfnamefont {M.}~\bibnamefont {Isi}}\ and\ \bibinfo {author} {\bibfnamefont {W.~M.}\ \bibnamefont {Farr}},\ }\bibfield  {title} {\bibinfo {title} {{Analyzing black-hole ringdowns}},\ }\Eprint {https://arxiv.org/abs/2107.05609} {arXiv:2107.05609 [gr-qc]}  (\bibinfo {year} {2021})\BibitemShut {NoStop}%
\bibitem [{\citenamefont {Dhani}(2021)}]{Dhani:2020nik}%
  \BibitemOpen
  \bibfield  {author} {\bibinfo {author} {\bibfnamefont {A.}~\bibnamefont {Dhani}},\ }\bibfield  {title} {\bibinfo {title} {{Importance of mirror modes in binary black hole ringdown waveform}},\ }\href {https://doi.org/10.1103/PhysRevD.103.104048} {\bibfield  {journal} {\bibinfo  {journal} {Phys. Rev. D}\ }\textbf {\bibinfo {volume} {103}},\ \bibinfo {pages} {104048} (\bibinfo {year} {2021})},\ \Eprint {https://arxiv.org/abs/2010.08602} {arXiv:2010.08602 [gr-qc]} \BibitemShut {NoStop}%
\bibitem [{\citenamefont {Li}\ \emph {et~al.}(2022)\citenamefont {Li}, \citenamefont {Sun}, \citenamefont {Lo}, \citenamefont {Payne},\ and\ \citenamefont {Chen}}]{Li:2021wgz}%
  \BibitemOpen
  \bibfield  {author} {\bibinfo {author} {\bibfnamefont {X.}~\bibnamefont {Li}}, \bibinfo {author} {\bibfnamefont {L.}~\bibnamefont {Sun}}, \bibinfo {author} {\bibfnamefont {R.~K.~L.}\ \bibnamefont {Lo}}, \bibinfo {author} {\bibfnamefont {E.}~\bibnamefont {Payne}},\ and\ \bibinfo {author} {\bibfnamefont {Y.}~\bibnamefont {Chen}},\ }\bibfield  {title} {\bibinfo {title} {{Angular emission patterns of remnant black holes}},\ }\href {https://doi.org/10.1103/PhysRevD.105.024016} {\bibfield  {journal} {\bibinfo  {journal} {Phys. Rev. D}\ }\textbf {\bibinfo {volume} {105}},\ \bibinfo {pages} {024016} (\bibinfo {year} {2022})},\ \Eprint {https://arxiv.org/abs/2110.03116} {arXiv:2110.03116 [gr-qc]} \BibitemShut {NoStop}%
\bibitem [{\citenamefont {Carullo}\ \emph {et~al.}(2023)\citenamefont {Carullo}, \citenamefont {Del~Pozzo},\ and\ \citenamefont {Veitch}}]{pyRing}%
  \BibitemOpen
  \bibfield  {author} {\bibinfo {author} {\bibfnamefont {G.}~\bibnamefont {Carullo}}, \bibinfo {author} {\bibfnamefont {W.}~\bibnamefont {Del~Pozzo}},\ and\ \bibinfo {author} {\bibfnamefont {J.}~\bibnamefont {Veitch}},\ }\href {https://doi.org/10.5281/zenodo.8165508} {\bibinfo {title} {\texttt{pyRing}: a time-domain ringdown analysis python package}},\ \bibinfo {howpublished} {\href{https://git.ligo.org/lscsoft/pyring}{git.ligo.org/lscsoft/pyring}} (\bibinfo {year} {2023})\BibitemShut {NoStop}%
\bibitem [{\citenamefont {Carullo}\ \emph {et~al.}(2018)\citenamefont {Carullo} \emph {et~al.}}]{Carullo:2018sfu}%
  \BibitemOpen
  \bibfield  {author} {\bibinfo {author} {\bibfnamefont {G.}~\bibnamefont {Carullo}} \emph {et~al.},\ }\bibfield  {title} {\bibinfo {title} {{Empirical tests of the black hole no-hair conjecture using gravitational-wave observations}},\ }\href {https://doi.org/10.1103/PhysRevD.98.104020} {\bibfield  {journal} {\bibinfo  {journal} {Phys. Rev. D}\ }\textbf {\bibinfo {volume} {98}},\ \bibinfo {pages} {104020} (\bibinfo {year} {2018})},\ \Eprint {https://arxiv.org/abs/1805.04760} {arXiv:1805.04760 [gr-qc]} \BibitemShut {NoStop}%
\bibitem [{\citenamefont {Giesler}\ \emph {et~al.}(2025)\citenamefont {Giesler} \emph {et~al.}}]{Giesler:2024hcr}%
  \BibitemOpen
  \bibfield  {author} {\bibinfo {author} {\bibfnamefont {M.}~\bibnamefont {Giesler}} \emph {et~al.},\ }\bibfield  {title} {\bibinfo {title} {{Overtones and nonlinearities in binary black hole ringdowns}},\ }\href {https://doi.org/10.1103/PhysRevD.111.084041} {\bibfield  {journal} {\bibinfo  {journal} {Phys. Rev. D}\ }\textbf {\bibinfo {volume} {111}},\ \bibinfo {pages} {084041} (\bibinfo {year} {2025})},\ \Eprint {https://arxiv.org/abs/2411.11269} {arXiv:2411.11269 [gr-qc]} \BibitemShut {NoStop}%
\bibitem [{\citenamefont {Isi}\ \emph {et~al.}(2019{\natexlab{b}})\citenamefont {Isi}, \citenamefont {Chatziioannou},\ and\ \citenamefont {Farr}}]{Isi:2019asy}%
  \BibitemOpen
  \bibfield  {author} {\bibinfo {author} {\bibfnamefont {M.}~\bibnamefont {Isi}}, \bibinfo {author} {\bibfnamefont {K.}~\bibnamefont {Chatziioannou}},\ and\ \bibinfo {author} {\bibfnamefont {W.~M.}\ \bibnamefont {Farr}},\ }\bibfield  {title} {\bibinfo {title} {{Hierarchical test of general relativity with gravitational waves}},\ }\href {https://doi.org/10.1103/PhysRevLett.123.121101} {\bibfield  {journal} {\bibinfo  {journal} {Phys. Rev. Lett.}\ }\textbf {\bibinfo {volume} {123}},\ \bibinfo {pages} {121101} (\bibinfo {year} {2019}{\natexlab{b}})},\ \Eprint {https://arxiv.org/abs/1904.08011} {arXiv:1904.08011 [gr-qc]} \BibitemShut {NoStop}%
\bibitem [{\citenamefont {Ma}\ \emph {et~al.}(2022)\citenamefont {Ma}, \citenamefont {Mitman}, \citenamefont {Sun}, \citenamefont {Deppe}, \citenamefont {H\'ebert}, \citenamefont {Kidder}, \citenamefont {Moxon}, \citenamefont {Throwe}, \citenamefont {Vu},\ and\ \citenamefont {Chen}}]{Ma:2022wpv}%
  \BibitemOpen
  \bibfield  {author} {\bibinfo {author} {\bibfnamefont {S.}~\bibnamefont {Ma}}, \bibinfo {author} {\bibfnamefont {K.}~\bibnamefont {Mitman}}, \bibinfo {author} {\bibfnamefont {L.}~\bibnamefont {Sun}}, \bibinfo {author} {\bibfnamefont {N.}~\bibnamefont {Deppe}}, \bibinfo {author} {\bibfnamefont {F.}~\bibnamefont {H\'ebert}}, \bibinfo {author} {\bibfnamefont {L.~E.}\ \bibnamefont {Kidder}}, \bibinfo {author} {\bibfnamefont {J.}~\bibnamefont {Moxon}}, \bibinfo {author} {\bibfnamefont {W.}~\bibnamefont {Throwe}}, \bibinfo {author} {\bibfnamefont {N.~L.}\ \bibnamefont {Vu}},\ and\ \bibinfo {author} {\bibfnamefont {Y.}~\bibnamefont {Chen}},\ }\bibfield  {title} {\bibinfo {title} {{Quasinormal-mode filters: A new approach to analyze the gravitational-wave ringdown of binary black-hole mergers}},\ }\href {https://doi.org/10.1103/PhysRevD.106.084036} {\bibfield  {journal} {\bibinfo  {journal} {Phys. Rev. D}\ }\textbf {\bibinfo {volume} {106}},\ \bibinfo {pages} {084036} (\bibinfo {year} {2022})},\ \Eprint {https://arxiv.org/abs/2207.10870} {arXiv:2207.10870 [gr-qc]} \BibitemShut {NoStop}%
\bibitem [{\citenamefont {Ma}\ \emph {et~al.}(2023{\natexlab{a}})\citenamefont {Ma}, \citenamefont {Sun},\ and\ \citenamefont {Chen}}]{Ma:2023cwe}%
  \BibitemOpen
  \bibfield  {author} {\bibinfo {author} {\bibfnamefont {S.}~\bibnamefont {Ma}}, \bibinfo {author} {\bibfnamefont {L.}~\bibnamefont {Sun}},\ and\ \bibinfo {author} {\bibfnamefont {Y.}~\bibnamefont {Chen}},\ }\bibfield  {title} {\bibinfo {title} {{Black Hole Spectroscopy by Mode Cleaning}},\ }\href {https://doi.org/10.1103/PhysRevLett.130.141401} {\bibfield  {journal} {\bibinfo  {journal} {Phys. Rev. Lett.}\ }\textbf {\bibinfo {volume} {130}},\ \bibinfo {pages} {141401} (\bibinfo {year} {2023}{\natexlab{a}})},\ \Eprint {https://arxiv.org/abs/2301.06705} {arXiv:2301.06705 [gr-qc]} \BibitemShut {NoStop}%
\bibitem [{\citenamefont {Ma}\ \emph {et~al.}(2023{\natexlab{b}})\citenamefont {Ma}, \citenamefont {Sun},\ and\ \citenamefont {Chen}}]{Ma:2023vvr}%
  \BibitemOpen
  \bibfield  {author} {\bibinfo {author} {\bibfnamefont {S.}~\bibnamefont {Ma}}, \bibinfo {author} {\bibfnamefont {L.}~\bibnamefont {Sun}},\ and\ \bibinfo {author} {\bibfnamefont {Y.}~\bibnamefont {Chen}},\ }\bibfield  {title} {\bibinfo {title} {{Using rational filters to uncover the first ringdown overtone in GW150914}},\ }\href {https://doi.org/10.1103/PhysRevD.107.084010} {\bibfield  {journal} {\bibinfo  {journal} {Phys. Rev. D}\ }\textbf {\bibinfo {volume} {107}},\ \bibinfo {pages} {084010} (\bibinfo {year} {2023}{\natexlab{b}})},\ \Eprint {https://arxiv.org/abs/2301.06639} {arXiv:2301.06639 [gr-qc]} \BibitemShut {NoStop}%
\bibitem [{\citenamefont {Lu}\ \emph {et~al.}(2025)\citenamefont {Lu}, \citenamefont {Ma}, \citenamefont {Piccinni}, \citenamefont {Sun},\ and\ \citenamefont {Finch}}]{Lu:2025mwp}%
  \BibitemOpen
  \bibfield  {author} {\bibinfo {author} {\bibfnamefont {N.}~\bibnamefont {Lu}}, \bibinfo {author} {\bibfnamefont {S.}~\bibnamefont {Ma}}, \bibinfo {author} {\bibfnamefont {O.~J.}\ \bibnamefont {Piccinni}}, \bibinfo {author} {\bibfnamefont {L.}~\bibnamefont {Sun}},\ and\ \bibinfo {author} {\bibfnamefont {E.}~\bibnamefont {Finch}},\ }\bibfield  {title} {\bibinfo {title} {{Statistical identification of ringdown modes with rational filters}},\ }\href@noop {} {\  (\bibinfo {year} {2025})},\ \Eprint {https://arxiv.org/abs/2505.18560} {arXiv:2505.18560 [gr-qc]} \BibitemShut {NoStop}%
\bibitem [{\citenamefont {Brito}\ \emph {et~al.}(2018)\citenamefont {Brito}, \citenamefont {Buonanno},\ and\ \citenamefont {Raymond}}]{Brito:2018rfr}%
  \BibitemOpen
  \bibfield  {author} {\bibinfo {author} {\bibfnamefont {R.}~\bibnamefont {Brito}}, \bibinfo {author} {\bibfnamefont {A.}~\bibnamefont {Buonanno}},\ and\ \bibinfo {author} {\bibfnamefont {V.}~\bibnamefont {Raymond}},\ }\bibfield  {title} {\bibinfo {title} {{Black-hole Spectroscopy by Making Full Use of Gravitational-Wave Modeling}},\ }\href {https://doi.org/10.1103/PhysRevD.98.084038} {\bibfield  {journal} {\bibinfo  {journal} {Phys. Rev. D}\ }\textbf {\bibinfo {volume} {98}},\ \bibinfo {pages} {084038} (\bibinfo {year} {2018})},\ \Eprint {https://arxiv.org/abs/1805.00293} {arXiv:1805.00293 [gr-qc]} \BibitemShut {NoStop}%
\bibitem [{\citenamefont {Pompili}\ \emph {et~al.}(2025)\citenamefont {Pompili}, \citenamefont {Maggio}, \citenamefont {Silva},\ and\ \citenamefont {Buonanno}}]{Pompili:2025cdc}%
  \BibitemOpen
  \bibfield  {author} {\bibinfo {author} {\bibfnamefont {L.}~\bibnamefont {Pompili}}, \bibinfo {author} {\bibfnamefont {E.}~\bibnamefont {Maggio}}, \bibinfo {author} {\bibfnamefont {H.~O.}\ \bibnamefont {Silva}},\ and\ \bibinfo {author} {\bibfnamefont {A.}~\bibnamefont {Buonanno}},\ }\bibfield  {title} {\bibinfo {title} {{Parametrized spin-precessing inspiral-merger-ringdown waveform model for tests of general relativity}},\ }\href {https://doi.org/10.1103/ng8w-98sz} {\bibfield  {journal} {\bibinfo  {journal} {Phys. Rev. D}\ }\textbf {\bibinfo {volume} {111}},\ \bibinfo {pages} {124040} (\bibinfo {year} {2025})},\ \Eprint {https://arxiv.org/abs/2504.10130} {arXiv:2504.10130 [gr-qc]} \BibitemShut {NoStop}%
\bibitem [{\citenamefont {Maggio}\ \emph {et~al.}(2023)\citenamefont {Maggio}, \citenamefont {Silva}, \citenamefont {Buonanno},\ and\ \citenamefont {Ghosh}}]{Maggio:2022hre}%
  \BibitemOpen
  \bibfield  {author} {\bibinfo {author} {\bibfnamefont {E.}~\bibnamefont {Maggio}}, \bibinfo {author} {\bibfnamefont {H.~O.}\ \bibnamefont {Silva}}, \bibinfo {author} {\bibfnamefont {A.}~\bibnamefont {Buonanno}},\ and\ \bibinfo {author} {\bibfnamefont {A.}~\bibnamefont {Ghosh}},\ }\bibfield  {title} {\bibinfo {title} {{Tests of general relativity in the nonlinear regime: A parametrized plunge-merger-ringdown gravitational waveform model}},\ }\href {https://doi.org/10.1103/PhysRevD.108.024043} {\bibfield  {journal} {\bibinfo  {journal} {Phys. Rev. D}\ }\textbf {\bibinfo {volume} {108}},\ \bibinfo {pages} {024043} (\bibinfo {year} {2023})},\ \Eprint {https://arxiv.org/abs/2212.09655} {arXiv:2212.09655 [gr-qc]} \BibitemShut {NoStop}%
\bibitem [{\citenamefont {Ghosh}\ \emph {et~al.}(2021)\citenamefont {Ghosh}, \citenamefont {Brito},\ and\ \citenamefont {Buonanno}}]{Ghosh:2021mrv}%
  \BibitemOpen
  \bibfield  {author} {\bibinfo {author} {\bibfnamefont {A.}~\bibnamefont {Ghosh}}, \bibinfo {author} {\bibfnamefont {R.}~\bibnamefont {Brito}},\ and\ \bibinfo {author} {\bibfnamefont {A.}~\bibnamefont {Buonanno}},\ }\bibfield  {title} {\bibinfo {title} {{Constraints on quasinormal-mode frequencies with LIGO-Virgo binary\textendash{}black-hole observations}},\ }\href {https://doi.org/10.1103/PhysRevD.103.124041} {\bibfield  {journal} {\bibinfo  {journal} {Phys. Rev. D}\ }\textbf {\bibinfo {volume} {103}},\ \bibinfo {pages} {124041} (\bibinfo {year} {2021})},\ \Eprint {https://arxiv.org/abs/2104.01906} {arXiv:2104.01906 [gr-qc]} \BibitemShut {NoStop}%
\bibitem [{\citenamefont {Ramos-Buades}\ \emph {et~al.}(2023)\citenamefont {Ramos-Buades}, \citenamefont {Buonanno}, \citenamefont {Estell\'es}, \citenamefont {Khalil}, \citenamefont {Mihaylov}, \citenamefont {Ossokine}, \citenamefont {Pompili},\ and\ \citenamefont {Shiferaw}}]{Ramos-Buades:2023ehm}%
  \BibitemOpen
  \bibfield  {author} {\bibinfo {author} {\bibfnamefont {A.}~\bibnamefont {Ramos-Buades}}, \bibinfo {author} {\bibfnamefont {A.}~\bibnamefont {Buonanno}}, \bibinfo {author} {\bibfnamefont {H.}~\bibnamefont {Estell\'es}}, \bibinfo {author} {\bibfnamefont {M.}~\bibnamefont {Khalil}}, \bibinfo {author} {\bibfnamefont {D.~P.}\ \bibnamefont {Mihaylov}}, \bibinfo {author} {\bibfnamefont {S.}~\bibnamefont {Ossokine}}, \bibinfo {author} {\bibfnamefont {L.}~\bibnamefont {Pompili}},\ and\ \bibinfo {author} {\bibfnamefont {M.}~\bibnamefont {Shiferaw}},\ }\bibfield  {title} {\bibinfo {title} {{Next generation of accurate and efficient multipolar precessing-spin effective-one-body waveforms for binary black holes}},\ }\href {https://doi.org/10.1103/PhysRevD.108.124037} {\bibfield  {journal} {\bibinfo  {journal} {Phys. Rev. D}\ }\textbf {\bibinfo {volume} {108}},\ \bibinfo {pages} {124037} (\bibinfo {year} {2023})},\ \Eprint {https://arxiv.org/abs/2303.18046} {arXiv:2303.18046 [gr-qc]} \BibitemShut {NoStop}%
\bibitem [{\citenamefont {Pompili}\ \emph {et~al.}(2023)\citenamefont {Pompili} \emph {et~al.}}]{Pompili:2023tna}%
  \BibitemOpen
  \bibfield  {author} {\bibinfo {author} {\bibfnamefont {L.}~\bibnamefont {Pompili}} \emph {et~al.},\ }\bibfield  {title} {\bibinfo {title} {{Laying the foundation of the effective-one-body waveform models SEOBNRv5: Improved accuracy and efficiency for spinning nonprecessing binary black holes}},\ }\href {https://doi.org/10.1103/PhysRevD.108.124035} {\bibfield  {journal} {\bibinfo  {journal} {Phys. Rev. D}\ }\textbf {\bibinfo {volume} {108}},\ \bibinfo {pages} {124035} (\bibinfo {year} {2023})},\ \Eprint {https://arxiv.org/abs/2303.18039} {arXiv:2303.18039 [gr-qc]} \BibitemShut {NoStop}%
\bibitem [{\citenamefont {Jiménez-Forteza}\ \emph {et~al.}(2017)\citenamefont {Jiménez-Forteza}, \citenamefont {Keitel}, \citenamefont {Husa}, \citenamefont {Hannam}, \citenamefont {Khan},\ and\ \citenamefont {Pürrer}}]{Jimenez-Forteza:2016oae}%
  \BibitemOpen
  \bibfield  {author} {\bibinfo {author} {\bibfnamefont {X.}~\bibnamefont {Jiménez-Forteza}}, \bibinfo {author} {\bibfnamefont {D.}~\bibnamefont {Keitel}}, \bibinfo {author} {\bibfnamefont {S.}~\bibnamefont {Husa}}, \bibinfo {author} {\bibfnamefont {M.}~\bibnamefont {Hannam}}, \bibinfo {author} {\bibfnamefont {S.}~\bibnamefont {Khan}},\ and\ \bibinfo {author} {\bibfnamefont {M.}~\bibnamefont {Pürrer}},\ }\bibfield  {title} {\bibinfo {title} {{Hierarchical data-driven approach to fitting numerical relativity data for nonprecessing binary black holes with an application to final spin and radiated energy}},\ }\href {https://doi.org/10.1103/PhysRevD.95.064024} {\bibfield  {journal} {\bibinfo  {journal} {Phys. Rev. D}\ }\textbf {\bibinfo {volume} {95}},\ \bibinfo {pages} {064024} (\bibinfo {year} {2017})},\ \Eprint {https://arxiv.org/abs/1611.00332} {arXiv:1611.00332 [gr-qc]} \BibitemShut {NoStop}%
\bibitem [{\citenamefont {Hofmann}\ \emph {et~al.}(2016)\citenamefont {Hofmann}, \citenamefont {Barausse},\ and\ \citenamefont {Rezzolla}}]{Hofmann:2016yih}%
  \BibitemOpen
  \bibfield  {author} {\bibinfo {author} {\bibfnamefont {F.}~\bibnamefont {Hofmann}}, \bibinfo {author} {\bibfnamefont {E.}~\bibnamefont {Barausse}},\ and\ \bibinfo {author} {\bibfnamefont {L.}~\bibnamefont {Rezzolla}},\ }\bibfield  {title} {\bibinfo {title} {{The final spin from binary black holes in quasi-circular orbits}},\ }\href {https://doi.org/10.3847/2041-8205/825/2/L19} {\bibfield  {journal} {\bibinfo  {journal} {Astrophys. J. Lett.}\ }\textbf {\bibinfo {volume} {825}},\ \bibinfo {pages} {L19} (\bibinfo {year} {2016})},\ \Eprint {https://arxiv.org/abs/1605.01938} {arXiv:1605.01938 [gr-qc]} \BibitemShut {NoStop}%
\bibitem [{\citenamefont {Baker}\ \emph {et~al.}(2008)\citenamefont {Baker}, \citenamefont {Boggs}, \citenamefont {Centrella}, \citenamefont {Kelly}, \citenamefont {McWilliams} \emph {et~al.}}]{Baker:2008mj}%
  \BibitemOpen
  \bibfield  {author} {\bibinfo {author} {\bibfnamefont {J.~G.}\ \bibnamefont {Baker}}, \bibinfo {author} {\bibfnamefont {W.~D.}\ \bibnamefont {Boggs}}, \bibinfo {author} {\bibfnamefont {J.}~\bibnamefont {Centrella}}, \bibinfo {author} {\bibfnamefont {B.~J.}\ \bibnamefont {Kelly}}, \bibinfo {author} {\bibfnamefont {S.~T.}\ \bibnamefont {McWilliams}}, \emph {et~al.},\ }\bibfield  {title} {\bibinfo {title} {{Mergers of non-spinning black-hole binaries: Gravitational radiation characteristics}},\ }\href {https://doi.org/10.1103/PhysRevD.78.044046} {\bibfield  {journal} {\bibinfo  {journal} {Phys. Rev. D}\ }\textbf {\bibinfo {volume} {78}},\ \bibinfo {pages} {044046} (\bibinfo {year} {2008})},\ \Eprint {https://arxiv.org/abs/0805.1428} {arXiv:0805.1428 [gr-qc]} \BibitemShut {NoStop}%
\bibitem [{\citenamefont {Damour}\ and\ \citenamefont {Nagar}(2014)}]{Damour:2014yha}%
  \BibitemOpen
  \bibfield  {author} {\bibinfo {author} {\bibfnamefont {T.}~\bibnamefont {Damour}}\ and\ \bibinfo {author} {\bibfnamefont {A.}~\bibnamefont {Nagar}},\ }\bibfield  {title} {\bibinfo {title} {{A new analytic representation of the ringdown waveform of coalescing spinning black hole binaries}},\ }\href {https://doi.org/10.1103/PhysRevD.90.024054} {\bibfield  {journal} {\bibinfo  {journal} {Phys. Rev. D}\ }\textbf {\bibinfo {volume} {90}},\ \bibinfo {pages} {024054} (\bibinfo {year} {2014})},\ \Eprint {https://arxiv.org/abs/1406.0401} {arXiv:1406.0401 [gr-qc]} \BibitemShut {NoStop}%
\bibitem [{\citenamefont {Blanchet}(2024)}]{Blanchet:2013haa}%
  \BibitemOpen
  \bibfield  {author} {\bibinfo {author} {\bibfnamefont {L.}~\bibnamefont {Blanchet}},\ }\bibfield  {title} {\bibinfo {title} {{Post-Newtonian theory for gravitational waves}},\ }\href {https://doi.org/10.1007/s41114-024-00050-z} {\bibfield  {journal} {\bibinfo  {journal} {Living Rev. Relativity}\ }\textbf {\bibinfo {volume} {27}},\ \bibinfo {pages} {4} (\bibinfo {year} {2024})},\ \Eprint {https://arxiv.org/abs/1310.1528} {arXiv:1310.1528 [gr-qc]} \BibitemShut {NoStop}%
\bibitem [{\citenamefont {Cotesta}\ \emph {et~al.}(2018)\citenamefont {Cotesta}, \citenamefont {Buonanno}, \citenamefont {Boh{\'e}}, \citenamefont {Taracchini}, \citenamefont {Hinder},\ and\ \citenamefont {Ossokine}}]{Cotesta:2018fcv}%
  \BibitemOpen
  \bibfield  {author} {\bibinfo {author} {\bibfnamefont {R.}~\bibnamefont {Cotesta}}, \bibinfo {author} {\bibfnamefont {A.}~\bibnamefont {Buonanno}}, \bibinfo {author} {\bibfnamefont {A.}~\bibnamefont {Boh{\'e}}}, \bibinfo {author} {\bibfnamefont {A.}~\bibnamefont {Taracchini}}, \bibinfo {author} {\bibfnamefont {I.}~\bibnamefont {Hinder}},\ and\ \bibinfo {author} {\bibfnamefont {S.}~\bibnamefont {Ossokine}},\ }\bibfield  {title} {\bibinfo {title} {{Enriching the Symphony of Gravitational Waves from Binary Black Holes by Tuning Higher Harmonics}},\ }\href {https://doi.org/10.1103/PhysRevD.98.084028} {\bibfield  {journal} {\bibinfo  {journal} {Phys. Rev. D}\ }\textbf {\bibinfo {volume} {98}},\ \bibinfo {pages} {084028} (\bibinfo {year} {2018})},\ \Eprint {https://arxiv.org/abs/1803.10701} {arXiv:1803.10701 [gr-qc]} \BibitemShut {NoStop}%
\bibitem [{\citenamefont {Mills}\ and\ \citenamefont {Fairhurst}(2021)}]{Mills:2020thr}%
  \BibitemOpen
  \bibfield  {author} {\bibinfo {author} {\bibfnamefont {C.}~\bibnamefont {Mills}}\ and\ \bibinfo {author} {\bibfnamefont {S.}~\bibnamefont {Fairhurst}},\ }\bibfield  {title} {\bibinfo {title} {{Measuring gravitational-wave higher-order multipoles}},\ }\href {https://doi.org/10.1103/PhysRevD.103.024042} {\bibfield  {journal} {\bibinfo  {journal} {Phys. Rev. D}\ }\textbf {\bibinfo {volume} {103}},\ \bibinfo {pages} {024042} (\bibinfo {year} {2021})},\ \Eprint {https://arxiv.org/abs/2007.04313} {arXiv:2007.04313 [gr-qc]} \BibitemShut {NoStop}%
\bibitem [{\citenamefont {Zhong}\ \emph {et~al.}(2024)\citenamefont {Zhong}, \citenamefont {Isi}, \citenamefont {Chatziioannou},\ and\ \citenamefont {Farr}}]{Zhong:2024pwb}%
  \BibitemOpen
  \bibfield  {author} {\bibinfo {author} {\bibfnamefont {H.}~\bibnamefont {Zhong}}, \bibinfo {author} {\bibfnamefont {M.}~\bibnamefont {Isi}}, \bibinfo {author} {\bibfnamefont {K.}~\bibnamefont {Chatziioannou}},\ and\ \bibinfo {author} {\bibfnamefont {W.~M.}\ \bibnamefont {Farr}},\ }\bibfield  {title} {\bibinfo {title} {{Multidimensional hierarchical tests of general relativity with gravitational waves}},\ }\href {https://doi.org/10.1103/PhysRevD.110.044053} {\bibfield  {journal} {\bibinfo  {journal} {Phys. Rev. D}\ }\textbf {\bibinfo {volume} {110}},\ \bibinfo {pages} {044053} (\bibinfo {year} {2024})},\ \Eprint {https://arxiv.org/abs/2405.19556} {arXiv:2405.19556 [gr-qc]} \BibitemShut {NoStop}%
\bibitem [{\citenamefont {Abac}\ \emph {et~al.}(2025{\natexlab{c}})\citenamefont {Abac} \emph {et~al.}}]{TGR-GWTC4}%
  \BibitemOpen
  \bibfield  {author} {\bibinfo {author} {\bibfnamefont {A.~G.}\ \bibnamefont {Abac}} \emph {et~al.},\ }\bibfield  {title} {\bibinfo {title} {{GWTC-4.0: Tests of General Relativity}},\ }\Eprint {https://arxiv.org/abs/2509.0000} {arXiv:2509.0000 [gr-qc]}  (\bibinfo {year} {2025}{\natexlab{c}})\BibitemShut {NoStop}%
\bibitem [{\citenamefont {Payne}\ \emph {et~al.}(2023)\citenamefont {Payne}, \citenamefont {Isi}, \citenamefont {Chatziioannou},\ and\ \citenamefont {Farr}}]{Payne:2023kwj}%
  \BibitemOpen
  \bibfield  {author} {\bibinfo {author} {\bibfnamefont {E.}~\bibnamefont {Payne}}, \bibinfo {author} {\bibfnamefont {M.}~\bibnamefont {Isi}}, \bibinfo {author} {\bibfnamefont {K.}~\bibnamefont {Chatziioannou}},\ and\ \bibinfo {author} {\bibfnamefont {W.~M.}\ \bibnamefont {Farr}},\ }\bibfield  {title} {\bibinfo {title} {{Fortifying gravitational-wave tests of general relativity against astrophysical assumptions}},\ }\href {https://doi.org/10.1103/PhysRevD.108.124060} {\bibfield  {journal} {\bibinfo  {journal} {Phys. Rev. D}\ }\textbf {\bibinfo {volume} {108}},\ \bibinfo {pages} {124060} (\bibinfo {year} {2023})},\ \Eprint {https://arxiv.org/abs/2309.04528} {arXiv:2309.04528 [gr-qc]} \BibitemShut {NoStop}%
\bibitem [{\citenamefont {Pacilio}\ \emph {et~al.}(2024)\citenamefont {Pacilio}, \citenamefont {Gerosa},\ and\ \citenamefont {Bhagwat}}]{Pacilio:2023uef}%
  \BibitemOpen
  \bibfield  {author} {\bibinfo {author} {\bibfnamefont {C.}~\bibnamefont {Pacilio}}, \bibinfo {author} {\bibfnamefont {D.}~\bibnamefont {Gerosa}},\ and\ \bibinfo {author} {\bibfnamefont {S.}~\bibnamefont {Bhagwat}},\ }\bibfield  {title} {\bibinfo {title} {{Catalog variance of testing general relativity with gravitational-wave data}},\ }\href {https://doi.org/10.1103/PhysRevD.109.L081302} {\bibfield  {journal} {\bibinfo  {journal} {Phys. Rev. D}\ }\textbf {\bibinfo {volume} {109}},\ \bibinfo {pages} {L081302} (\bibinfo {year} {2024})},\ \Eprint {https://arxiv.org/abs/2310.03811} {arXiv:2310.03811 [gr-qc]} \BibitemShut {NoStop}%
\bibitem [{\citenamefont {Del~Pozzo}\ and\ \citenamefont {Nagar}(2017)}]{DelPozzo:2016kmd}%
  \BibitemOpen
  \bibfield  {author} {\bibinfo {author} {\bibfnamefont {W.}~\bibnamefont {Del~Pozzo}}\ and\ \bibinfo {author} {\bibfnamefont {A.}~\bibnamefont {Nagar}},\ }\bibfield  {title} {\bibinfo {title} {{Analytic family of post-merger template waveforms}},\ }\href {https://doi.org/10.1103/PhysRevD.95.124034} {\bibfield  {journal} {\bibinfo  {journal} {Phys. Rev. D}\ }\textbf {\bibinfo {volume} {95}},\ \bibinfo {pages} {124034} (\bibinfo {year} {2017})},\ \Eprint {https://arxiv.org/abs/1606.03952} {arXiv:1606.03952 [gr-qc]} \BibitemShut {NoStop}%
\bibitem [{\citenamefont {Nagar}\ \emph {et~al.}(2020)\citenamefont {Nagar}, \citenamefont {Riemenschneider}, \citenamefont {Pratten}, \citenamefont {Rettegno},\ and\ \citenamefont {Messina}}]{Nagar:2020pcj}%
  \BibitemOpen
  \bibfield  {author} {\bibinfo {author} {\bibfnamefont {A.}~\bibnamefont {Nagar}}, \bibinfo {author} {\bibfnamefont {G.}~\bibnamefont {Riemenschneider}}, \bibinfo {author} {\bibfnamefont {G.}~\bibnamefont {Pratten}}, \bibinfo {author} {\bibfnamefont {P.}~\bibnamefont {Rettegno}},\ and\ \bibinfo {author} {\bibfnamefont {F.}~\bibnamefont {Messina}},\ }\bibfield  {title} {\bibinfo {title} {{Multipolar effective one body waveform model for spin-aligned black hole binaries}},\ }\href {https://doi.org/10.1103/PhysRevD.102.024077} {\bibfield  {journal} {\bibinfo  {journal} {Phys. Rev. D}\ }\textbf {\bibinfo {volume} {102}},\ \bibinfo {pages} {024077} (\bibinfo {year} {2020})},\ \Eprint {https://arxiv.org/abs/2001.09082} {arXiv:2001.09082 [gr-qc]} \BibitemShut {NoStop}%
\bibitem [{\citenamefont {Yunes}\ \emph {et~al.}(2016)\citenamefont {Yunes}, \citenamefont {Yagi},\ and\ \citenamefont {Pretorius}}]{Yunes:2016jcc}%
  \BibitemOpen
  \bibfield  {author} {\bibinfo {author} {\bibfnamefont {N.}~\bibnamefont {Yunes}}, \bibinfo {author} {\bibfnamefont {K.}~\bibnamefont {Yagi}},\ and\ \bibinfo {author} {\bibfnamefont {F.}~\bibnamefont {Pretorius}},\ }\bibfield  {title} {\bibinfo {title} {{Theoretical Physics Implications of the Binary Black-Hole Mergers GW150914 and GW151226}},\ }\href {https://doi.org/10.1103/PhysRevD.94.084002} {\bibfield  {journal} {\bibinfo  {journal} {Phys. Rev. D}\ }\textbf {\bibinfo {volume} {94}},\ \bibinfo {pages} {084002} (\bibinfo {year} {2016})},\ \Eprint {https://arxiv.org/abs/1603.08955} {arXiv:1603.08955 [gr-qc]} \BibitemShut {NoStop}%
\bibitem [{\citenamefont {Maselli}\ \emph {et~al.}(2020)\citenamefont {Maselli}, \citenamefont {Pani}, \citenamefont {Gualtieri},\ and\ \citenamefont {Berti}}]{Maselli:2019mjd}%
  \BibitemOpen
  \bibfield  {author} {\bibinfo {author} {\bibfnamefont {A.}~\bibnamefont {Maselli}}, \bibinfo {author} {\bibfnamefont {P.}~\bibnamefont {Pani}}, \bibinfo {author} {\bibfnamefont {L.}~\bibnamefont {Gualtieri}},\ and\ \bibinfo {author} {\bibfnamefont {E.}~\bibnamefont {Berti}},\ }\bibfield  {title} {\bibinfo {title} {{Parametrized ringdown spin expansion coefficients: a data-analysis framework for black-hole spectroscopy with multiple events}},\ }\href {https://doi.org/10.1103/PhysRevD.101.024043} {\bibfield  {journal} {\bibinfo  {journal} {Phys. Rev. D}\ }\textbf {\bibinfo {volume} {101}},\ \bibinfo {pages} {024043} (\bibinfo {year} {2020})},\ \Eprint {https://arxiv.org/abs/1910.12893} {arXiv:1910.12893 [gr-qc]} \BibitemShut {NoStop}%
\bibitem [{\citenamefont {Silva}\ \emph {et~al.}(2023)\citenamefont {Silva}, \citenamefont {Ghosh},\ and\ \citenamefont {Buonanno}}]{Silva:2022srr}%
  \BibitemOpen
  \bibfield  {author} {\bibinfo {author} {\bibfnamefont {H.~O.}\ \bibnamefont {Silva}}, \bibinfo {author} {\bibfnamefont {A.}~\bibnamefont {Ghosh}},\ and\ \bibinfo {author} {\bibfnamefont {A.}~\bibnamefont {Buonanno}},\ }\bibfield  {title} {\bibinfo {title} {{Black-hole ringdown as a probe of higher-curvature gravity theories}},\ }\href {https://doi.org/10.1103/PhysRevD.107.044030} {\bibfield  {journal} {\bibinfo  {journal} {Phys. Rev. D}\ }\textbf {\bibinfo {volume} {107}},\ \bibinfo {pages} {044030} (\bibinfo {year} {2023})},\ \Eprint {https://arxiv.org/abs/2205.05132} {arXiv:2205.05132 [gr-qc]} \BibitemShut {NoStop}%
\bibitem [{\citenamefont {S\"anger}\ \emph {et~al.}(2024)\citenamefont {S\"anger} \emph {et~al.}}]{Sanger:2024axs}%
  \BibitemOpen
  \bibfield  {author} {\bibinfo {author} {\bibfnamefont {E.~M.}\ \bibnamefont {S\"anger}} \emph {et~al.},\ }\bibfield  {title} {\bibinfo {title} {{Tests of General Relativity with GW230529: a neutron star merging with a lower mass-gap compact object}},\ }\Eprint {https://arxiv.org/abs/2406.03568} {arXiv:2406.03568 [gr-qc]}  (\bibinfo {year} {2024})\BibitemShut {NoStop}%
\bibitem [{\citenamefont {Chung}\ and\ \citenamefont {Yunes}(2025)}]{Chung:2025wbg}%
  \BibitemOpen
  \bibfield  {author} {\bibinfo {author} {\bibfnamefont {A.~K.-W.}\ \bibnamefont {Chung}}\ and\ \bibinfo {author} {\bibfnamefont {N.}~\bibnamefont {Yunes}},\ }\bibfield  {title} {\bibinfo {title} {{Probing quadratic gravity with black-hole ringdown gravitational waves measured by LIGO-Virgo-KAGRA detectors}},\ }\href@noop {} {\  (\bibinfo {year} {2025})},\ \Eprint {https://arxiv.org/abs/2506.14695} {arXiv:2506.14695 [gr-qc]} \BibitemShut {NoStop}%
\bibitem [{\citenamefont {Cardoso}\ \emph {et~al.}(2016)\citenamefont {Cardoso}, \citenamefont {Hopper}, \citenamefont {Macedo}, \citenamefont {Palenzuela},\ and\ \citenamefont {Pani}}]{Cardoso:2016oxy}%
  \BibitemOpen
  \bibfield  {author} {\bibinfo {author} {\bibfnamefont {V.}~\bibnamefont {Cardoso}}, \bibinfo {author} {\bibfnamefont {S.}~\bibnamefont {Hopper}}, \bibinfo {author} {\bibfnamefont {C.~F.~B.}\ \bibnamefont {Macedo}}, \bibinfo {author} {\bibfnamefont {C.}~\bibnamefont {Palenzuela}},\ and\ \bibinfo {author} {\bibfnamefont {P.}~\bibnamefont {Pani}},\ }\bibfield  {title} {\bibinfo {title} {{Gravitational-wave signatures of exotic compact objects and of quantum corrections at the horizon scale}},\ }\href {https://doi.org/10.1103/PhysRevD.94.084031} {\bibfield  {journal} {\bibinfo  {journal} {Phys. Rev. D}\ }\textbf {\bibinfo {volume} {94}},\ \bibinfo {pages} {084031} (\bibinfo {year} {2016})},\ \Eprint {https://arxiv.org/abs/1608.08637} {arXiv:1608.08637 [gr-qc]} \BibitemShut {NoStop}%
\bibitem [{\citenamefont {Maggio}\ \emph {et~al.}(2020)\citenamefont {Maggio}, \citenamefont {Buoninfante}, \citenamefont {Mazumdar},\ and\ \citenamefont {Pani}}]{Maggio:2020jml}%
  \BibitemOpen
  \bibfield  {author} {\bibinfo {author} {\bibfnamefont {E.}~\bibnamefont {Maggio}}, \bibinfo {author} {\bibfnamefont {L.}~\bibnamefont {Buoninfante}}, \bibinfo {author} {\bibfnamefont {A.}~\bibnamefont {Mazumdar}},\ and\ \bibinfo {author} {\bibfnamefont {P.}~\bibnamefont {Pani}},\ }\bibfield  {title} {\bibinfo {title} {{How does a dark compact object ringdown?}},\ }\href {https://doi.org/10.1103/PhysRevD.102.064053} {\bibfield  {journal} {\bibinfo  {journal} {Phys. Rev. D}\ }\textbf {\bibinfo {volume} {102}},\ \bibinfo {pages} {064053} (\bibinfo {year} {2020})},\ \Eprint {https://arxiv.org/abs/2006.14628} {arXiv:2006.14628 [gr-qc]} \BibitemShut {NoStop}%
\bibitem [{\citenamefont {Alexander}\ and\ \citenamefont {Yunes}(2009)}]{Alexander:2009tp}%
  \BibitemOpen
  \bibfield  {author} {\bibinfo {author} {\bibfnamefont {S.}~\bibnamefont {Alexander}}\ and\ \bibinfo {author} {\bibfnamefont {N.}~\bibnamefont {Yunes}},\ }\bibfield  {title} {\bibinfo {title} {{Chern-Simons Modified General Relativity}},\ }\href {https://doi.org/10.1016/j.physrep.2009.07.002} {\bibfield  {journal} {\bibinfo  {journal} {Phys. Rept.}\ }\textbf {\bibinfo {volume} {480}},\ \bibinfo {pages} {1} (\bibinfo {year} {2009})},\ \Eprint {https://arxiv.org/abs/0907.2562} {arXiv:0907.2562 [hep-th]} \BibitemShut {NoStop}%
\bibitem [{\citenamefont {Chung}\ \emph {et~al.}(2025)\citenamefont {Chung}, \citenamefont {Lam},\ and\ \citenamefont {Yunes}}]{Chung:2025gyg}%
  \BibitemOpen
  \bibfield  {author} {\bibinfo {author} {\bibfnamefont {A.~K.-W.}\ \bibnamefont {Chung}}, \bibinfo {author} {\bibfnamefont {K.~K.-H.}\ \bibnamefont {Lam}},\ and\ \bibinfo {author} {\bibfnamefont {N.}~\bibnamefont {Yunes}},\ }\bibfield  {title} {\bibinfo {title} {{Quasinormal mode frequencies and gravitational perturbations of spinning black holes in modified gravity through METRICS: The dynamical Chern-Simons gravity case}},\ }\href {https://doi.org/10.1103/g83n-rrlj} {\bibfield  {journal} {\bibinfo  {journal} {Phys. Rev. D}\ }\textbf {\bibinfo {volume} {111}},\ \bibinfo {pages} {124052} (\bibinfo {year} {2025})},\ \Eprint {https://arxiv.org/abs/2503.11759} {arXiv:2503.11759 [gr-qc]} \BibitemShut {NoStop}%
\bibitem [{\citenamefont {Blanchet}(2014)}]{Blanchet:2014}%
  \BibitemOpen
  \bibfield  {author} {\bibinfo {author} {\bibfnamefont {L.}~\bibnamefont {Blanchet}},\ }\bibfield  {title} {\bibinfo {title} {Gravitational radiation from post-{N}ewtonian sources and inspiralling compact binaries},\ }\href@noop {} {\bibfield  {journal} {\bibinfo  {journal} {Living Rev. Rel.}\ }\textbf {\bibinfo {volume} {17}},\ \bibinfo {pages} {2} (\bibinfo {year} {2014})},\ \Eprint {https://arxiv.org/abs/gr-qc/0202016} {gr-qc/0202016} \BibitemShut {NoStop}%
\bibitem [{\citenamefont {Blanchet}\ and\ \citenamefont {Sathyaprakash}(1995)}]{Blanchet:1994ez}%
  \BibitemOpen
  \bibfield  {author} {\bibinfo {author} {\bibfnamefont {L.}~\bibnamefont {Blanchet}}\ and\ \bibinfo {author} {\bibfnamefont {B.~S.}\ \bibnamefont {Sathyaprakash}},\ }\bibfield  {title} {\bibinfo {title} {{Detecting the tail effect in gravitational wave experiments}},\ }\href {https://doi.org/10.1103/PhysRevLett.74.1067} {\bibfield  {journal} {\bibinfo  {journal} {Phys.\ Rev.\ Lett.}\ }\textbf {\bibinfo {volume} {74}},\ \bibinfo {pages} {1067} (\bibinfo {year} {1995})}\BibitemShut {NoStop}%
\bibitem [{\citenamefont {Blanchet}\ and\ \citenamefont {Sathyaprakash}(1994)}]{Blanchet:1994ex}%
  \BibitemOpen
  \bibfield  {author} {\bibinfo {author} {\bibfnamefont {L.}~\bibnamefont {Blanchet}}\ and\ \bibinfo {author} {\bibfnamefont {B.~S.}\ \bibnamefont {Sathyaprakash}},\ }\bibfield  {title} {\bibinfo {title} {{Signal analysis of gravitational wave tails}},\ }\href {https://doi.org/10.1088/0264-9381/11/11/020} {\bibfield  {journal} {\bibinfo  {journal} {Class. Quantum Grav.}\ }\textbf {\bibinfo {volume} {11}},\ \bibinfo {pages} {2807} (\bibinfo {year} {1994})}\BibitemShut {NoStop}%
\bibitem [{\citenamefont {Arun}\ \emph {et~al.}(2006)\citenamefont {Arun}, \citenamefont {Iyer}, \citenamefont {Qusailah},\ and\ \citenamefont {Sathyaprakash}}]{Arun:2006hn}%
  \BibitemOpen
  \bibfield  {author} {\bibinfo {author} {\bibfnamefont {K.~G.}\ \bibnamefont {Arun}}, \bibinfo {author} {\bibfnamefont {B.~R.}\ \bibnamefont {Iyer}}, \bibinfo {author} {\bibfnamefont {M.~S.~S.}\ \bibnamefont {Qusailah}},\ and\ \bibinfo {author} {\bibfnamefont {B.~S.}\ \bibnamefont {Sathyaprakash}},\ }\bibfield  {title} {\bibinfo {title} {{Probing the non-linear structure of general relativity with black hole binaries}},\ }\href {https://doi.org/10.1103/PhysRevD.74.024006} {\bibfield  {journal} {\bibinfo  {journal} {Phys. Rev. D}\ }\textbf {\bibinfo {volume} {74}},\ \bibinfo {pages} {024006} (\bibinfo {year} {2006})},\ \Eprint {https://arxiv.org/abs/gr-qc/0604067} {arXiv:gr-qc/0604067} \BibitemShut {NoStop}%
\bibitem [{\citenamefont {Yunes}\ and\ \citenamefont {Pretorius}(2009)}]{Yunes:2009ke}%
  \BibitemOpen
  \bibfield  {author} {\bibinfo {author} {\bibfnamefont {N.}~\bibnamefont {Yunes}}\ and\ \bibinfo {author} {\bibfnamefont {F.}~\bibnamefont {Pretorius}},\ }\bibfield  {title} {\bibinfo {title} {{Fundamental Theoretical Bias in Gravitational Wave Astrophysics and the Parameterized Post-Einsteinian Framework}},\ }\href {https://doi.org/10.1103/PhysRevD.80.122003} {\bibfield  {journal} {\bibinfo  {journal} {Phys. Rev. D}\ }\textbf {\bibinfo {volume} {80}},\ \bibinfo {pages} {122003} (\bibinfo {year} {2009})},\ \Eprint {https://arxiv.org/abs/0909.3328} {arXiv:0909.3328 [gr-qc]} \BibitemShut {NoStop}%
\bibitem [{\citenamefont {Mishra}\ \emph {et~al.}(2010)\citenamefont {Mishra}, \citenamefont {Arun}, \citenamefont {Iyer},\ and\ \citenamefont {Sathyaprakash}}]{Mishra:2010tp}%
  \BibitemOpen
  \bibfield  {author} {\bibinfo {author} {\bibfnamefont {C.~K.}\ \bibnamefont {Mishra}}, \bibinfo {author} {\bibfnamefont {K.~G.}\ \bibnamefont {Arun}}, \bibinfo {author} {\bibfnamefont {B.~R.}\ \bibnamefont {Iyer}},\ and\ \bibinfo {author} {\bibfnamefont {B.~S.}\ \bibnamefont {Sathyaprakash}},\ }\bibfield  {title} {\bibinfo {title} {{Parametrized tests of post-Newtonian theory using Advanced LIGO and Einstein Telescope}},\ }\href {https://doi.org/10.1103/PhysRevD.82.064010} {\bibfield  {journal} {\bibinfo  {journal} {Phys. Rev. D}\ }\textbf {\bibinfo {volume} {82}},\ \bibinfo {pages} {064010} (\bibinfo {year} {2010})},\ \Eprint {https://arxiv.org/abs/1005.0304} {arXiv:1005.0304 [gr-qc]} \BibitemShut {NoStop}%
\bibitem [{\citenamefont {Li}\ \emph {et~al.}(2012)\citenamefont {Li}, \citenamefont {Del~Pozzo}, \citenamefont {Vitale}, \citenamefont {Van Den~Broeck}, \citenamefont {Agathos} \emph {et~al.}}]{Li:2011cg}%
  \BibitemOpen
  \bibfield  {author} {\bibinfo {author} {\bibfnamefont {T.~G.~F.}\ \bibnamefont {Li}}, \bibinfo {author} {\bibfnamefont {W.}~\bibnamefont {Del~Pozzo}}, \bibinfo {author} {\bibfnamefont {S.}~\bibnamefont {Vitale}}, \bibinfo {author} {\bibfnamefont {C.}~\bibnamefont {Van Den~Broeck}}, \bibinfo {author} {\bibfnamefont {M.}~\bibnamefont {Agathos}}, \emph {et~al.},\ }\bibfield  {title} {\bibinfo {title} {{Towards a generic test of the strong field dynamics of general relativity using compact binary coalescence}},\ }\href {https://doi.org/10.1103/PhysRevD.85.082003} {\bibfield  {journal} {\bibinfo  {journal} {Phys. Rev. D}\ }\textbf {\bibinfo {volume} {85}},\ \bibinfo {pages} {082003} (\bibinfo {year} {2012})},\ \Eprint {https://arxiv.org/abs/1110.0530} {arXiv:1110.0530 [gr-qc]} \BibitemShut {NoStop}%
\bibitem [{\citenamefont {Agathos}\ \emph {et~al.}(2014)\citenamefont {Agathos}, \citenamefont {Del~Pozzo}, \citenamefont {Li}, \citenamefont {Van Den~Broeck}, \citenamefont {Veitch},\ and\ \citenamefont {Vitale}}]{Agathos:2013upa}%
  \BibitemOpen
  \bibfield  {author} {\bibinfo {author} {\bibfnamefont {M.}~\bibnamefont {Agathos}}, \bibinfo {author} {\bibfnamefont {W.}~\bibnamefont {Del~Pozzo}}, \bibinfo {author} {\bibfnamefont {T.~G.~F.}\ \bibnamefont {Li}}, \bibinfo {author} {\bibfnamefont {C.}~\bibnamefont {Van Den~Broeck}}, \bibinfo {author} {\bibfnamefont {J.}~\bibnamefont {Veitch}},\ and\ \bibinfo {author} {\bibfnamefont {S.}~\bibnamefont {Vitale}},\ }\bibfield  {title} {\bibinfo {title} {{TIGER: A data analysis pipeline for testing the strong-field dynamics of general relativity with gravitational wave signals from coalescing compact binaries}},\ }\href {https://doi.org/10.1103/PhysRevD.89.082001} {\bibfield  {journal} {\bibinfo  {journal} {Phys. Rev. D}\ }\textbf {\bibinfo {volume} {89}},\ \bibinfo {pages} {082001} (\bibinfo {year} {2014})},\ \Eprint {https://arxiv.org/abs/1311.0420} {arXiv:1311.0420 [gr-qc]} \BibitemShut {NoStop}%
\bibitem [{\citenamefont {Mehta}\ \emph {et~al.}(2023)\citenamefont {Mehta}, \citenamefont {Buonanno}, \citenamefont {Cotesta}, \citenamefont {Ghosh}, \citenamefont {Sennett},\ and\ \citenamefont {Steinhoff}}]{Mehta:2022pcn}%
  \BibitemOpen
  \bibfield  {author} {\bibinfo {author} {\bibfnamefont {A.~K.}\ \bibnamefont {Mehta}}, \bibinfo {author} {\bibfnamefont {A.}~\bibnamefont {Buonanno}}, \bibinfo {author} {\bibfnamefont {R.}~\bibnamefont {Cotesta}}, \bibinfo {author} {\bibfnamefont {A.}~\bibnamefont {Ghosh}}, \bibinfo {author} {\bibfnamefont {N.}~\bibnamefont {Sennett}},\ and\ \bibinfo {author} {\bibfnamefont {J.}~\bibnamefont {Steinhoff}},\ }\bibfield  {title} {\bibinfo {title} {{Tests of general relativity with gravitational-wave observations using a flexible theory-independent method}},\ }\href {https://doi.org/10.1103/PhysRevD.107.044020} {\bibfield  {journal} {\bibinfo  {journal} {Phys. Rev. D}\ }\textbf {\bibinfo {volume} {107}},\ \bibinfo {pages} {044020} (\bibinfo {year} {2023})},\ \Eprint {https://arxiv.org/abs/2203.13937} {arXiv:2203.13937 [gr-qc]} \BibitemShut {NoStop}%
\bibitem [{\citenamefont {{Agathos}}\ \emph {et~al.}(2014)\citenamefont {{Agathos}}, \citenamefont {{Del Pozzo}}, \citenamefont {{Li}}, \citenamefont {{Van Den Broeck}}, \citenamefont {{Veitch}},\ and\ \citenamefont {{Vitale}}}]{AgathosEtAl:2014}%
  \BibitemOpen
  \bibfield  {author} {\bibinfo {author} {\bibfnamefont {M.}~\bibnamefont {{Agathos}}}, \bibinfo {author} {\bibfnamefont {W.}~\bibnamefont {{Del Pozzo}}}, \bibinfo {author} {\bibfnamefont {T.~G.~F.}\ \bibnamefont {{Li}}}, \bibinfo {author} {\bibfnamefont {C.}~\bibnamefont {{Van Den Broeck}}}, \bibinfo {author} {\bibfnamefont {J.}~\bibnamefont {{Veitch}}},\ and\ \bibinfo {author} {\bibfnamefont {S.}~\bibnamefont {{Vitale}}},\ }\bibfield  {title} {\bibinfo {title} {{TIGER: A data analysis pipeline for testing the strong-field dynamics of general relativity with gravitational wave signals from coalescing compact binaries}},\ }\href {https://doi.org/10.1103/PhysRevD.89.082001} {\bibfield  {journal} {\bibinfo  {journal} {\prd}\ }\textbf {\bibinfo {volume} {89}},\ \bibinfo {eid} {082001} (\bibinfo {year} {2014})},\ \Eprint {https://arxiv.org/abs/1311.0420} {arXiv:1311.0420 [gr-qc]} \BibitemShut {NoStop}%
\bibitem [{\citenamefont {Meidam}\ \emph {et~al.}(2018)\citenamefont {Meidam} \emph {et~al.}}]{Meidam:2017dgf}%
  \BibitemOpen
  \bibfield  {author} {\bibinfo {author} {\bibfnamefont {J.}~\bibnamefont {Meidam}} \emph {et~al.},\ }\bibfield  {title} {\bibinfo {title} {{Parametrized tests of the strong-field dynamics of general relativity using gravitational wave signals from coalescing binary black holes: Fast likelihood calculations and sensitivity of the method}},\ }\href {https://doi.org/10.1103/PhysRevD.97.044033} {\bibfield  {journal} {\bibinfo  {journal} {Phys. Rev. D}\ }\textbf {\bibinfo {volume} {97}},\ \bibinfo {pages} {044033} (\bibinfo {year} {2018})},\ \Eprint {https://arxiv.org/abs/1712.08772} {arXiv:1712.08772 [gr-qc]} \BibitemShut {NoStop}%
\bibitem [{\citenamefont {Roy}\ \emph {et~al.}(2025)\citenamefont {Roy}, \citenamefont {Haney}, \citenamefont {Pratten}, \citenamefont {Pang},\ and\ \citenamefont {Van Den~Broeck}}]{Roy:2025gzv}%
  \BibitemOpen
  \bibfield  {author} {\bibinfo {author} {\bibfnamefont {S.}~\bibnamefont {Roy}}, \bibinfo {author} {\bibfnamefont {M.}~\bibnamefont {Haney}}, \bibinfo {author} {\bibfnamefont {G.}~\bibnamefont {Pratten}}, \bibinfo {author} {\bibfnamefont {P.~T.~H.}\ \bibnamefont {Pang}},\ and\ \bibinfo {author} {\bibfnamefont {C.}~\bibnamefont {Van Den~Broeck}},\ }\bibfield  {title} {\bibinfo {title} {{An improved parametrized test of general relativity using the IMRPhenomX waveform family: Including higher harmonics and precession}},\ }\href@noop {} {\  (\bibinfo {year} {2025})},\ \Eprint {https://arxiv.org/abs/2504.21147} {arXiv:2504.21147 [gr-qc]} \BibitemShut {NoStop}%
\bibitem [{\citenamefont {Pratten}\ \emph {et~al.}(2021)\citenamefont {Pratten} \emph {et~al.}}]{Pratten:2020ceb}%
  \BibitemOpen
  \bibfield  {author} {\bibinfo {author} {\bibfnamefont {G.}~\bibnamefont {Pratten}} \emph {et~al.},\ }\bibfield  {title} {\bibinfo {title} {{Computationally efficient models for the dominant and subdominant harmonic modes of precessing binary black holes}},\ }\href {https://doi.org/10.1103/PhysRevD.103.104056} {\bibfield  {journal} {\bibinfo  {journal} {Phys. Rev. D}\ }\textbf {\bibinfo {volume} {103}},\ \bibinfo {pages} {104056} (\bibinfo {year} {2021})},\ \Eprint {https://arxiv.org/abs/2004.06503} {arXiv:2004.06503 [gr-qc]} \BibitemShut {NoStop}%
\bibitem [{\citenamefont {Colleoni}\ \emph {et~al.}(2025)\citenamefont {Colleoni}, \citenamefont {Vidal}, \citenamefont {Garc\'\i{}a-Quir\'os}, \citenamefont {Ak\c{c}ay},\ and\ \citenamefont {Bera}}]{Colleoni:2024knd}%
  \BibitemOpen
  \bibfield  {author} {\bibinfo {author} {\bibfnamefont {M.}~\bibnamefont {Colleoni}}, \bibinfo {author} {\bibfnamefont {F.~A.~R.}\ \bibnamefont {Vidal}}, \bibinfo {author} {\bibfnamefont {C.}~\bibnamefont {Garc\'\i{}a-Quir\'os}}, \bibinfo {author} {\bibfnamefont {S.}~\bibnamefont {Ak\c{c}ay}},\ and\ \bibinfo {author} {\bibfnamefont {S.}~\bibnamefont {Bera}},\ }\bibfield  {title} {\bibinfo {title} {{Fast frequency-domain gravitational waveforms for precessing binaries with a new twist}},\ }\href {https://doi.org/10.1103/PhysRevD.111.104019} {\bibfield  {journal} {\bibinfo  {journal} {Phys. Rev. D}\ }\textbf {\bibinfo {volume} {111}},\ \bibinfo {pages} {104019} (\bibinfo {year} {2025})},\ \Eprint {https://arxiv.org/abs/2412.16721} {arXiv:2412.16721 [gr-qc]} \BibitemShut {NoStop}%
\bibitem [{\citenamefont {Pai}\ and\ \citenamefont {Arun}(2013)}]{Pai:2012mv}%
  \BibitemOpen
  \bibfield  {author} {\bibinfo {author} {\bibfnamefont {A.}~\bibnamefont {Pai}}\ and\ \bibinfo {author} {\bibfnamefont {K.~G.}\ \bibnamefont {Arun}},\ }\bibfield  {title} {\bibinfo {title} {{Singular value decomposition in parametrised tests of post-Newtonian theory}},\ }\href {https://doi.org/10.1088/0264-9381/30/2/025011} {\bibfield  {journal} {\bibinfo  {journal} {Class. Quantum Grav.}\ }\textbf {\bibinfo {volume} {30}},\ \bibinfo {pages} {025011} (\bibinfo {year} {2013})},\ \Eprint {https://arxiv.org/abs/1207.1943} {arXiv:1207.1943 [gr-qc]} \BibitemShut {NoStop}%
\bibitem [{\citenamefont {Shoom}\ \emph {et~al.}(2023)\citenamefont {Shoom}, \citenamefont {Gupta}, \citenamefont {Krishnan}, \citenamefont {Nielsen},\ and\ \citenamefont {Capano}}]{Shoom:2021mdj}%
  \BibitemOpen
  \bibfield  {author} {\bibinfo {author} {\bibfnamefont {A.~A.}\ \bibnamefont {Shoom}}, \bibinfo {author} {\bibfnamefont {P.~K.}\ \bibnamefont {Gupta}}, \bibinfo {author} {\bibfnamefont {B.}~\bibnamefont {Krishnan}}, \bibinfo {author} {\bibfnamefont {A.~B.}\ \bibnamefont {Nielsen}},\ and\ \bibinfo {author} {\bibfnamefont {C.~D.}\ \bibnamefont {Capano}},\ }\bibfield  {title} {\bibinfo {title} {{Testing the post-Newtonian expansion with GW170817}},\ }\href {https://doi.org/10.1007/s10714-023-03100-z} {\bibfield  {journal} {\bibinfo  {journal} {Gen. Rel. Grav.}\ }\textbf {\bibinfo {volume} {55}},\ \bibinfo {pages} {55} (\bibinfo {year} {2023})},\ \Eprint {https://arxiv.org/abs/2105.02191} {arXiv:2105.02191 [gr-qc]} \BibitemShut {NoStop}%
\bibitem [{\citenamefont {Saleem}\ \emph {et~al.}(2022)\citenamefont {Saleem}, \citenamefont {Datta}, \citenamefont {Arun},\ and\ \citenamefont {Sathyaprakash}}]{Saleem:2021nsb}%
  \BibitemOpen
  \bibfield  {author} {\bibinfo {author} {\bibfnamefont {M.}~\bibnamefont {Saleem}}, \bibinfo {author} {\bibfnamefont {S.}~\bibnamefont {Datta}}, \bibinfo {author} {\bibfnamefont {K.~G.}\ \bibnamefont {Arun}},\ and\ \bibinfo {author} {\bibfnamefont {B.~S.}\ \bibnamefont {Sathyaprakash}},\ }\bibfield  {title} {\bibinfo {title} {{Parametrized tests of post-Newtonian theory using principal component analysis}},\ }\href {https://doi.org/10.1103/PhysRevD.105.084062} {\bibfield  {journal} {\bibinfo  {journal} {Phys. Rev. D}\ }\textbf {\bibinfo {volume} {105}},\ \bibinfo {pages} {084062} (\bibinfo {year} {2022})},\ \Eprint {https://arxiv.org/abs/2110.10147} {arXiv:2110.10147 [gr-qc]} \BibitemShut {NoStop}%
\bibitem [{\citenamefont {Mahapatra}\ \emph {et~al.}(2025)\citenamefont {Mahapatra} \emph {et~al.}}]{Mahapatra:2025cwk}%
  \BibitemOpen
  \bibfield  {author} {\bibinfo {author} {\bibfnamefont {P.}~\bibnamefont {Mahapatra}} \emph {et~al.},\ }\bibfield  {title} {\bibinfo {title} {{Confronting General Relativity with Principal Component Analysis: Simulations and Results from GWTC-3 Events}},\ }\href@noop {} {\  (\bibinfo {year} {2025})},\ \Eprint {https://arxiv.org/abs/2508.06862} {arXiv:2508.06862 [gr-qc]} \BibitemShut {NoStop}%
\bibitem [{\citenamefont {Hughes}\ and\ \citenamefont {Menou}(2005)}]{Hughes:2004vw}%
  \BibitemOpen
  \bibfield  {author} {\bibinfo {author} {\bibfnamefont {S.~A.}\ \bibnamefont {Hughes}}\ and\ \bibinfo {author} {\bibfnamefont {K.}~\bibnamefont {Menou}},\ }\bibfield  {title} {\bibinfo {title} {{Golden binaries for LISA: Robust probes of strong-field gravity}},\ }\href {https://doi.org/10.1086/428826} {\bibfield  {journal} {\bibinfo  {journal} {Astrophys. J.}\ }\textbf {\bibinfo {volume} {623}},\ \bibinfo {pages} {689} (\bibinfo {year} {2005})},\ \Eprint {https://arxiv.org/abs/astro-ph/0410148} {arXiv:astro-ph/0410148 [astro-ph]} \BibitemShut {NoStop}%
\bibitem [{\citenamefont {Ghosh}\ \emph {et~al.}(2016)\citenamefont {Ghosh} \emph {et~al.}}]{Ghosh:2016qgn}%
  \BibitemOpen
  \bibfield  {author} {\bibinfo {author} {\bibfnamefont {A.}~\bibnamefont {Ghosh}} \emph {et~al.},\ }\bibfield  {title} {\bibinfo {title} {{Testing general relativity using golden black-hole binaries}},\ }\href {https://doi.org/10.1103/PhysRevD.94.021101} {\bibfield  {journal} {\bibinfo  {journal} {Phys. Rev. D}\ }\textbf {\bibinfo {volume} {94}},\ \bibinfo {pages} {021101(R)} (\bibinfo {year} {2016})},\ \Eprint {https://arxiv.org/abs/1602.02453} {arXiv:1602.02453 [gr-qc]} \BibitemShut {NoStop}%
\bibitem [{\citenamefont {Ghosh}\ \emph {et~al.}(2018)\citenamefont {Ghosh}, \citenamefont {Johnson-McDaniel}, \citenamefont {Ghosh}, \citenamefont {Mishra}, \citenamefont {Ajith}, \citenamefont {Del~Pozzo}, \citenamefont {Berry}, \citenamefont {Nielsen},\ and\ \citenamefont {London}}]{Ghosh:2017gfp}%
  \BibitemOpen
  \bibfield  {author} {\bibinfo {author} {\bibfnamefont {A.}~\bibnamefont {Ghosh}}, \bibinfo {author} {\bibfnamefont {N.~K.}\ \bibnamefont {Johnson-McDaniel}}, \bibinfo {author} {\bibfnamefont {A.}~\bibnamefont {Ghosh}}, \bibinfo {author} {\bibfnamefont {C.~K.}\ \bibnamefont {Mishra}}, \bibinfo {author} {\bibfnamefont {P.}~\bibnamefont {Ajith}}, \bibinfo {author} {\bibfnamefont {W.}~\bibnamefont {Del~Pozzo}}, \bibinfo {author} {\bibfnamefont {C.~P.~L.}\ \bibnamefont {Berry}}, \bibinfo {author} {\bibfnamefont {A.~B.}\ \bibnamefont {Nielsen}},\ and\ \bibinfo {author} {\bibfnamefont {L.}~\bibnamefont {London}},\ }\bibfield  {title} {\bibinfo {title} {{Testing general relativity using gravitational wave signals from the inspiral, merger and ringdown of binary black holes}},\ }\href {https://doi.org/10.1088/1361-6382/aa972e} {\bibfield  {journal} {\bibinfo  {journal} {Class. Quantum Grav.}\ }\textbf {\bibinfo {volume} {35}},\ \bibinfo {pages} {014002} (\bibinfo {year} {2018})},\ \Eprint {https://arxiv.org/abs/1704.06784} {arXiv:1704.06784 [gr-qc]} \BibitemShut {NoStop}%
\bibitem [{\citenamefont {Bardeen}\ \emph {et~al.}(1972)\citenamefont {Bardeen}, \citenamefont {Press},\ and\ \citenamefont {Teukolsky}}]{Bardeen:1972fi}%
  \BibitemOpen
  \bibfield  {author} {\bibinfo {author} {\bibfnamefont {J.~M.}\ \bibnamefont {Bardeen}}, \bibinfo {author} {\bibfnamefont {W.~H.}\ \bibnamefont {Press}},\ and\ \bibinfo {author} {\bibfnamefont {S.~A.}\ \bibnamefont {Teukolsky}},\ }\bibfield  {title} {\bibinfo {title} {{Rotating black holes: Locally nonrotating frames, energy extraction, and scalar synchrotron radiation}},\ }\href {https://doi.org/10.1086/151796} {\bibfield  {journal} {\bibinfo  {journal} {Astrophys. J.}\ }\textbf {\bibinfo {volume} {178}},\ \bibinfo {pages} {347} (\bibinfo {year} {1972})}\BibitemShut {NoStop}%
\bibitem [{\citenamefont {Healy}\ and\ \citenamefont {Lousto}(2017)}]{Healy:2016lce}%
  \BibitemOpen
  \bibfield  {author} {\bibinfo {author} {\bibfnamefont {J.}~\bibnamefont {Healy}}\ and\ \bibinfo {author} {\bibfnamefont {C.~O.}\ \bibnamefont {Lousto}},\ }\bibfield  {title} {\bibinfo {title} {{Remnant of binary black-hole mergers: New simulations and peak luminosity studies}},\ }\href {https://doi.org/10.1103/PhysRevD.95.024037} {\bibfield  {journal} {\bibinfo  {journal} {Phys. Rev. D}\ }\textbf {\bibinfo {volume} {95}},\ \bibinfo {pages} {024037} (\bibinfo {year} {2017})},\ \Eprint {https://arxiv.org/abs/1610.09713} {arXiv:1610.09713 [gr-qc]} \BibitemShut {NoStop}%
\bibitem [{\citenamefont {Jim\'enez-Forteza}\ \emph {et~al.}(2017)\citenamefont {Jim\'enez-Forteza}, \citenamefont {Keitel}, \citenamefont {Husa}, \citenamefont {Hannam}, \citenamefont {Khan},\ and\ \citenamefont {P\"urrer}}]{PhysRevD.95.064024}%
  \BibitemOpen
  \bibfield  {author} {\bibinfo {author} {\bibfnamefont {X.}~\bibnamefont {Jim\'enez-Forteza}}, \bibinfo {author} {\bibfnamefont {D.}~\bibnamefont {Keitel}}, \bibinfo {author} {\bibfnamefont {S.}~\bibnamefont {Husa}}, \bibinfo {author} {\bibfnamefont {M.}~\bibnamefont {Hannam}}, \bibinfo {author} {\bibfnamefont {S.}~\bibnamefont {Khan}},\ and\ \bibinfo {author} {\bibfnamefont {M.}~\bibnamefont {P\"urrer}},\ }\bibfield  {title} {\bibinfo {title} {Hierarchical data-driven approach to fitting numerical relativity data for nonprecessing binary black holes with an application to final spin and radiated energy},\ }\href {https://doi.org/10.1103/PhysRevD.95.064024} {\bibfield  {journal} {\bibinfo  {journal} {Phys. Rev. D}\ }\textbf {\bibinfo {volume} {95}},\ \bibinfo {pages} {064024} (\bibinfo {year} {2017})}\BibitemShut {NoStop}%
\bibitem [{\citenamefont {Hawking}(1971)}]{Hawking:1971tu}%
  \BibitemOpen
  \bibfield  {author} {\bibinfo {author} {\bibfnamefont {S.~W.}\ \bibnamefont {Hawking}},\ }\bibfield  {title} {\bibinfo {title} {{Gravitational radiation from colliding black holes}},\ }\href {https://doi.org/10.1103/PhysRevLett.26.1344} {\bibfield  {journal} {\bibinfo  {journal} {Phys. Rev. Lett.}\ }\textbf {\bibinfo {volume} {26}},\ \bibinfo {pages} {1344} (\bibinfo {year} {1971})}\BibitemShut {NoStop}%
\bibitem [{\citenamefont {Cornish}\ \emph {et~al.}(2011)\citenamefont {Cornish}, \citenamefont {Sampson}, \citenamefont {Yunes},\ and\ \citenamefont {Pretorius}}]{Cornish:2011ys}%
  \BibitemOpen
  \bibfield  {author} {\bibinfo {author} {\bibfnamefont {N.}~\bibnamefont {Cornish}}, \bibinfo {author} {\bibfnamefont {L.}~\bibnamefont {Sampson}}, \bibinfo {author} {\bibfnamefont {N.}~\bibnamefont {Yunes}},\ and\ \bibinfo {author} {\bibfnamefont {F.}~\bibnamefont {Pretorius}},\ }\bibfield  {title} {\bibinfo {title} {{Gravitational Wave Tests of General Relativity with the Parameterized Post-Einsteinian Framework}},\ }\href {https://doi.org/10.1103/PhysRevD.84.062003} {\bibfield  {journal} {\bibinfo  {journal} {\prd}\ }\textbf {\bibinfo {volume} {84}},\ \bibinfo {pages} {062003} (\bibinfo {year} {2011})},\ \Eprint {https://arxiv.org/abs/1105.2088} {arXiv:1105.2088 [gr-qc]} \BibitemShut {NoStop}%
\bibitem [{\citenamefont {Vallisneri}(2012)}]{Vallisneri:2012qq}%
  \BibitemOpen
  \bibfield  {author} {\bibinfo {author} {\bibfnamefont {M.}~\bibnamefont {Vallisneri}},\ }\bibfield  {title} {\bibinfo {title} {{Testing general relativity with gravitational waves: a reality check}},\ }\href {https://doi.org/10.1103/PhysRevD.86.082001} {\bibfield  {journal} {\bibinfo  {journal} {Phys. Rev. D}\ }\textbf {\bibinfo {volume} {86}},\ \bibinfo {pages} {082001} (\bibinfo {year} {2012})},\ \Eprint {https://arxiv.org/abs/1207.4759} {arXiv:1207.4759 [gr-qc]} \BibitemShut {NoStop}%
\bibitem [{\citenamefont {Jia}\ \emph {et~al.}(2024)\citenamefont {Jia} \emph {et~al.}}]{membersoftheLIGOScientific:2024elc}%
  \BibitemOpen
  \bibfield  {author} {\bibinfo {author} {\bibfnamefont {W.}~\bibnamefont {Jia}} \emph {et~al.} (\bibinfo {collaboration} {members of the LIGO Scientific Collaboration}),\ }\bibfield  {title} {\bibinfo {title} {{Squeezing the quantum noise of a gravitational-wave detector below the standard quantum limit}},\ }\href {https://doi.org/10.1126/science.ado8069} {\bibfield  {journal} {\bibinfo  {journal} {Science}\ }\textbf {\bibinfo {volume} {385}},\ \bibinfo {pages} {1318} (\bibinfo {year} {2024})},\ \Eprint {https://arxiv.org/abs/2404.14569} {arXiv:2404.14569 [gr-qc]} \BibitemShut {NoStop}%
\bibitem [{\citenamefont {Ganapathy}\ \emph {et~al.}(2022)\citenamefont {Ganapathy}, \citenamefont {Xu}, \citenamefont {Jia}, \citenamefont {Whittle}, \citenamefont {Tse}, \citenamefont {Barsotti}, \citenamefont {Evans},\ and\ \citenamefont {McCuller}}]{Ganapathy:2022hgu}%
  \BibitemOpen
  \bibfield  {author} {\bibinfo {author} {\bibfnamefont {D.}~\bibnamefont {Ganapathy}}, \bibinfo {author} {\bibfnamefont {V.}~\bibnamefont {Xu}}, \bibinfo {author} {\bibfnamefont {W.}~\bibnamefont {Jia}}, \bibinfo {author} {\bibfnamefont {C.}~\bibnamefont {Whittle}}, \bibinfo {author} {\bibfnamefont {M.}~\bibnamefont {Tse}}, \bibinfo {author} {\bibfnamefont {L.}~\bibnamefont {Barsotti}}, \bibinfo {author} {\bibfnamefont {M.}~\bibnamefont {Evans}},\ and\ \bibinfo {author} {\bibfnamefont {L.}~\bibnamefont {McCuller}},\ }\bibfield  {title} {\bibinfo {title} {{Probing squeezing for gravitational-wave detectors with an audio-band field}},\ }\href {https://doi.org/10.1103/PhysRevD.105.122005} {\bibfield  {journal} {\bibinfo  {journal} {Phys. Rev. D}\ }\textbf {\bibinfo {volume} {105}},\ \bibinfo {pages} {122005} (\bibinfo {year} {2022})},\ \Eprint {https://arxiv.org/abs/2203.03849} {arXiv:2203.03849 [astro-ph.IM]} \BibitemShut {NoStop}%
\bibitem [{\citenamefont {Capote}\ \emph {et~al.}(2024)\citenamefont {Capote} \emph {et~al.}}]{Capote:2024rmo}%
  \BibitemOpen
  \bibfield  {author} {\bibinfo {author} {\bibfnamefont {E.}~\bibnamefont {Capote}} \emph {et~al.},\ }\bibfield  {title} {\bibinfo {title} {{Advanced LIGO detector performance in the fourth observing run}},\ }\href@noop {} {\  (\bibinfo {year} {2024})},\ \Eprint {https://arxiv.org/abs/2411.14607} {arXiv:2411.14607 [gr-qc]} \BibitemShut {NoStop}%
\bibitem [{LIG(2025{\natexlab{b}})}]{LIGOScientific:2025pvj}%
  \BibitemOpen
  \bibfield  {title} {\bibinfo {title} {{GWTC-4.0: Population Properties of Merging Compact Binaries}},\ }\href@noop {} {\  (\bibinfo {year} {2025}{\natexlab{b}})},\ \Eprint {https://arxiv.org/abs/2508.18083} {arXiv:2508.18083 [astro-ph.HE]} \BibitemShut {NoStop}%
\bibitem [{\citenamefont {Abbott}\ \emph {et~al.}(2016{\natexlab{c}})\citenamefont {Abbott} \emph {et~al.}}]{LIGOScientific:2016aoc}%
  \BibitemOpen
  \bibfield  {author} {\bibinfo {author} {\bibfnamefont {B.~P.}\ \bibnamefont {Abbott}} \emph {et~al.} (\bibinfo {collaboration} {LIGO Scientific, Virgo}),\ }\bibfield  {title} {\bibinfo {title} {{Observation of Gravitational Waves from a Binary Black Hole Merger}},\ }\href {https://doi.org/10.1103/PhysRevLett.116.061102} {\bibfield  {journal} {\bibinfo  {journal} {Phys. Rev. Lett.}\ }\textbf {\bibinfo {volume} {116}},\ \bibinfo {pages} {061102} (\bibinfo {year} {2016}{\natexlab{c}})},\ \Eprint {https://arxiv.org/abs/1602.03837} {arXiv:1602.03837 [gr-qc]} \BibitemShut {NoStop}%
\bibitem [{\citenamefont {Abbott}\ \emph {et~al.}(2020)\citenamefont {Abbott} \emph {et~al.}}]{Aasi:2013wya}%
  \BibitemOpen
  \bibfield  {author} {\bibinfo {author} {\bibfnamefont {B.~P.}\ \bibnamefont {Abbott}} \emph {et~al.} (\bibinfo {collaboration} {KAGRA Collaboration, LIGO Scientific Collaboration, and Virgo Collaboration}),\ }\bibfield  {title} {\bibinfo {title} {{Prospects for Observing and Localizing Gravitational-Wave Transients with Advanced LIGO, Advanced Virgo and KAGRA}},\ }\href {https://doi.org/10.1007/s41114-020-00026-9} {\bibfield  {journal} {\bibinfo  {journal} {Living Rev. Relativity}\ }\textbf {\bibinfo {volume} {23}},\ \bibinfo {pages} {3} (\bibinfo {year} {2020})},\ \Eprint {https://arxiv.org/abs/1304.0670} {arXiv:1304.0670 [gr-qc]} \BibitemShut {NoStop}%
\bibitem [{\citenamefont {{LIGO Scientific Collaboration}}\ \emph {et~al.}(2025{\natexlab{a}})\citenamefont {{LIGO Scientific Collaboration}}, \citenamefont {{Virgo Collaboration}},\ and\ \citenamefont {{KAGRA Collaboration}}}]{GWOSC_page}%
  \BibitemOpen
  \bibfield  {author} {\bibinfo {author} {\bibnamefont {{LIGO Scientific Collaboration}}}, \bibinfo {author} {\bibnamefont {{Virgo Collaboration}}},\ and\ \bibinfo {author} {\bibnamefont {{KAGRA Collaboration}}},\ }\href {https://doi.org/10.7935/1g4j-2028} {\bibinfo {title} {Gw250114 gwosc page}} (\bibinfo {year} {2025}{\natexlab{a}})\BibitemShut {NoStop}%
\bibitem [{\citenamefont {{LIGO Scientific Collaboration}}\ \emph {et~al.}(2025{\natexlab{b}})\citenamefont {{LIGO Scientific Collaboration}}, \citenamefont {{Virgo Collaboration}},\ and\ \citenamefont {{KAGRA Collaboration}}}]{zenodo_data_release}%
  \BibitemOpen
  \bibfield  {author} {\bibinfo {author} {\bibnamefont {{LIGO Scientific Collaboration}}}, \bibinfo {author} {\bibnamefont {{Virgo Collaboration}}},\ and\ \bibinfo {author} {\bibnamefont {{KAGRA Collaboration}}},\ }\href {https://doi.org/10.5281/zenodo.17018009} {\bibinfo {title} {Black hole spectroscopy and tests of general relativity with gw250114: Data release}} (\bibinfo {year} {2025}{\natexlab{b}})\BibitemShut {NoStop}%
\bibitem [{\citenamefont {Kumar}\ \emph {et~al.}(2019)\citenamefont {Kumar}, \citenamefont {Carroll}, \citenamefont {Hartikainen},\ and\ \citenamefont {Martin}}]{arviz_2019}%
  \BibitemOpen
  \bibfield  {author} {\bibinfo {author} {\bibfnamefont {R.}~\bibnamefont {Kumar}}, \bibinfo {author} {\bibfnamefont {C.}~\bibnamefont {Carroll}}, \bibinfo {author} {\bibfnamefont {A.}~\bibnamefont {Hartikainen}},\ and\ \bibinfo {author} {\bibfnamefont {O.}~\bibnamefont {Martin}},\ }\bibfield  {title} {\bibinfo {title} {Arviz a unified library for exploratory analysis of bayesian models in python},\ }\href {https://doi.org/10.21105/joss.01143} {\bibfield  {journal} {\bibinfo  {journal} {Journal of Open Source Software}\ }\textbf {\bibinfo {volume} {4}},\ \bibinfo {pages} {1143} (\bibinfo {year} {2019})}\BibitemShut {NoStop}%
\bibitem [{\citenamefont {Williams}\ \emph {et~al.}(2023)\citenamefont {Williams}, \citenamefont {Veitch}, \citenamefont {Chiofalo}, \citenamefont {Schmidt}, \citenamefont {Udall}, \citenamefont {Vajpeji},\ and\ \citenamefont {Hoy}}]{Williams:2022pgn}%
  \BibitemOpen
  \bibfield  {author} {\bibinfo {author} {\bibfnamefont {D.}~\bibnamefont {Williams}}, \bibinfo {author} {\bibfnamefont {J.}~\bibnamefont {Veitch}}, \bibinfo {author} {\bibfnamefont {M.~L.}\ \bibnamefont {Chiofalo}}, \bibinfo {author} {\bibfnamefont {P.}~\bibnamefont {Schmidt}}, \bibinfo {author} {\bibfnamefont {R.~P.}\ \bibnamefont {Udall}}, \bibinfo {author} {\bibfnamefont {A.}~\bibnamefont {Vajpeji}},\ and\ \bibinfo {author} {\bibfnamefont {C.}~\bibnamefont {Hoy}},\ }\bibfield  {title} {\bibinfo {title} {{Asimov: A framework for coordinating parameter estimation workflows}},\ }\href {https://doi.org/10.21105/joss.04170} {\bibfield  {journal} {\bibinfo  {journal} {J. Open Source Softw.}\ }\textbf {\bibinfo {volume} {8}},\ \bibinfo {pages} {4170} (\bibinfo {year} {2023})},\ \Eprint {https://arxiv.org/abs/2207.01468} {arXiv:2207.01468 [gr-qc]} \BibitemShut {NoStop}%
\bibitem [{\citenamefont {{Robitaille}}\ \emph {et~al.}(2013)\citenamefont {{Robitaille}} \emph {et~al.}}]{astropy:2013}%
  \BibitemOpen
  \bibfield  {author} {\bibinfo {author} {\bibfnamefont {T.~P.}\ \bibnamefont {{Robitaille}}} \emph {et~al.} (\bibinfo {collaboration} {Astropy Collaboration}),\ }\bibfield  {title} {\bibinfo {title} {{Astropy: A community Python package for astronomy}},\ }\href {https://doi.org/10.1051/0004-6361/201322068} {\bibfield  {journal} {\bibinfo  {journal} {Astron. and Astrophys.}\ }\textbf {\bibinfo {volume} {558}},\ \bibinfo {eid} {A33} (\bibinfo {year} {2013})},\ \Eprint {https://arxiv.org/abs/1307.6212} {arXiv:1307.6212 [astro-ph.IM]} \BibitemShut {NoStop}%
\bibitem [{\citenamefont {{Price-Whelan}}\ \emph {et~al.}(2018)\citenamefont {{Price-Whelan}} \emph {et~al.}}]{astropy:2018}%
  \BibitemOpen
  \bibfield  {author} {\bibinfo {author} {\bibfnamefont {A.~M.}\ \bibnamefont {{Price-Whelan}}} \emph {et~al.} (\bibinfo {collaboration} {Astropy Collaboration}),\ }\bibfield  {title} {\bibinfo {title} {{The Astropy Project: Building an Open-science Project and Status of the v2.0 Core Package}},\ }\href {https://doi.org/10.3847/1538-3881/aabc4f} {\bibfield  {journal} {\bibinfo  {journal} {The Astronomical Journal}\ }\textbf {\bibinfo {volume} {156}},\ \bibinfo {eid} {123} (\bibinfo {year} {2018})},\ \Eprint {https://arxiv.org/abs/1801.02634} {arXiv:1801.02634 [astro-ph.IM]} \BibitemShut {NoStop}%
\bibitem [{\citenamefont {{Price-Whelan}}\ \emph {et~al.}(2022)\citenamefont {{Price-Whelan}} \emph {et~al.}}]{astropy:2022}%
  \BibitemOpen
  \bibfield  {author} {\bibinfo {author} {\bibfnamefont {A.~M.}\ \bibnamefont {{Price-Whelan}}} \emph {et~al.} (\bibinfo {collaboration} {Astropy Collaboration}),\ }\bibfield  {title} {\bibinfo {title} {{The Astropy Project: Sustaining and Growing a Community-oriented Open-source Project and the Latest Major Release (v5.0) of the Core Package}},\ }\href {https://doi.org/10.3847/1538-4357/ac7c74} {\bibfield  {journal} {\bibinfo  {journal} {\apj}\ }\textbf {\bibinfo {volume} {935}},\ \bibinfo {eid} {167} (\bibinfo {year} {2022})},\ \Eprint {https://arxiv.org/abs/2206.14220} {arXiv:2206.14220 [astro-ph.IM]} \BibitemShut {NoStop}%
\bibitem [{\citenamefont {Cornish}\ \emph {et~al.}(2021)\citenamefont {Cornish}, \citenamefont {Littenberg}, \citenamefont {B\'ecsy}, \citenamefont {Chatziioannou}, \citenamefont {Clark}, \citenamefont {Ghonge},\ and\ \citenamefont {Millhouse}}]{Cornish:2020dwh}%
  \BibitemOpen
  \bibfield  {author} {\bibinfo {author} {\bibfnamefont {N.~J.}\ \bibnamefont {Cornish}}, \bibinfo {author} {\bibfnamefont {T.~B.}\ \bibnamefont {Littenberg}}, \bibinfo {author} {\bibfnamefont {B.}~\bibnamefont {B\'ecsy}}, \bibinfo {author} {\bibfnamefont {K.}~\bibnamefont {Chatziioannou}}, \bibinfo {author} {\bibfnamefont {J.~A.}\ \bibnamefont {Clark}}, \bibinfo {author} {\bibfnamefont {S.}~\bibnamefont {Ghonge}},\ and\ \bibinfo {author} {\bibfnamefont {M.}~\bibnamefont {Millhouse}},\ }\bibfield  {title} {\bibinfo {title} {{BayesWave analysis pipeline in the era of gravitational wave observations}},\ }\href {https://doi.org/10.1103/PhysRevD.103.044006} {\bibfield  {journal} {\bibinfo  {journal} {Phys. Rev. D}\ }\textbf {\bibinfo {volume} {103}},\ \bibinfo {pages} {044006} (\bibinfo {year} {2021})},\ \Eprint {https://arxiv.org/abs/2011.09494} {arXiv:2011.09494 [gr-qc]} \BibitemShut {NoStop}%
\bibitem [{\citenamefont {Ashton}\ \emph {et~al.}(2019)\citenamefont {Ashton} \emph {et~al.}}]{Ashton:2018jfp}%
  \BibitemOpen
  \bibfield  {author} {\bibinfo {author} {\bibfnamefont {G.}~\bibnamefont {Ashton}} \emph {et~al.},\ }\bibfield  {title} {\bibinfo {title} {{Bilby: A user-friendly Bayesian inference library for gravitational-wave astronomy}},\ }\href {https://doi.org/10.3847/1538-4365/ab06fc} {\bibfield  {journal} {\bibinfo  {journal} {Astrophys. J. Suppl. Ser.}\ }\textbf {\bibinfo {volume} {241}},\ \bibinfo {pages} {27} (\bibinfo {year} {2019})},\ \Eprint {https://arxiv.org/abs/1811.02042} {arXiv:1811.02042 [astro-ph.IM]} \BibitemShut {NoStop}%
\bibitem [{\citenamefont {Romero-Shaw}\ \emph {et~al.}(2020)\citenamefont {Romero-Shaw} \emph {et~al.}}]{Romero-Shaw:2020owr}%
  \BibitemOpen
  \bibfield  {author} {\bibinfo {author} {\bibfnamefont {I.~M.}\ \bibnamefont {Romero-Shaw}} \emph {et~al.},\ }\bibfield  {title} {\bibinfo {title} {{Bayesian inference for compact binary coalescences with bilby: validation and application to the first LIGO\textendash{}Virgo gravitational-wave transient catalogue}},\ }\href {https://doi.org/10.1093/mnras/staa2850} {\bibfield  {journal} {\bibinfo  {journal} {Mon. Not. Roy. Astron. Soc.}\ }\textbf {\bibinfo {volume} {499}},\ \bibinfo {pages} {3295} (\bibinfo {year} {2020})},\ \Eprint {https://arxiv.org/abs/2006.00714} {arXiv:2006.00714 [astro-ph.IM]} \BibitemShut {NoStop}%
\bibitem [{\citenamefont {Ashton}\ \emph {et~al.}(2025)\citenamefont {Ashton}, \citenamefont {Talbot}, \citenamefont {Roy}, \citenamefont {Pratten}, \citenamefont {Pang}, \citenamefont {Agathos}, \citenamefont {Baka}, \citenamefont {Sänger}, \citenamefont {Mehta}, \citenamefont {Steinhoff}, \citenamefont {Maggio}, \citenamefont {Ghosh}, \citenamefont {Vijaykumar}, \citenamefont {Enficiaud},\ and\ \citenamefont {Pompili}}]{ashton_2025_15676285}%
  \BibitemOpen
  \bibfield  {author} {\bibinfo {author} {\bibfnamefont {G.}~\bibnamefont {Ashton}}, \bibinfo {author} {\bibfnamefont {C.}~\bibnamefont {Talbot}}, \bibinfo {author} {\bibfnamefont {S.}~\bibnamefont {Roy}}, \bibinfo {author} {\bibfnamefont {G.}~\bibnamefont {Pratten}}, \bibinfo {author} {\bibfnamefont {T.-H.}\ \bibnamefont {Pang}}, \bibinfo {author} {\bibfnamefont {M.}~\bibnamefont {Agathos}}, \bibinfo {author} {\bibfnamefont {T.}~\bibnamefont {Baka}}, \bibinfo {author} {\bibfnamefont {E.}~\bibnamefont {Sänger}}, \bibinfo {author} {\bibfnamefont {A.}~\bibnamefont {Mehta}}, \bibinfo {author} {\bibfnamefont {J.}~\bibnamefont {Steinhoff}}, \bibinfo {author} {\bibfnamefont {E.}~\bibnamefont {Maggio}}, \bibinfo {author} {\bibfnamefont {A.}~\bibnamefont {Ghosh}}, \bibinfo {author} {\bibfnamefont {A.}~\bibnamefont {Vijaykumar}}, \bibinfo {author} {\bibfnamefont {R.}~\bibnamefont {Enficiaud}},\ and\ \bibinfo {author} {\bibfnamefont {L.}~\bibnamefont {Pompili}},\ }\href {https://doi.org/10.5281/zenodo.15676285} {\bibinfo {title} {Bilby {TGR}}} (\bibinfo {year} {2025})\BibitemShut {NoStop}%
\bibitem [{\citenamefont {Del~Pozzo}\ and\ \citenamefont {Veitch}(2025)}]{cpnest}%
  \BibitemOpen
  \bibfield  {author} {\bibinfo {author} {\bibfnamefont {W.}~\bibnamefont {Del~Pozzo}}\ and\ \bibinfo {author} {\bibfnamefont {J.}~\bibnamefont {Veitch}},\ }\href@noop {} {\bibinfo {title} {{\texttt{CPNest}: an efficient python parallelizable nested sampling algorithm}}},\ \bibinfo {howpublished} {\url{https://github.com/johnveitch/cpnest}} (\bibinfo {year} {2025})\BibitemShut {NoStop}%
\bibitem [{\citenamefont {Speagle}(2020)}]{Speagle:2019ivv}%
  \BibitemOpen
  \bibfield  {author} {\bibinfo {author} {\bibfnamefont {J.~S.}\ \bibnamefont {Speagle}},\ }\bibfield  {title} {\bibinfo {title} {{dynesty: a dynamic nested sampling package for estimating Bayesian posteriors and evidences}},\ }\href {https://doi.org/10.1093/mnras/staa278} {\bibfield  {journal} {\bibinfo  {journal} {Mon. Not. Roy. Astron. Soc.}\ }\textbf {\bibinfo {volume} {493}},\ \bibinfo {pages} {3132} (\bibinfo {year} {2020})},\ \Eprint {https://arxiv.org/abs/1904.02180} {arXiv:1904.02180 [astro-ph.IM]} \BibitemShut {NoStop}%
\bibitem [{\citenamefont {{Macleod}}\ \emph {et~al.}(2021)\citenamefont {{Macleod}}, \citenamefont {{Areeda}}, \citenamefont {{Coughlin}}, \citenamefont {{Massinger}},\ and\ \citenamefont {{Urban}}}]{gwpy}%
  \BibitemOpen
  \bibfield  {author} {\bibinfo {author} {\bibfnamefont {D.~M.}\ \bibnamefont {{Macleod}}}, \bibinfo {author} {\bibfnamefont {J.~S.}\ \bibnamefont {{Areeda}}}, \bibinfo {author} {\bibfnamefont {S.~B.}\ \bibnamefont {{Coughlin}}}, \bibinfo {author} {\bibfnamefont {T.~J.}\ \bibnamefont {{Massinger}}},\ and\ \bibinfo {author} {\bibfnamefont {A.~L.}\ \bibnamefont {{Urban}}},\ }\bibfield  {title} {\bibinfo {title} {{GWpy: A Python package for gravitational-wave astrophysics}},\ }\href {https://doi.org/10.1016/j.softx.2021.100657} {\bibfield  {journal} {\bibinfo  {journal} {SoftwareX}\ }\textbf {\bibinfo {volume} {13}},\ \bibinfo {pages} {100657} (\bibinfo {year} {2021})}\BibitemShut {NoStop}%
\bibitem [{\citenamefont {Macleod}\ \emph {et~al.}(2024)\citenamefont {Macleod}, \citenamefont {Coughlin}, \citenamefont {Southgate}, \citenamefont {Davis}, \citenamefont {Pitkin}, \citenamefont {Areeda}, \citenamefont {George}, \citenamefont {Altin}, \citenamefont {Godwin}, \citenamefont {Singer} \emph {et~al.}}]{duncan_macleod_2024_12734623}%
  \BibitemOpen
  \bibfield  {author} {\bibinfo {author} {\bibfnamefont {D.}~\bibnamefont {Macleod}}, \bibinfo {author} {\bibfnamefont {S.}~\bibnamefont {Coughlin}}, \bibinfo {author} {\bibfnamefont {A.}~\bibnamefont {Southgate}}, \bibinfo {author} {\bibfnamefont {D.}~\bibnamefont {Davis}}, \bibinfo {author} {\bibfnamefont {M.}~\bibnamefont {Pitkin}}, \bibinfo {author} {\bibfnamefont {J.}~\bibnamefont {Areeda}}, \bibinfo {author} {\bibfnamefont {R.~N.}\ \bibnamefont {George}}, \bibinfo {author} {\bibfnamefont {P.}~\bibnamefont {Altin}}, \bibinfo {author} {\bibfnamefont {P.}~\bibnamefont {Godwin}}, \bibinfo {author} {\bibfnamefont {L.}~\bibnamefont {Singer}}, \emph {et~al.},\ }\href {https://doi.org/10.5281/zenodo.12734623} {\bibinfo {title} {gwpy/gwpy: Gwpy 3.0.9}} (\bibinfo {year} {2024})\BibitemShut {NoStop}%
\bibitem [{\citenamefont {Collette}(2013)}]{collette_python_hdf5_2014}%
  \BibitemOpen
  \bibfield  {author} {\bibinfo {author} {\bibfnamefont {A.}~\bibnamefont {Collette}},\ }\href@noop {} {\emph {\bibinfo {title} {Python and HDF5}}}\ (\bibinfo  {publisher} {O'Reilly},\ \bibinfo {year} {2013})\BibitemShut {NoStop}%
\bibitem [{\citenamefont {Perez}\ and\ \citenamefont {Granger}(2007)}]{ipython_2007}%
  \BibitemOpen
  \bibfield  {author} {\bibinfo {author} {\bibfnamefont {F.}~\bibnamefont {Perez}}\ and\ \bibinfo {author} {\bibfnamefont {B.~E.}\ \bibnamefont {Granger}},\ }\bibfield  {title} {\bibinfo {title} {Ipython: A system for interactive scientific computing},\ }\href {https://doi.org/10.1109/MCSE.2007.53} {\bibfield  {journal} {\bibinfo  {journal} {Computing in Science \& Engineering}\ }\textbf {\bibinfo {volume} {9}},\ \bibinfo {pages} {21} (\bibinfo {year} {2007})}\BibitemShut {NoStop}%
\bibitem [{\citenamefont {{Kluyver}}\ \emph {et~al.}(2016)\citenamefont {{Kluyver}} \emph {et~al.}}]{jupyter_2016}%
  \BibitemOpen
  \bibfield  {author} {\bibinfo {author} {\bibfnamefont {T.}~\bibnamefont {{Kluyver}}} \emph {et~al.} (\bibinfo {collaboration} {Jupyter Development Team}),\ }\bibfield  {title} {\bibinfo {title} {{Jupyter Notebooks{\textemdash}a publishing format for reproducible computational workflows}},\ }in\ \href {https://doi.org/10.3233/978-1-61499-649-1-87} {\emph {\bibinfo {booktitle} {IOS Press}}}\ (\bibinfo  {publisher} {IOS Press},\ \bibinfo {year} {2016})\ pp.\ \bibinfo {pages} {87--90}\BibitemShut {NoStop}%
\bibitem [{\citenamefont {Beg}\ \emph {et~al.}(2021)\citenamefont {Beg}, \citenamefont {Taka}, \citenamefont {Kluyver}, \citenamefont {Konovalov}, \citenamefont {Ragan-Kelley}, \citenamefont {Thiéry},\ and\ \citenamefont {Fangohr}}]{jupyter_2021}%
  \BibitemOpen
  \bibfield  {author} {\bibinfo {author} {\bibfnamefont {M.}~\bibnamefont {Beg}}, \bibinfo {author} {\bibfnamefont {J.}~\bibnamefont {Taka}}, \bibinfo {author} {\bibfnamefont {T.}~\bibnamefont {Kluyver}}, \bibinfo {author} {\bibfnamefont {A.}~\bibnamefont {Konovalov}}, \bibinfo {author} {\bibfnamefont {M.}~\bibnamefont {Ragan-Kelley}}, \bibinfo {author} {\bibfnamefont {N.~M.}\ \bibnamefont {Thiéry}},\ and\ \bibinfo {author} {\bibfnamefont {H.}~\bibnamefont {Fangohr}},\ }\bibfield  {title} {\bibinfo {title} {Using jupyter for reproducible scientific workflows},\ }\href {https://doi.org/10.1109/MCSE.2021.3052101} {\bibfield  {journal} {\bibinfo  {journal} {Computing in Science \& Engineering}\ }\textbf {\bibinfo {volume} {23}},\ \bibinfo {pages} {36} (\bibinfo {year} {2021})}\BibitemShut {NoStop}%
\bibitem [{\citenamefont {{LIGO Scientific Collaboration}}\ \emph {et~al.}(2018)\citenamefont {{LIGO Scientific Collaboration}}, \citenamefont {{Virgo Collaboration}},\ and\ \citenamefont {{KAGRA Collaboration}}}]{lalsuite}%
  \BibitemOpen
  \bibfield  {author} {\bibinfo {author} {\bibnamefont {{LIGO Scientific Collaboration}}}, \bibinfo {author} {\bibnamefont {{Virgo Collaboration}}},\ and\ \bibinfo {author} {\bibnamefont {{KAGRA Collaboration}}},\ }\href {https://doi.org/10.7935/GT1W-FZ16} {\bibinfo {title} {{LVK} {A}lgorithm {L}ibrary - {LALS}uite}},\ \bibinfo {howpublished} {Free software (GPL)} (\bibinfo {year} {2018})\BibitemShut {NoStop}%
\bibitem [{\citenamefont {Wette}(2020)}]{swiglal}%
  \BibitemOpen
  \bibfield  {author} {\bibinfo {author} {\bibfnamefont {K.}~\bibnamefont {Wette}},\ }\bibfield  {title} {\bibinfo {title} {{SWIGLAL: Python and Octave interfaces to the LALSuite gravitational-wave data analysis libraries}},\ }\href {https://doi.org/10.1016/j.softx.2020.100634} {\bibfield  {journal} {\bibinfo  {journal} {SoftwareX}\ }\textbf {\bibinfo {volume} {12}},\ \bibinfo {pages} {100634} (\bibinfo {year} {2020})},\ \Eprint {https://arxiv.org/abs/2012.09552} {arXiv:2012.09552 [astro-ph.IM]} \BibitemShut {NoStop}%
\bibitem [{\citenamefont {Caswell}\ \emph {et~al.}(2023)\citenamefont {Caswell}, \citenamefont {de~Andrade}, \citenamefont {Lee}, \citenamefont {Droettboom}, \citenamefont {Hoffmann}, \citenamefont {Klymak}, \citenamefont {Hunter}, \citenamefont {Firing}, \citenamefont {Stansby}, \citenamefont {Varoquaux} \emph {et~al.}}]{matplotlib_3_7_3_zenodo}%
  \BibitemOpen
  \bibfield  {author} {\bibinfo {author} {\bibfnamefont {T.~A.}\ \bibnamefont {Caswell}}, \bibinfo {author} {\bibfnamefont {E.~S.}\ \bibnamefont {de~Andrade}}, \bibinfo {author} {\bibfnamefont {A.}~\bibnamefont {Lee}}, \bibinfo {author} {\bibfnamefont {M.}~\bibnamefont {Droettboom}}, \bibinfo {author} {\bibfnamefont {T.}~\bibnamefont {Hoffmann}}, \bibinfo {author} {\bibfnamefont {J.}~\bibnamefont {Klymak}}, \bibinfo {author} {\bibfnamefont {J.}~\bibnamefont {Hunter}}, \bibinfo {author} {\bibfnamefont {E.}~\bibnamefont {Firing}}, \bibinfo {author} {\bibfnamefont {D.}~\bibnamefont {Stansby}}, \bibinfo {author} {\bibfnamefont {N.}~\bibnamefont {Varoquaux}}, \emph {et~al.},\ }\href {https://doi.org/10.5281/zenodo.8336761} {\bibinfo {title} {matplotlib/matplotlib: Rel: v3.7.3}} (\bibinfo {year} {2023})\BibitemShut {NoStop}%
\bibitem [{\citenamefont {Hunter}(2007)}]{Hunter:2007}%
  \BibitemOpen
  \bibfield  {author} {\bibinfo {author} {\bibfnamefont {J.~D.}\ \bibnamefont {Hunter}},\ }\bibfield  {title} {\bibinfo {title} {Matplotlib: A 2d graphics environment},\ }\href {https://doi.org/10.1109/MCSE.2007.55} {\bibfield  {journal} {\bibinfo  {journal} {Computing in Science \& Engineering}\ }\textbf {\bibinfo {volume} {9}},\ \bibinfo {pages} {90} (\bibinfo {year} {2007})}\BibitemShut {NoStop}%
\bibitem [{\citenamefont {Harris}\ \emph {et~al.}(2020)\citenamefont {Harris}, \citenamefont {Millman}, \citenamefont {van~der Walt}, \citenamefont {Gommers}, \citenamefont {Virtanen}, \citenamefont {Cournapeau}, \citenamefont {Wieser}, \citenamefont {Taylor}, \citenamefont {Berg}, \citenamefont {Smith} \emph {et~al.}}]{numpy}%
  \BibitemOpen
  \bibfield  {author} {\bibinfo {author} {\bibfnamefont {C.~R.}\ \bibnamefont {Harris}}, \bibinfo {author} {\bibfnamefont {K.~J.}\ \bibnamefont {Millman}}, \bibinfo {author} {\bibfnamefont {S.~J.}\ \bibnamefont {van~der Walt}}, \bibinfo {author} {\bibfnamefont {R.}~\bibnamefont {Gommers}}, \bibinfo {author} {\bibfnamefont {P.}~\bibnamefont {Virtanen}}, \bibinfo {author} {\bibfnamefont {D.}~\bibnamefont {Cournapeau}}, \bibinfo {author} {\bibfnamefont {E.}~\bibnamefont {Wieser}}, \bibinfo {author} {\bibfnamefont {J.}~\bibnamefont {Taylor}}, \bibinfo {author} {\bibfnamefont {S.}~\bibnamefont {Berg}}, \bibinfo {author} {\bibfnamefont {N.~J.}\ \bibnamefont {Smith}}, \emph {et~al.},\ }\bibfield  {title} {\bibinfo {title} {Array programming with {NumPy}},\ }\href {https://doi.org/10.1038/s41586-020-2649-2} {\bibfield  {journal} {\bibinfo  {journal} {Nature}\ }\textbf {\bibinfo {volume} {585}},\ \bibinfo {pages} {357} (\bibinfo {year} {2020})}\BibitemShut {NoStop}%
\bibitem [{\citenamefont {McKinney}(2010)}]{mckinney-proc-scipy-2010}%
  \BibitemOpen
  \bibfield  {author} {\bibinfo {author} {\bibfnamefont {W.}~\bibnamefont {McKinney}},\ }\bibfield  {title} {\bibinfo {title} {{D}ata {S}tructures for {S}tatistical {C}omputing in {P}ython},\ }in\ \href {https://doi.org/10.25080/Majora-92bf1922-00a} {\emph {\bibinfo {booktitle} {{P}roceedings of the 9th {P}ython in {S}cience {C}onference}}},\ \bibinfo {editor} {edited by\ \bibinfo {editor} {\bibnamefont {{S}t\'efan van~der {W}alt}}\ and\ \bibinfo {editor} {\bibnamefont {{J}arrod {M}illman}}}\ (\bibinfo {year} {2010})\ pp.\ \bibinfo {pages} {56 -- 61}\BibitemShut {NoStop}%
\bibitem [{\citenamefont {{The pandas development team}}(2024)}]{pandas_13819579}%
  \BibitemOpen
  \bibfield  {author} {\bibinfo {author} {\bibnamefont {{The pandas development team}}},\ }\href {https://doi.org/10.5281/zenodo.13819579} {\bibinfo {title} {pandas-dev/pandas: Pandas}} (\bibinfo {year} {2024})\BibitemShut {NoStop}%
\bibitem [{\citenamefont {Hoy}\ and\ \citenamefont {Raymond}(2021)}]{Hoy:2020vys}%
  \BibitemOpen
  \bibfield  {author} {\bibinfo {author} {\bibfnamefont {C.}~\bibnamefont {Hoy}}\ and\ \bibinfo {author} {\bibfnamefont {V.}~\bibnamefont {Raymond}},\ }\bibfield  {title} {\bibinfo {title} {{PESummary: the code agnostic Parameter Estimation Summary page builder}},\ }\href {https://doi.org/10.1016/j.softx.2021.100765} {\bibfield  {journal} {\bibinfo  {journal} {SoftwareX}\ }\textbf {\bibinfo {volume} {15}},\ \bibinfo {pages} {100765} (\bibinfo {year} {2021})},\ \Eprint {https://arxiv.org/abs/2006.06639} {arXiv:2006.06639 [astro-ph.IM]} \BibitemShut {NoStop}%
\bibitem [{\citenamefont {Mihaylov}\ \emph {et~al.}(2025)\citenamefont {Mihaylov}, \citenamefont {Ossokine}, \citenamefont {Buonanno}, \citenamefont {Estelles}, \citenamefont {Pompili}, \citenamefont {P\"urrer},\ and\ \citenamefont {Ramos-Buades}}]{Mihaylov:2023bkc}%
  \BibitemOpen
  \bibfield  {author} {\bibinfo {author} {\bibfnamefont {D.~P.}\ \bibnamefont {Mihaylov}}, \bibinfo {author} {\bibfnamefont {S.}~\bibnamefont {Ossokine}}, \bibinfo {author} {\bibfnamefont {A.}~\bibnamefont {Buonanno}}, \bibinfo {author} {\bibfnamefont {H.}~\bibnamefont {Estelles}}, \bibinfo {author} {\bibfnamefont {L.}~\bibnamefont {Pompili}}, \bibinfo {author} {\bibfnamefont {M.}~\bibnamefont {P\"urrer}},\ and\ \bibinfo {author} {\bibfnamefont {A.}~\bibnamefont {Ramos-Buades}},\ }\bibfield  {title} {\bibinfo {title} {{pySEOBNR: a software package for the next generation of effective-one-body multipolar waveform models}},\ }\href {https://doi.org/10.1016/j.softx.2025.102080} {\bibfield  {journal} {\bibinfo  {journal} {SoftwareX}\ }\textbf {\bibinfo {volume} {30}},\ \bibinfo {pages} {102080} (\bibinfo {year} {2025})},\ \Eprint {https://arxiv.org/abs/2303.18203} {arXiv:2303.18203 [gr-qc]} \BibitemShut {NoStop}%
\bibitem [{\citenamefont {Van~Rossum}\ and\ \citenamefont {Drake}(2009)}]{python}%
  \BibitemOpen
  \bibfield  {author} {\bibinfo {author} {\bibfnamefont {G.}~\bibnamefont {Van~Rossum}}\ and\ \bibinfo {author} {\bibfnamefont {F.~L.}\ \bibnamefont {Drake}},\ }\href@noop {} {\emph {\bibinfo {title} {Python 3 Reference Manual}}}\ (\bibinfo  {publisher} {CreateSpace},\ \bibinfo {address} {Scotts Valley, CA},\ \bibinfo {year} {2009})\BibitemShut {NoStop}%
\bibitem [{\citenamefont {Stein}(2019)}]{Stein:2019mop}%
  \BibitemOpen
  \bibfield  {author} {\bibinfo {author} {\bibfnamefont {L.~C.}\ \bibnamefont {Stein}},\ }\bibfield  {title} {\bibinfo {title} {{qnm: A Python package for calculating Kerr quasinormal modes, separation constants, and spherical-spheroidal mixing coefficients}},\ }\href {https://doi.org/10.21105/joss.01683} {\bibfield  {journal} {\bibinfo  {journal} {J. Open Source Softw.}\ }\textbf {\bibinfo {volume} {4}},\ \bibinfo {pages} {1683} (\bibinfo {year} {2019})},\ \Eprint {https://arxiv.org/abs/1908.10377} {arXiv:1908.10377 [gr-qc]} \BibitemShut {NoStop}%
\bibitem [{\citenamefont {Isi}\ and\ \citenamefont {Farr}(2024)}]{ringdown}%
  \BibitemOpen
  \bibfield  {author} {\bibinfo {author} {\bibfnamefont {M.}~\bibnamefont {Isi}}\ and\ \bibinfo {author} {\bibfnamefont {W.~M.}\ \bibnamefont {Farr}},\ }\href {https://doi.org/10.5281/zenodo.5094067} {\bibinfo {title} {\textsc{ringdown} package}},\ \bibinfo {howpublished} {\href{https://ringdown.readthedocs.io/en/latest/}{ringdown.readthedocs.io}} (\bibinfo {year} {2024})\BibitemShut {NoStop}%
\bibitem [{\citenamefont {Virtanen}\ \emph {et~al.}(2020)\citenamefont {Virtanen} \emph {et~al.}}]{2020SciPy-NMeth}%
  \BibitemOpen
  \bibfield  {author} {\bibinfo {author} {\bibfnamefont {P.}~\bibnamefont {Virtanen}} \emph {et~al.} (\bibinfo {collaboration} {SciPy 1.0 Contributors}),\ }\bibfield  {title} {\bibinfo {title} {{{SciPy} 1.0: Fundamental Algorithms for Scientific Computing in Python}},\ }\href {https://doi.org/10.1038/s41592-019-0686-2} {\bibfield  {journal} {\bibinfo  {journal} {Nature Methods}\ }\textbf {\bibinfo {volume} {17}},\ \bibinfo {pages} {261} (\bibinfo {year} {2020})}\BibitemShut {NoStop}%
\bibitem [{\citenamefont {Gommers}\ \emph {et~al.}(2024)\citenamefont {Gommers}, \citenamefont {Virtanen}, \citenamefont {Haberland}, \citenamefont {Burovski}, \citenamefont {Weckesser}, \citenamefont {Reddy}, \citenamefont {Oliphant}, \citenamefont {Cournapeau}, \citenamefont {Nelson} \emph {et~al.}}]{scipy_10543017}%
  \BibitemOpen
  \bibfield  {author} {\bibinfo {author} {\bibfnamefont {R.}~\bibnamefont {Gommers}}, \bibinfo {author} {\bibfnamefont {P.}~\bibnamefont {Virtanen}}, \bibinfo {author} {\bibfnamefont {M.}~\bibnamefont {Haberland}}, \bibinfo {author} {\bibfnamefont {E.}~\bibnamefont {Burovski}}, \bibinfo {author} {\bibfnamefont {W.}~\bibnamefont {Weckesser}}, \bibinfo {author} {\bibfnamefont {T.}~\bibnamefont {Reddy}}, \bibinfo {author} {\bibfnamefont {T.~E.}\ \bibnamefont {Oliphant}}, \bibinfo {author} {\bibfnamefont {D.}~\bibnamefont {Cournapeau}}, \bibinfo {author} {\bibfnamefont {A.}~\bibnamefont {Nelson}}, \emph {et~al.},\ }\href {https://doi.org/10.5281/zenodo.10543017} {\bibinfo {title} {scipy/scipy: Scipy 1.12.0}} (\bibinfo {year} {2024})\BibitemShut {NoStop}%
\bibitem [{\citenamefont {Waskom}(2021)}]{Waskom2021}%
  \BibitemOpen
  \bibfield  {author} {\bibinfo {author} {\bibfnamefont {M.~L.}\ \bibnamefont {Waskom}},\ }\bibfield  {title} {\bibinfo {title} {seaborn: statistical data visualization},\ }\href {https://doi.org/10.21105/joss.03021} {\bibfield  {journal} {\bibinfo  {journal} {Journal of Open Source Software}\ }\textbf {\bibinfo {volume} {6}},\ \bibinfo {pages} {3021} (\bibinfo {year} {2021})}\BibitemShut {NoStop}%
\bibitem [{\citenamefont {Khalil}\ \emph {et~al.}(2023)\citenamefont {Khalil}, \citenamefont {Buonanno}, \citenamefont {Estelles}, \citenamefont {Mihaylov}, \citenamefont {Ossokine}, \citenamefont {Pompili},\ and\ \citenamefont {Ramos-Buades}}]{Khalil:2023kep}%
  \BibitemOpen
  \bibfield  {author} {\bibinfo {author} {\bibfnamefont {M.}~\bibnamefont {Khalil}}, \bibinfo {author} {\bibfnamefont {A.}~\bibnamefont {Buonanno}}, \bibinfo {author} {\bibfnamefont {H.}~\bibnamefont {Estelles}}, \bibinfo {author} {\bibfnamefont {D.~P.}\ \bibnamefont {Mihaylov}}, \bibinfo {author} {\bibfnamefont {S.}~\bibnamefont {Ossokine}}, \bibinfo {author} {\bibfnamefont {L.}~\bibnamefont {Pompili}},\ and\ \bibinfo {author} {\bibfnamefont {A.}~\bibnamefont {Ramos-Buades}},\ }\bibfield  {title} {\bibinfo {title} {{Theoretical groundwork supporting the precessing-spin two-body dynamics of the effective-one-body waveform models SEOBNRv5}},\ }\href {https://doi.org/10.1103/PhysRevD.108.124036} {\bibfield  {journal} {\bibinfo  {journal} {Phys. Rev. D}\ }\textbf {\bibinfo {volume} {108}},\ \bibinfo {pages} {124036} (\bibinfo {year} {2023})},\ \Eprint {https://arxiv.org/abs/2303.18143} {arXiv:2303.18143 [gr-qc]} \BibitemShut {NoStop}%
\bibitem [{\citenamefont {van~de Meent}\ \emph {et~al.}(2023)\citenamefont {van~de Meent}, \citenamefont {Buonanno}, \citenamefont {Mihaylov}, \citenamefont {Ossokine}, \citenamefont {Pompili}, \citenamefont {Warburton}, \citenamefont {Pound}, \citenamefont {Wardell}, \citenamefont {Durkan},\ and\ \citenamefont {Miller}}]{vandeMeent:2023ols}%
  \BibitemOpen
  \bibfield  {author} {\bibinfo {author} {\bibfnamefont {M.}~\bibnamefont {van~de Meent}}, \bibinfo {author} {\bibfnamefont {A.}~\bibnamefont {Buonanno}}, \bibinfo {author} {\bibfnamefont {D.~P.}\ \bibnamefont {Mihaylov}}, \bibinfo {author} {\bibfnamefont {S.}~\bibnamefont {Ossokine}}, \bibinfo {author} {\bibfnamefont {L.}~\bibnamefont {Pompili}}, \bibinfo {author} {\bibfnamefont {N.}~\bibnamefont {Warburton}}, \bibinfo {author} {\bibfnamefont {A.}~\bibnamefont {Pound}}, \bibinfo {author} {\bibfnamefont {B.}~\bibnamefont {Wardell}}, \bibinfo {author} {\bibfnamefont {L.}~\bibnamefont {Durkan}},\ and\ \bibinfo {author} {\bibfnamefont {J.}~\bibnamefont {Miller}},\ }\bibfield  {title} {\bibinfo {title} {{Enhancing the SEOBNRv5 effective-one-body waveform model with second-order gravitational self-force fluxes}},\ }\href {https://doi.org/10.1103/PhysRevD.108.124038} {\bibfield  {journal} {\bibinfo  {journal} {Phys. Rev. D}\ }\textbf {\bibinfo {volume} {108}},\ \bibinfo {pages} {124038} (\bibinfo {year} {2023})},\ \Eprint {https://arxiv.org/abs/2303.18026} {arXiv:2303.18026 [gr-qc]} \BibitemShut {NoStop}%
\bibitem [{\citenamefont {Boyle}\ \emph {et~al.}(2025)\citenamefont {Boyle}, \citenamefont {Mitman}, \citenamefont {Scheel},\ and\ \citenamefont {Stein}}]{boyle_2025_16277847}%
  \BibitemOpen
  \bibfield  {author} {\bibinfo {author} {\bibfnamefont {M.}~\bibnamefont {Boyle}}, \bibinfo {author} {\bibfnamefont {K.}~\bibnamefont {Mitman}}, \bibinfo {author} {\bibfnamefont {M.}~\bibnamefont {Scheel}},\ and\ \bibinfo {author} {\bibfnamefont {L.}~\bibnamefont {Stein}},\ }\href {https://doi.org/10.5281/zenodo.16277847} {\bibinfo {title} {The {SXS} package}} (\bibinfo {year} {2025})\BibitemShut {NoStop}%
\bibitem [{\citenamefont {da~Costa-Luis}\ \emph {et~al.}(2024)\citenamefont {da~Costa-Luis}, \citenamefont {Larroque}, \citenamefont {Altendorf}, \citenamefont {Mary}, \citenamefont {Sheridan}, \citenamefont {Korobov}, \citenamefont {Yorav-Raphael}, \citenamefont {Ivanov}, \citenamefont {Bargull}, \citenamefont {Rodrigues} \emph {et~al.}}]{tqdm_14002015}%
  \BibitemOpen
  \bibfield  {author} {\bibinfo {author} {\bibfnamefont {C.}~\bibnamefont {da~Costa-Luis}}, \bibinfo {author} {\bibfnamefont {S.~K.}\ \bibnamefont {Larroque}}, \bibinfo {author} {\bibfnamefont {K.}~\bibnamefont {Altendorf}}, \bibinfo {author} {\bibfnamefont {H.}~\bibnamefont {Mary}}, \bibinfo {author} {\bibfnamefont {R.}~\bibnamefont {Sheridan}}, \bibinfo {author} {\bibfnamefont {M.}~\bibnamefont {Korobov}}, \bibinfo {author} {\bibfnamefont {N.}~\bibnamefont {Yorav-Raphael}}, \bibinfo {author} {\bibfnamefont {I.}~\bibnamefont {Ivanov}}, \bibinfo {author} {\bibfnamefont {M.}~\bibnamefont {Bargull}}, \bibinfo {author} {\bibfnamefont {N.}~\bibnamefont {Rodrigues}}, \emph {et~al.},\ }\href {https://doi.org/10.5281/zenodo.14002015} {\bibinfo {title} {tqdm: A fast, extensible progress bar for {Python} and {CLI}}} (\bibinfo {year} {2024})\BibitemShut {NoStop}%
\bibitem [{\citenamefont {Talbot}\ \emph {et~al.}(2025)\citenamefont {Talbot} \emph {et~al.}}]{Talbot:2025vth}%
  \BibitemOpen
  \bibfield  {author} {\bibinfo {author} {\bibfnamefont {C.}~\bibnamefont {Talbot}} \emph {et~al.},\ }\bibfield  {title} {\bibinfo {title} {{Inference with finite time series II: the window strikes back}},\ }\href@noop {} {\  (\bibinfo {year} {2025})},\ \Eprint {https://arxiv.org/abs/2508.11091} {arXiv:2508.11091 [gr-qc]} \BibitemShut {NoStop}%
\bibitem [{\citenamefont {Abac}\ \emph {et~al.}(2025{\natexlab{d}})\citenamefont {Abac} \emph {et~al.}}]{LIGOScientific:2025yae}%
  \BibitemOpen
  \bibfield  {author} {\bibinfo {author} {\bibfnamefont {A.~G.}\ \bibnamefont {Abac}} \emph {et~al.} (\bibinfo {collaboration} {LIGO Scientific, VIRGO, KAGRA}),\ }\bibfield  {title} {\bibinfo {title} {{GWTC-4.0: Methods for Identifying and Characterizing Gravitational-wave Transients}},\ }\href@noop {} {\  (\bibinfo {year} {2025}{\natexlab{d}})},\ \Eprint {https://arxiv.org/abs/2508.18081} {arXiv:2508.18081 [gr-qc]} \BibitemShut {NoStop}%
\bibitem [{\citenamefont {Baibhav}\ \emph {et~al.}(2023)\citenamefont {Baibhav}, \citenamefont {Cheung}, \citenamefont {Berti}, \citenamefont {Cardoso}, \citenamefont {Carullo}, \citenamefont {Cotesta}, \citenamefont {Del~Pozzo},\ and\ \citenamefont {Duque}}]{Baibhav:2023clw}%
  \BibitemOpen
  \bibfield  {author} {\bibinfo {author} {\bibfnamefont {V.}~\bibnamefont {Baibhav}}, \bibinfo {author} {\bibfnamefont {M.~H.-Y.}\ \bibnamefont {Cheung}}, \bibinfo {author} {\bibfnamefont {E.}~\bibnamefont {Berti}}, \bibinfo {author} {\bibfnamefont {V.}~\bibnamefont {Cardoso}}, \bibinfo {author} {\bibfnamefont {G.}~\bibnamefont {Carullo}}, \bibinfo {author} {\bibfnamefont {R.}~\bibnamefont {Cotesta}}, \bibinfo {author} {\bibfnamefont {W.}~\bibnamefont {Del~Pozzo}},\ and\ \bibinfo {author} {\bibfnamefont {F.}~\bibnamefont {Duque}},\ }\bibfield  {title} {\bibinfo {title} {{Agnostic black hole spectroscopy: Quasinormal mode content of numerical relativity waveforms and limits of validity of linear perturbation theory}},\ }\href {https://doi.org/10.1103/PhysRevD.108.104020} {\bibfield  {journal} {\bibinfo  {journal} {Phys. Rev. D}\ }\textbf {\bibinfo {volume} {108}},\ \bibinfo {pages} {104020} (\bibinfo {year} {2023})},\ \Eprint {https://arxiv.org/abs/2302.03050} {arXiv:2302.03050 [gr-qc]} \BibitemShut {NoStop}%
\bibitem [{\citenamefont {Mitman}\ \emph {et~al.}(2025)\citenamefont {Mitman} \emph {et~al.}}]{Mitman:2025hgy}%
  \BibitemOpen
  \bibfield  {author} {\bibinfo {author} {\bibfnamefont {K.}~\bibnamefont {Mitman}} \emph {et~al.},\ }\bibfield  {title} {\bibinfo {title} {{Probing the ringdown perturbation in binary black hole coalescences with an improved quasi-normal mode extraction algorithm}},\ }\Eprint {https://arxiv.org/abs/2503.09678} {arXiv:2503.09678 [gr-qc]}  (\bibinfo {year} {2025})\BibitemShut {NoStop}%
\bibitem [{\citenamefont {{SXS Collaboration}}()}]{SXS}%
  \BibitemOpen
  \bibfield  {author} {\bibinfo {author} {\bibnamefont {{SXS Collaboration}}},\ }\href@noop {} {\bibinfo {title} {{SXS Gravitational Waveform Database}}},\ \bibinfo {howpublished} {\url{http://www.black-holes.org/waveforms/}}\BibitemShut {NoStop}%
\bibitem [{\citenamefont {Scheel}\ \emph {et~al.}(2025)\citenamefont {Scheel} \emph {et~al.}}]{Scheel:2025jct}%
  \BibitemOpen
  \bibfield  {author} {\bibinfo {author} {\bibfnamefont {M.~A.}\ \bibnamefont {Scheel}} \emph {et~al.},\ }\bibfield  {title} {\bibinfo {title} {{The {SXS} Collaboration's third catalog of binary black hole simulations}},\ }\href@noop {} {\  (\bibinfo {year} {2025})},\ \Eprint {https://arxiv.org/abs/2505.13378} {arXiv:2505.13378 [gr-qc]} \BibitemShut {NoStop}%
\bibitem [{\citenamefont {Cook}(2020)}]{Cook:2020otn}%
  \BibitemOpen
  \bibfield  {author} {\bibinfo {author} {\bibfnamefont {G.~B.}\ \bibnamefont {Cook}},\ }\bibfield  {title} {\bibinfo {title} {{Aspects of multimode Kerr ringdown fitting}},\ }\href {https://doi.org/10.1103/PhysRevD.102.024027} {\bibfield  {journal} {\bibinfo  {journal} {Phys. Rev. D}\ }\textbf {\bibinfo {volume} {102}},\ \bibinfo {pages} {024027} (\bibinfo {year} {2020})},\ \Eprint {https://arxiv.org/abs/2004.08347} {arXiv:2004.08347 [gr-qc]} \BibitemShut {NoStop}%
\bibitem [{\citenamefont {Morey}\ \emph {et~al.}(2016)\citenamefont {Morey}, \citenamefont {Romeijn},\ and\ \citenamefont {Rouder}}]{Morey2016}%
  \BibitemOpen
  \bibfield  {author} {\bibinfo {author} {\bibfnamefont {R.~D.}\ \bibnamefont {Morey}}, \bibinfo {author} {\bibfnamefont {J.-W.}\ \bibnamefont {Romeijn}},\ and\ \bibinfo {author} {\bibfnamefont {J.~N.}\ \bibnamefont {Rouder}},\ }\bibfield  {title} {\bibinfo {title} {The philosophy of bayes’ factors and the quantification of statistical evidence},\ }\href {https://doi.org/10.1016/j.jmp.2015.11.001} {\bibfield  {journal} {\bibinfo  {journal} {Journal of Mathematical Psychology}\ }\textbf {\bibinfo {volume} {72}},\ \bibinfo {pages} {6} (\bibinfo {year} {2016})}\BibitemShut {NoStop}%
\bibitem [{\citenamefont {{Dickey}}(1971)}]{Dickey1971}%
  \BibitemOpen
  \bibfield  {author} {\bibinfo {author} {\bibfnamefont {J.~M.}\ \bibnamefont {{Dickey}}},\ }\bibfield  {title} {\bibinfo {title} {{The Weighted Likelihood Ratio, Linear Hypotheses on Normal Location Parameters}},\ }\href {https://doi.org/https://doi.org/10.1214/aoms/1177693507} {\bibfield  {journal} {\bibinfo  {journal} {Ann. Math. Statist.}\ }\textbf {\bibinfo {volume} {42}},\ \bibinfo {pages} {204} (\bibinfo {year} {1971})}\BibitemShut {NoStop}%
\bibitem [{\citenamefont {Zimmerman}\ \emph {et~al.}(2019)\citenamefont {Zimmerman}, \citenamefont {Haster},\ and\ \citenamefont {Chatziioannou}}]{Zimmerman:2019wzo}%
  \BibitemOpen
  \bibfield  {author} {\bibinfo {author} {\bibfnamefont {A.}~\bibnamefont {Zimmerman}}, \bibinfo {author} {\bibfnamefont {C.-J.}\ \bibnamefont {Haster}},\ and\ \bibinfo {author} {\bibfnamefont {K.}~\bibnamefont {Chatziioannou}},\ }\bibfield  {title} {\bibinfo {title} {{On combining information from multiple gravitational wave sources}},\ }\href {https://doi.org/10.1103/PhysRevD.99.124044} {\bibfield  {journal} {\bibinfo  {journal} {Phys. Rev. D}\ }\textbf {\bibinfo {volume} {99}},\ \bibinfo {pages} {124044} (\bibinfo {year} {2019})},\ \Eprint {https://arxiv.org/abs/1903.11008} {arXiv:1903.11008 [astro-ph.IM]} \BibitemShut {NoStop}%
\bibitem [{\citenamefont {Aky{\"u}z}\ \emph {et~al.}(2025)\citenamefont {Aky{\"u}z}, \citenamefont {Correia}, \citenamefont {Garofalo}, \citenamefont {Kacanja}, \citenamefont {Roy}, \citenamefont {Soni}, \citenamefont {Tan}, \citenamefont {Y}, \citenamefont {Nitz},\ and\ \citenamefont {Capano}}]{Akyuz:2025seg}%
  \BibitemOpen
  \bibfield  {author} {\bibinfo {author} {\bibfnamefont {A.}~\bibnamefont {Aky{\"u}z}}, \bibinfo {author} {\bibfnamefont {A.}~\bibnamefont {Correia}}, \bibinfo {author} {\bibfnamefont {J.}~\bibnamefont {Garofalo}}, \bibinfo {author} {\bibfnamefont {K.}~\bibnamefont {Kacanja}}, \bibinfo {author} {\bibfnamefont {L.}~\bibnamefont {Roy}}, \bibinfo {author} {\bibfnamefont {K.}~\bibnamefont {Soni}}, \bibinfo {author} {\bibfnamefont {H.}~\bibnamefont {Tan}}, \bibinfo {author} {\bibfnamefont {V.~J.}\ \bibnamefont {Y}}, \bibinfo {author} {\bibfnamefont {A.~H.}\ \bibnamefont {Nitz}},\ and\ \bibinfo {author} {\bibfnamefont {C.~D.}\ \bibnamefont {Capano}},\ }\bibfield  {title} {\bibinfo {title} {{Potential science with GW250114 -- the loudest binary black hole merger detected to date}},\ }\href@noop {} {\bibfield  {journal} {\bibinfo  {journal} {arXiv eprints}\ } (\bibinfo {year} {2025})},\ \Eprint {https://arxiv.org/abs/2507.08789} {arXiv:2507.08789 [gr-qc]} \BibitemShut {NoStop}%
\bibitem [{\citenamefont {Xin}\ \emph {et~al.}(2025)\citenamefont {Xin}, \citenamefont {Isi}, \citenamefont {Farr},\ and\ \citenamefont {Haiman}}]{Xin:2025voy}%
  \BibitemOpen
  \bibfield  {author} {\bibinfo {author} {\bibfnamefont {C.}~\bibnamefont {Xin}}, \bibinfo {author} {\bibfnamefont {M.}~\bibnamefont {Isi}}, \bibinfo {author} {\bibfnamefont {W.~M.}\ \bibnamefont {Farr}},\ and\ \bibinfo {author} {\bibfnamefont {Z.}~\bibnamefont {Haiman}},\ }\bibfield  {title} {\bibinfo {title} {{Identifying Compact Chirping SMBHBs in LSST using Bayesian Analysis}},\ }\href@noop {} {\bibfield  {journal} {\bibinfo  {journal} {arXiv eprints}\ } (\bibinfo {year} {2025})},\ \Eprint {https://arxiv.org/abs/2506.10846} {arXiv:2506.10846 [astro-ph.HE]} \BibitemShut {NoStop}%
\end{thebibliography}%

\end{document}